\documentclass[footinbib,a4paper,aps,prx,superscriptaddress,reprint,twocolumn,preprintnumbers,amsmath,amssymb,nobalancelastpage,10pt,longbibliography]{revtex4-2}
\usepackage{dutchcal}
\usepackage{amsmath, amsthm, amsfonts, amssymb,amstext}
\usepackage{mathptmx}
\usepackage{mathrsfs}
\usepackage[utf8]{inputenc}
\usepackage[english]{babel}
\usepackage[T1]{fontenc}
\usepackage{amsmath}
\usepackage{hyperref}
\usepackage{standalone}
\usepackage{import}
\usepackage{multirow}
\usepackage{diagbox}
\usepackage{tikz}
\usepackage{adjustbox}
\usepackage{braket}
\usepackage[caption=false]{subfig}

\usepackage{forest}
\usepackage[table]{xcolor}
\definecolor{myblue}{rgb}{0.2,0.2,0.8}
\definecolor{myred}{rgb}{1,0.,0.3}
\usepackage{hyperref}
\hypersetup{
  colorlinks   = true, 
  urlcolor     = blue, 
  linkcolor    = blue, 
  citecolor   = blue
}

\usetikzlibrary{quantikz2}
\usetikzlibrary{external}
\usepackage{algpseudocode}
\algnewcommand\algorithmicnot{\textbf{not}}
\algdef{SE}[IF]{IfNot}{EndIf}[1]{\algorithmicif\ \algorithmicnot\ #1\ \algorithmicthen}{\algorithmicend\ \algorithmicif}

 \def\ee{\mathord{\rm e}}
 
 \def\ii{\mathord{\rm i}}
\def\min{\mathord{\rm min}}

\def\beq{\begin{equation}}
\def\eeq{\end{equation}}
\def\barray{\begin{eqnarray}}
\def\earray{\end{eqnarray}}

\begin{document}
\title{ Scaling roadmap for  modular trapped-ion QEC  and lattice-surgery  teleportation}
\author{César Benito}
\affiliation{Instituto de F\'isica Te\'orica UAM-CSIC, Universidad Aut\'onoma de Madrid, Cantoblanco, 28049, Madrid, Spain}
\author{Alfredo Ricci Vasquez}
\affiliation{Institute for Quantum Electronics, ETH Z\"{u}rich, 8093 Z\"{u}rich, Switzerland}
\author{Jonathan Home}
\affiliation{Institute for Quantum Electronics, ETH Z\"{u}rich, 8093 Z\"{u}rich, Switzerland}
\author{Karan K. Mehta}
\affiliation{School of Electrical and Computer Engineering, Cornell University, Ithaca, NY 14853}
\author{Thomas Monz}
\affiliation{Universit\"{a}t Innsbruck, Institut f\"{u}r Experimentalphysik, Innsbruck, Austria}
\author{Markus M\"{u}ller}
\affiliation{Institute for Quantum Information, RWTH Aachen University, Aachen, Germany}
\affiliation{Institute for Theoretical Nanoelectronics (PGI-2), Forschungszentrum J\"{u}lich, J\"{u}lich, Germany}
\author{Alejandro Bermudez}
\affiliation{Instituto de F\'isica Te\'orica UAM-CSIC, Universidad Aut\'onoma de Madrid, Cantoblanco, 28049, Madrid, Spain} 

\begin{abstract}
    We  present a   footprint study for the scaling of modular quantum error correction (QEC) protocols designed for  triangular color codes, including  a lattice-surgery-based logical teleportation gadget, and  compare the performance of   various possible architectures based on trapped ions. The differences in these architectures arise from the technology   that enables the   connectivity between  physical qubits and the modularity required for the QEC gadgets, which is either based on laser-beam deflectors focused to independent modules hosting mid-size ion crystals, or  integrated photonics guided to  segmented modules of the trap and allowing for the manipulation of smaller ion crystals. Our approach integrates the transpilation of the QEC gadgets into native trapped-ion primitives   
    and a detailed  account of the specific laser addressing  and ion transport  leading to  different amounts of  crosstalk errors, motional excitation and idle qubit errors. Combining a microscopically-informed noise model with an efficient Pauli-frame simulator and different scalable decoders, we assess the near-term performance of the color-code memory and teleportation protocols on these  architectures. Our analysis demonstrates that modular color-code teleportation  is achievable in these near-term  trapped-ion architectures, and identifies the integrated-photonics connectivity as the most promising route for longer-term scaling. 
\end{abstract}
\maketitle

\setcounter{tocdepth}{2}
\begingroup
\hypersetup{linkcolor=black}
\tableofcontents
\endgroup

\section{\bf Introduction}
 Quantum error correction (QEC)  based on stabilizer codes~\cite{gottesman1997stabilizercodesquantumerror} is one of the leading approaches towards large-scale fault-tolerant (FT) quantum computers~\cite{Nielsen_Chuang_2010}. 
For instance, the surface code~\cite{KITAEV20032,bravyi1998quantumcodeslatticeboundary,10.1063/1.1499754,Fowler2012} has become an archetype of topological QEC~\cite{Terhal_2015,Campbell2017}, as it offers a high error threshold under circuit-level noise~\cite{PhysRevA.89.022321,PhysRevA.80.052312,10.5555/2011362.2011368}, while simplifying technological requirements by  requiring only nearest-neighbor connectivities for stabilizer readout. These features have motivated several experimental realizations~\cite{Krinner2022,Acharya2023,PhysRevLett.131.210603,Bluvstein2024,Acharya2025,remm2025experimentallyinformeddecodingstabilizer,bluvstein2025architecturalmechanismsuniversalfaulttolerant,vezvaee2025surfacecodescalingheavyhex}, including an  experimental hallmark in QEC: the sustained suppression of errors of a logical qubit that is redundantly encoded in a topological quantum memory~\cite{Acharya2025,bluvstein2025architecturalmechanismsuniversalfaulttolerant}. This is achieved through a linear increase in the code distance $d$, which  has  a quadratic cost in the number of physical qubits. Other topological  codes provide complementary advantages when going beyond   a quantum memory, such as  simplifying some of the logical operations on the encoded logical qubits. For instance,  the two-dimensional triangular color codes~\cite{Bombin2007,bombin2013introductiontopologicalquantumcodes} support a richer set of transversal  gates~\cite{Bombin2007,bombin2015gaugecolorcodesoptimal,PhysRevA.91.032330}, enabling a  direct implementation of the whole Clifford group at the logical level~\cite{PhysRevLett.102.110502,PhysRevLett.110.170503,PhysRevLett.78.2252}, which can lead to  considerable savings in resources~\cite{PhysRevResearch.6.043125}. These specific advantages come at the cost of deeper FT circuits for stabilizer readout~\cite{548464,PhysRevLett.77.3260,Knill2005,landahl2011faulttolerant,Chamberland_2020,PRXQuantum.2.020341,PRXQuantum.5.020336,gidney2023newcircuitsopensource}, as well as  more complex decoding strategies~\cite{wang2009graphicalalgorithmsthresholderror,PhysRevA.89.012317,chubb2021generaltensornetworkdecoding,PRXQuantum.2.020341,Kubica2023efficientcolorcode}, particularly under circuit-level noise~\cite{landahl2011faulttolerant,stephens2014efficientfaulttolerantdecodingtopological,Sahay2022,gidney2023newcircuitsopensource,PRXQuantum.5.020336,zhang2024facilitatingpracticalfaulttolerantquantum,Lee2025colorcodedecoder}. This overhead contrasts  the simplicity  and efficiency of the surface code readout~\cite{PhysRevA.90.062320}, decoding~\cite{wang2009graphicalalgorithmsthresholderror,deMartiiOlius2024decodingalgorithms} and estimated  resource counts~\cite{Litinski2019gameofsurfacecodes}. Recent experiments have  also demonstrated the potential of various aspects of color-code based QEC~\cite{Nigg2014,PhysRevX.11.041058,Postler2022,ryananderson2022implementingfaulttolerantentanglinggates,paetznick2024demonstrationlogicalqubitsrepeated,ryananderson2024highfidelityfaulttolerantteleportationlogical,PRXQuantum.5.030326,mayer2024benchmarkinglogicalthreequbitquantum,Bluvstein2024,Lacroix2025,Pogorelov2025,bluvstein2025architecturalmechanismsuniversalfaulttolerant}, including historical hallmarks in QEC such as the first realization of a topological QEC code~\cite{Nigg2014}, the first realization of a universal gate set at the logical level~\cite{Postler2022}, and also the  suppression of errors by increasing code distance~\cite{Lacroix2025}.

Recent works in color-code QEC have leveraged simplified approaches for FT syndrome extraction, which are particularly relevant  in the low-distance regime that can be achieved in current quantum processing units (QPUs). Flag-qubit schemes~\cite{Chao_prl_2018,Chao2018,Reichardt_2021,Chamberland2018flagfaulttolerant}  enable a FT stabilizer readout using minimal ancilla resources, which are used to detect the propagation of
potentially-dangerous high-weight errors during the stabilizer readout that are in conflict with fault tolerance. In contrast to previous FT schemes~\cite{doi:10.1137/S0097539795293172,PhysRevLett.77.3260,Knill2005}, flag-qubit methods with just two ancilla qubits suffice to detect and correct arbitrary  faults at a prescribed FT level for small-distance codes~\cite{Chao_prl_2018,Chao2018,Reichardt_2021,Chamberland2018flagfaulttolerant}, and can also be  generalized to larger distances~\cite{Chamberland2018flagfaulttolerant,Kubica2023efficientcolorcode}. In the context of trapped\rm -ion  QPUs~\cite{HAFFNER2008155,10.1063/1.5088164,10.1116/1.5126186}, flag qubits can provide an advantage~\cite{PhysRevA.100.062307,PhysRevA.99.022330,ParradoRodriguez2021} over other FT syndrome extractions~\cite{PhysRevX.7.041061}.  Trapped-ion experiments  realizing the flag-qubit scheme were conducted in parallel: Hilder~\emph{et al.} demonstrated a FT flag-based measurement of a weight-4 parity check  in a shuttling-based QCCD trapped-ion device~\cite{Hilder2022}, while Ryan-Anderson~\emph{et al.} presented major advances by integrating flag-based QEC for a full $d=3$ color code with mid-circuit measurements,  real-time decoding, and conditional logic~\cite{PhysRevX.11.041058}.  Postler~\emph{et al.}~\cite{Postler2022} and  Ryan-Anderson~\emph{et al.}~\cite{ryananderson2022implementingfaulttolerantentanglinggates} implemented transversal entangling gates  between two $d=3$ color codes using flag-based encoding~\cite{Goto2016}, as well as  a universal set of FT operations~\cite{Postler2022}, including the preparation  of a magic $|T\rangle$-state using flag-based methods~\cite{Goto2016,chamberland2019faulttolerantmagic}.

Another key technique to perform FT logical operations in topological QEC codes is \emph{lattice surgery} (LS)~\cite{Horsman_2012}, which exploits local measurements of joint multi-qubit operators along shared boundaries of neighboring code  blocks to perform entangling  gates at the logical level. These techniques can likewise be exploited for the teleportation of arbitrary logical states. Originally developed for surface codes~\cite{Horsman_2012,Litinski2018latticesurgery,fowler2019lowoverheadquantumcomputation,Litinski2019gameofsurfacecodes,PRXQuantum.3.010331}, LS can also be adapted to color codes by merging and splitting local  patches of the logical qubits~\cite{Landahl2014,PhysRevResearch.6.043125},  enabling the  implementation of  logical gates  in a fully modular fashion. By means of  joint-measurement protocols,  code deformation and LS have also been experimentally demonstrated with two single-plaquette patches of the surface code~\cite{Erhard2021}, and two distance $d=5$ surface-code logical qubits~~\cite{bluvstein2025architecturalmechanismsuniversalfaulttolerant}. Fault-tolerant LS  and logical-state teleportation between color-code qubits have been recently reported in superconducting QPUs,  demonstrating the transfer of encoded logical states with fidelities up to $90\%$~\cite{Lacroix2025}. In the context of trapped-ion color codes,  Ryan-Anderson~\emph{et al.}~\cite{ryananderson2024highfidelityfaulttolerantteleportationlogical} have reported the similar fidelities in the teleportation of a logical qubit between two $d=3$ color codes. This was achieved by a single measurement of the joint logical operator which, as we show below, must be carefully assessed in light of fault tolerance.  Moreover, when scaled to larger codes, we show in Sec.~\ref{sec:ls-teleportation-scheme} that the direct measurement approach has a worse performance in comparison to the LS strategy.

Scaling up these QEC demonstrations towards larger-distance codes can benefit from a {modular approach with self-contained logical units providing independent capabilities  for encoding, syndrome extraction, or error correction,  and interacting via specific  FT interfaces using LS or other ancilla-mediated operations. Likewise,  exploring hardware-efficient strategies for these modular approaches that optimizes the required quantum resources is important, which requires a realistic modeling of the device capabilities as well as a detailed account of the noise sources. Considering the available qubit resources in light of current capabilities, it will also be important to operate well below the corresponding error thresholds, such that one can reach target logical fidelities with a reasonable overhead of physical qubits, the so-called QEC footprint. Trapped-ion  QPUs are naturally suited to this research and development program, as they offer high-fidelity operations~\cite{PhysRevLett.113.220501,PhysRevLett.117.060504,PhysRevLett.117.060505,PhysRevLett.127.130505}, and a flexible entangling-gate connectivity via ion transport or by individual addressing of ion pairs in larger Coulomb crystals. This can be combined with longer-range connectivity via photonic interconnects~\cite{PhysRevA.89.022317} or inter-module transport~\cite{Akhtar2023}. The quantum  charge-coupled device (QCCD) architecture~\cite{Kielpinski2002} enables reconfigurable modular systems via segmented linear traps and ion shuttling~\cite{Lekitsch2017}.  Home \emph{et al.} demonstrated for the first time that all the basic building blocks for a scalable trapped-ion QCCD could be integrated experimentally in a segmented linear Paul trap~\cite{doi:10.1126/science.1177077}. Later, Pino \emph{et al.} combined multi-zone ion transport and ion swaps  with high-fidelity gates to achieve a higher qubit connectivity in
a cryogenic surface trap~\cite{Pino2021}.

Finding the optimal hardware combination of shuttling vs addressing is actually complicated by the fact that there are also multiple variants of the basic QEC gadgets, such as state preparation and syndrome extraction. Running all possible combinations in real hardware with the different ion-trap designs required to combine these modular  approaches  would entail a significant overhead in time, resources and costs. Therefore, developing realistic  noisy simulations to make comparative studies of the various QEC strategies and modular architectures is a useful and valuable tool to make comparative studies assessing the QEC performance,  guide hardware development towards efficient QEC, and  identify the most promising strategies to only execute  the best-performing circuits experimentally. We focus  on logical state teleportation, as it represents a fundamental primitive for more complex FT protocols. As will become clearer below, logical teleportation underlies key tasks such as LS–based quantum logic, entanglement distribution, and modular interconnects between logical qubits all of which are key for the progress of scaling of FT quantum computing.

The goal of this work is to update the microscopic noise modeling and  performance assessment presented in~\cite{PhysRevX.7.041061,PhysRevA.100.062307,PhysRevA.99.022330,ParradoRodriguez2021} to explore modular approaches for flag-qubit QEC and LS gadgets, making a detailed comparison of shuttling-based and laser-addressing  schemes, and extending previous studies towards larger register sizes. This will allow us on the one hand to identify the best strategy in the near term, and on the other hand to estimate the respective QEC footprints for a trapped-ion color-code memory and logical  QPUs when scaling up the respective code distance.  

To achieve this goal, we start by reviewing the color-code approach to quantum memories in Sec.\ref{sec:color_code}, pointing to some overlooked aspects regarding FT in flag-based QEC. We discuss FT state-preparation circuits, various syndrome-extraction strategies with dynamic circuits, and decoding methods ranging from lookup tables to belief-propagation and restricted minimum-weight perfect matching decoders, while identifying subtle caveats with flag-based fault tolerance. In Sec.~\ref{sec:teleportation}, we extend these ideas to lattice-surgery teleportation between color-code blocks, comparing direct joint-measurement and merge–split approaches and quantifying their respective QEC footprints. This section also introduces and benchmarks different modular trapped-ion architectures—ranging from beam-deflector–based linear traps to segmented and junction-based devices—and the corresponding circuit transpilation and microscopic noise modeling used to assess performance. Finally, in Sec.~\ref{sec:conclusions}, we summarize the results and discuss their implications for the scaling roadmap of fault-tolerant trapped-ion QEC.

\section{\bf The color code as a quantum memory}
\label{sec:color_code}

QEC codes can be defined by parameters $[[n, k, d]]$ encoding $k$ logical qubits into a Hilbert space of $n$ physical qubits, and providing protection against  $t = \lfloor (d-1)/2 \rfloor$ arbitrary single-qubit errors~\cite{Nielsen_Chuang_2010}. In the stabilizer formalism~\cite{gottesman1997stabilizercodesquantumerror}, the QEC code can be described using an Abelian subgroup $\mathcal{S} \subset \mathcal{P}_{n}$ of the $n$-qubit Pauli group generated by $\mathcal{P}_{n} = \langle  \ii , X, Z \rangle^{\otimes n}$, which includes all tensor products of single-qubit Pauli operators with a possible complex phase. The generators are a set of $G = n - k$  commuting operators $\{g_1, g_2, \dots, g_G\}$, each  satisfying $g_i^2 = I$ and defining a subspace $\mathcal{V}_\mathcal{C} \subset \mathcal{H}=\mathbb{C}^{2n}$  by
\beq
\mathcal{V}_{\mathcal{C}} = \left\{ |\psi\rangle \in \mathbb{C}^{2n} \ : \ g_i |\psi\rangle = |\psi\rangle, \ \forall\, i = 1, \dots, G \right\}.
\eeq
The dimension of $\mathcal{V}_{\mathcal{C}}$ is $2^{n-G}$, allowing it to encode $k=n-G$ logical qubits. Operators that preserve $\mathcal{V}_{\mathcal{C}}$ but are not elements of $\mathcal{S}$ act as logical gates $\{ X_{L,\ell}, {Z}_{L,\ell} \}_{\ell=1}^k$, commuting with all stabilizer generators. They act within the code space and obey the algebra $
{X}_{L,\ell}^2 ={Z}_{L,\ell}^2 = I, \quad {X}_{L,\ell} {Z}_{L,\ell'} = (-1)^{\delta_{\ell\ell'}} {Z}_{L,\ell'} {X}_{L,\ell}$.

In the case of topological stabilizer codes, such as the triangular color codes~\cite{Bombin2007,bombin2013introductiontopologicalquantumcodes}, qubits are arranged on a lattice with locality constraints on stabilizer support. Specifically, in the color code defined on a trivalent, three-colorable planar tiling, each plaquette \(p\) of the lattice defines two stabilizer generators: one composed of Pauli-\(X\) operators, and another of Pauli-\(Z\) operators, each acting on the set of vertices \(v(p)\) that define the plaquette
\beq
\label{eq:stabilizers}
S^X_{p} = \prod_{q \in v(p)} X_q, \qquad S^Z_{p} = \prod_{q \in v(p)} Z_q.
\eeq
These plaquette operators define a  QEC code with topological protection, where logical information is stored globally across the lattice in the form of long-range  entanglement. We will focus on a family of triangular color codes defined over a uniform hexagonal lattice (see Fig.~\ref{fig:colorcodes}). This choice ensures that the stabilizer weight is at most six, in contrast to other tilings~\cite{landahl2011faulttolerant}. For triangular color codes, boundaries can always be colored with the complementary color to that of the plaquettes, and a single $k=1$ logical qubit can be hosted in the code subspace $\mathcal{V}_{\mathcal{C}}={\rm span}\{\ket{0}_L,\ket{1}_L\}$  with logical states specified by string-net-type correlations $Z_L\ket{0}_L=\ket{0}_L,\,Z_L\ket{1}_L=-\ket{1}_L$ connecting the different boundaries $Z_L=Z_0\otimes Z_1\otimes Z_4$~\cite{PhysRevLett.97.180501}. Likewise, the logical Hadamard states $\ket{\pm}_L=\tfrac{1}{\sqrt{2}}(\ket{0}_L\pm\ket{1}_L)$  are specified by string-net-type correlations $X_L\ket{+}_L=\ket{+}_L, X_L\ket{-}_L=-\ket{-}_L$ with $X_L=X_0\otimes X_1\otimes X_4$ (see Fig.~\ref{fig:colorcodes})~

\begin{figure}
\includegraphics[scale=0.55]{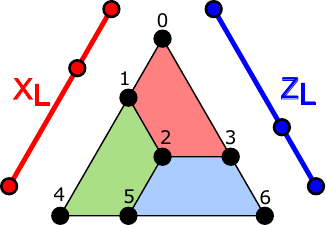}\qquad
\includegraphics[scale=0.55]{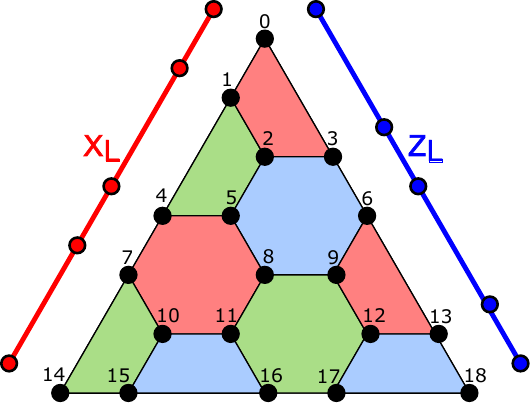}
\caption{{\bf Hexagonal color code family}: Qubit layout of the first two instances of the hexagonal color code. Plaquettes in the lattice are tri-colored such that adjacent plaquettes have different colors. Each plaquette defines both an $X-$type and $Z-$type stabilizer. Chains of $X$ or $Z$ operators along a boundary side define logical $X$ and $Z$ operators, respectively. The [[7, 1, 3]] code, also called the Steane code~\cite{PhysRevLett.77.793}, only contains weight-4 stabilizers. The [[19,1,5]] and bigger color codes include as well hexagonal plaquette stabilizers in the bulk.}
\label{fig:colorcodes}
\end{figure}

The color code is a  CSS code~\cite{Calderbank1996,Steane1996} with purely 
$X$- and  
$Z$-type stabilizers~\eqref{eq:stabilizers}. The parity check matrices   $H_{X}=[h_{ij}^x],H_{Z}=[h_{ij}^z]$ set the valid codewords through a collection of linear constraints, and   have $\mathbb{Z}_2$-valued matrix elements defined through $g_i^x=\bigotimes_j(X_j)^{h_{ij}^x},g_i^z=\bigotimes_j(Z_j)^{h_{ij}^z}$. The  parity checks correspond to two classical linear codes that must fulfill $H_{X}^{\phantom{t}}H_{Z}^{\rm T}=0$ in order for the two types of stabilizers to commute. Since the color code is of CSS type,
the transversal CNOT gate between two color-code blocks implements the logical CNOT gate  directly. In addition,  2D color codes are self-dual $H_X=H_Z$, providing a transversal Hadamard gate ${H}_L=\otimes_q({X}_q+{Z}_q)/\sqrt{2}$. This offers a contrast to the surface code~\cite{KITAEV20032,bravyi1998quantumcodeslatticeboundary,10.1063/1.1499754,Fowler2012}, in which the  $X$- and $Z$-  stabilizers have a different support, and the qubit-wise Hadamard maps onto the dual code. This requires a redefinition of the stabilizers and an adaptive decoding strategy, posing further real-time constraints. Color codes in $D$ dimensions also allow a logical transversal rotation about the $z$ axis $R_{{Z}_L}(\theta)=\otimes_q{\rm exp}\{\ii\theta{Z}_q/2\}$ of angle $\theta=2\pi/2^{D}$~\cite{Bombin2007,bombin2015gaugecolorcodesoptimal,PhysRevA.91.032330}. Thus, the whole Clifford group is transversal for the $D=2$ color code as this $\theta=\pi/2$ rotation leads to an ${S}_L$ gate, lowering  the use of expensive magic states with respect to the surface code to only the $T$ gate. Note that the ${S}_L$ gate is transversal for hexagonal color codes of Fig.~\ref{fig:colorcodes}, but it cannot be implemented as a global $S^\dagger$ gate affecting all qubits~\cite{PhysRevLett.97.180501,Bombín_2015}, as occurs for the square-octagon tiling. Instead, it requires individual addressing to apply $S$ instead of $S^\dagger$ in some qubits. To complete a universal gate set, a $\overline{T}=R_{{Z}_L}(-\pi/4)$ gate is also required, which can be achieved by either magic state injection~\cite{PhysRevA.71.022316}, or  by code switching~\cite{PRXQuantum.5.020345,Pogorelov2025,dck4-x9c2} to the 3D color code}, for which the $T$ gate, but not the $H$ gate,  is transversal~\cite{bombin2015gaugecolorcodesoptimal,PhysRevA.91.032330}.

Once we have reminded the reader about the color-code approach, let us discuss a few  aspects of the specific circuits, our Pauli-frame simulation toolkit, and mention several caveats with FT in CSS codes that, to the best of our knowledge,  have passed unnoticed in previous literature.  In particular, we find that some flag-based QEC gadgets  remain fault-tolerant in isolation, but compromise FT when followed by transversal gates.  Some  weight-2 errors that can be corrected by exploiting the  CSS nature of the codes, transform into uncorrectable correlated errors  by subsequent transversal gates, breaking fault tolerance unless a full QEC round is applied after these transversal operations.

\subsection{Flag-type verified logical state preparation}
The general method to prepare a CSS code in the $\ket{0}_L$ state involves initializing all physical qubits in $\ket{\psi}=\otimes_q\ket{0_q}$, such that the state is already  a $+1$ eigenstate of all $Z$-type stabilizers as well as ${Z}_L$, followed by a measurement of all $X$-type stabilizers~\cite{gottesman1997stabilizercodesquantumerror}, namely the $X$-type plaquette operators in the color code~\eqref{eq:stabilizers}. The measurement outcomes will be $\pm 1$ randomly, but they always project the code to a simultaneous eigenstate of all stabilizers as well as the logical ${Z}_L$ . The $\ket{0}_L$ state is then obtained after applying  the corresponding Pauli corrections or, simply, by a redefinition of the Pauli frame~\cite{Knill2005,Terhal_2015} as a classical record to be used in classical post-processing or future decoding. A similar procedure can also be done to prepare  $\ket{+}_L$  by initializing all physical qubits in $\ket{\psi}=\otimes_q\ket{+_q}$ and measuring the $Z$-type stabilizers. From these states, other logical states can be achieved  by applying transversal gates. For example, in the 2D color code, ${Y}_L$ eigenstates can be prepared by applying a transversal $S$ gate to $\ket{+}_L$. The use of the transversal ${X}_L$ and ${Z}_L$ allow for the generation of the remaining logical states in the set of 6 cardinal states.

\begin{figure}
\includegraphics[width=\linewidth]{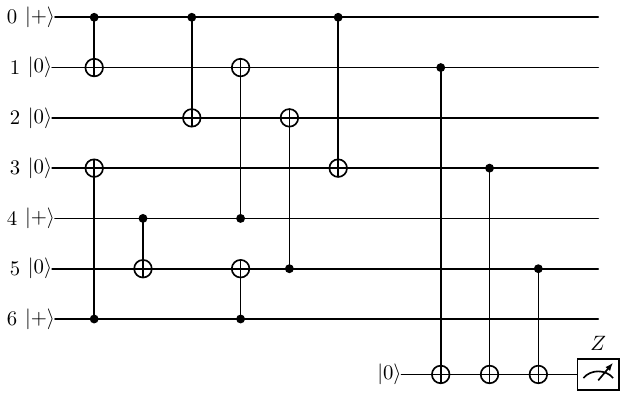}
\caption{{\bf Flag-type verified logical state preparation:} preparation of a logical $\ket{0}$ state for the Steane code~\cite{Goto2016}. The circuit uses a non-FT encoding to create the desired state, followed by a verification step. The verification circuit implements a logical $Z$ measurement which detects any fault that propagated to more than one data qubit, causing an uncorrectable error. If the measurement outcome is flipped, the prepared qubit is discarded.}
\label{fig:steaneprep}
\end{figure}

While the above method is valid for  arbitrary CSS codes, it requires a FT measurement of the corresponding stabilizers, which can increase the complexity of the circuits and  qubit resources~\cite{PhysRevLett.78.2252,Knill2005}. Otherwise, correlated errors could spread into the same logical block, lowering the effective distance of the code, or even invalidating the  corrections of subsequent QEC rounds. It is also possible to use a non-FT circuit, but this must then be followed by a verification step~\cite{548464,doi:10.1137/S0097539795293172,PhysRevLett.77.3260,PhysRevLett.98.020501}, ensuring thus that no dangerous error  spreads into multiple physical qubits,  which would  otherwise  reduce the code distance. If the verification fails, the circuit run must be discarded,  preserving the FT level of the QEC code at the cost of increasing the runtime. In the specific case of $d=3$ color code, which is the stabilizer version of Steane's 7-qubit code~\cite{PhysRevLett.77.793,doi:10.1098/rspa.1996.0136}, a non-FT circuit for the preparation  of $\ket{0}_L$ is presented in Fig.~\ref{fig:steaneprep}. The verification step uses a single flag-type qubit to identify the potential occurrence of dangerous errors, discarding the corresponding runs where the flag-qubit readout is raised~\cite{Goto2016}. We note that automated learning algorithms to find  FT circuits for logical state preparation   of other codes have also been   developed recently~\cite{zen2024quantumcircuitdiscoveryfaulttolerant,PRXQuantum.6.020330}, which might also be generalized to larger distances.  We also note that additional measurements conditioned on the flag-qubit readout can  be applied to achieve FT without  post-selection~\cite{PhysRevA.99.022330}. The FT preparation of  $\ket{+}_L$ can be achieved by applying a transversal qubit-wise Hadamard gate.

The gates in the encoding circuit, as well as the idle periods and the measurements, will be subject to noise and control errors. For circuit-level noise, deviations from these ideal  operations are typically modeled using Pauli channels: completely-positive trace-preserving maps~\cite{Nielsen_Chuang_2010} that follow the ideal operations and act on the resulting density matrix as 
\begin{equation}
\label{eq:pauli_channel}
\mathcal{E}_{\rm P}(\rho) = \sum_{\alpha=0}^{4^n - 1} p_\alpha E_\alpha \rho E_\alpha^\dagger, \quad \text{with} \quad \sum_{\alpha} p_\alpha = 1.
\end{equation}
Here,  the noise operators are drawn from the Pauli group
\begin{equation}
E_\alpha \in \{I, X, Y, Z\}^{\otimes n}, \quad \text{with} \quad E_0 = I \otimes I \otimes \cdots \otimes I,
\end{equation}
and the error rates $p_\alpha\in[0,1]$ determine the probability with which a specific type of error occurs, such that $p_0=1-\sum_{\alpha\neq0}p_\alpha$ is the chance that no error  occurs at all. 

The   standard circuit-level error model (SCEM) used to evaluate the performance of a particular QEC code  is based on a uniform depolarizing channel, which does not single out any specific noise direction, nor any particular circuit operation. Noisy single-qubit  gates, idle periods, or qubit-wise measurements are modeled by  applying a single-qubit $n=1$ Pauli channel  with $p_x=p_y=p_z=p/3$ after the ideal operation, where $p\in[0,1]$ is the overall depolarizing rate. For entangling two-qubit gates, the depolarizing channel now  contains all Pauli errors in the set $\left\{I,X,Y,Z\right\}^{\otimes2}\setminus{I\otimes I}$. Each error has probability $p_{\alpha}=p/15$, adding up to a total depolarizing probability $p=1-p_0$. We thus see that this noise model quantifies the errors with a single parameter $p$ regardless of the specific nature of the quantum operation. This is a very useful model to test for fault-tolerance at the circuit level, and for making generic comparisons of various QEC strategies or even across different QEC codes. On the other hand, it falls short of a realistic account of errors in actual  QPUs, and must be superseded by more elaborate error models if one aims at quantifying the real performance of a certain QEC code in a certain QPU.

In our numerical simulations, we utilize a Pauli-frame approach tailored for Clifford circuits~\cite{Gidney2021stimfaststabilizer}, which improves upon stabilizer simulations as one does not operate on the full stabilizer group table representing the quantum state, but only updates  the error frame. 
The specific tool that we have developed  in this context is  a sparse variation of a high-performance stochastic simulator of Clifford noisy circuits called  stim~\cite{Gidney2021stimfaststabilizer}, the details of which are  described in Appendix~\ref{sec:sparsesim}. Our method  only tracks  those circuit runs that are actually affected by errors. This reduces considerably the computational overhead, as we do not spend time or memory updating error-free circuit runs, i.e. shots, particularly for the low physical error rates required for beneficial QEC.  In analogy to stim, noise events are sampled from a geometric probability distribution associated   to the noise channel error rate $p$ of the quantum operation, instead of sampling over the Bernoulli trials for every single shot. In this way, we  can capture  the number of error-free runs before the next error occurs. Using a sparse representation of the errors enables the efficient sorting of error occurrences among all of the circuit runs involved in a certain QEC protocol. Each sampled error is then applied as a random Pauli operator drawn uniformly from the non-identity Pauli set of the SCEM, and implemented through  a simple Pauli-frame update after propagating it through the Clifford gates involved in the circuit. Furthermore, our simulation toolkit also upgrades stim by incorporating  dynamic-circuit branching with in-sequence logic and feed-forward operations, allowing adaptive QEC  based on flag qubits to be faithfully simulated. Details on the noise sampling algorithm, Pauli frame updates, and the structured representation of complex tree-like logical structures are given  in Appendix~\ref{sec:sparsesim}.

\begin{figure}
\includegraphics[width=\linewidth]{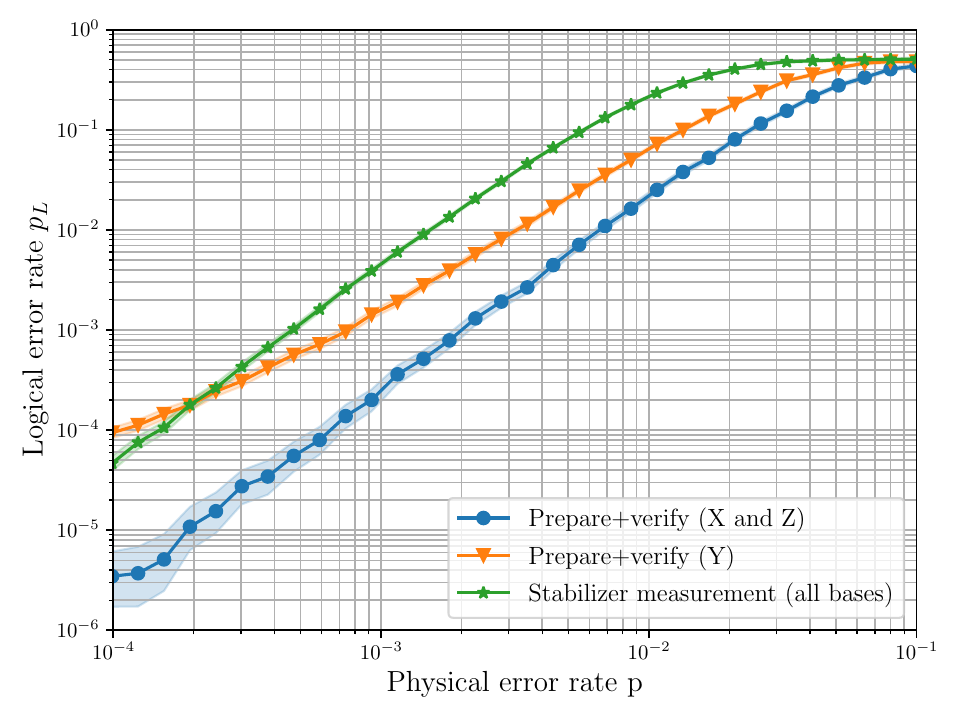}
\caption{{\bf Color code state preparation fidelity}: logical error rates achieved by different state preparation protocols for the color code. Error rates are determined under standard circuit-level noise and lookup table decoding. For any CSS code, a codeword can be prepared by measuring the stabilizers to project to the code space. The green line shows the performance of this approach for the color code. A state preparation based on preparation+verification is also benchmarked. This latter method provides higher fidelities for the X and Z basis eigenstate preparation (blue line), but it is non-FT when preparing a $\ket{{\rm +i}}$ state (orange line). For that state, the syndrome extraction scheme has to be used, incurring in a performance cost.}\label{fig:preparationcomparison}
\end{figure}

In Fig.~\ref{fig:preparationcomparison}, we present the results of our sparse stim simulator  for the error rate  
of  Steane's 7-qubit code encoding into  3 characteristic cardinal states    using  two different  strategies. The green curve represents the stabilizer measurement-based approach, which yields equal state-preparation fidelity for  $\ket{0}_L,\ket{+}_L$ when considering the self-dual CSS code and the depolarizing SCEM. For the performance of the flag-type verification  approach, represented  in blue, the preparation fidelity of $\ket{0}_L,\ket{+}_L$  as a function of the depolarizing rate is also equal, showing the same slope with $p$ as the stabilizer measurement-based approach. This confirms the FT character of the flag-based verification. From an application perspective, this figure also shows the  considerable advantage of   the flag-based verification method, i.e. roughly two orders of magnitude at realistic $p=0.1\%$ physical errors. As will be explained in more detail in the following subsection, this is a consequence of the  overhead of quantum operations required to achieve a FT readout of the stabilizers, even when using a flag-based approach.  

Let us now consider the preparation of the logical states $\ket{\pm{\ii}}_L=\tfrac{1}{\sqrt{2}}(\ket{0}_L\pm\ii\ket{1}_L)$, which can  appear to be straightforward by exploiting the transversality of the ${S}_L$ gate in this $d=3$ color code. One would naively apply a qubit-wise operation to the previous flag-based preparation of the $\ket{+}_L$ logical state.
However, as we will explain in Sec.~\ref{sec:ftcaveats} in more detail, the flag-based  method is actually not FT for the $\ket{\rm +i}$ state, which can be seen by noticing the different orange slope of the figure. This is a consequence of the general fact that the spread of multiple errors  during the encoding CNOT gates, that might appear as  mild due to the  CSS nature of a code, can actually change upon application of further transversal gates. This would make them dangerous  at a later stage of a circuit and effectively reduce the distance of the code.

\begin{figure*}
\subfloat[\label{fig:xxxx}]{\includegraphics[width=.3\linewidth]{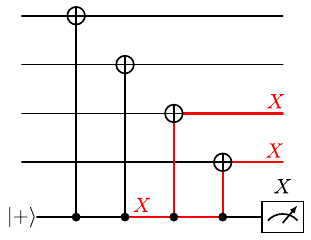}}
\subfloat[\label{fig:flagged}]{\includegraphics[width=.33\linewidth]{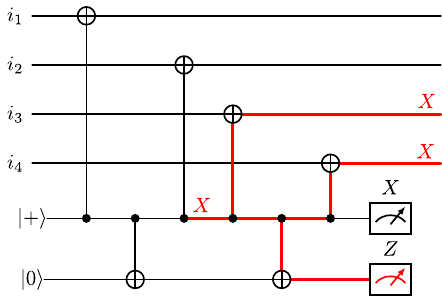}}\quad
\subfloat[\label{fig:superdensecircuit}]{\includegraphics[width=.33\linewidth]{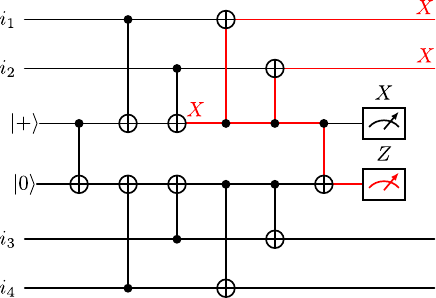}}
\caption{{\bf Syndrome extraction circuits for the color code.} a) Syndrome extraction circuit for a weight-4 $X-$type stabilizer with a single ancilla qubit. A single error happening in the ancilla qubit can propagate to multiple data qubits, causing an uncorrectable error, as indicated by the red wires. b) Flagged syndrome extraction circuit. An error in the ancilla qubit could propagate to two data qubits causing a non-correctable error. By adding a flag qubit, the weight-2 error can be detected and corrected. c) Superdense syndrome extraction circuit: fault tolerant circuit for simultaneous extraction of $X$ and $Z$ stabilizers, using an ancilla Bell pair~\cite{gidney2023newcircuitsopensource}. Any dangerous error happening during the circuit is detected by either ancilla qubit.}
\end{figure*}

\subsection{Syndrome extraction and dynamic circuits}
\label{sec:syndrome_extraction}

Syndrome extraction requires the measurement of all the code stabilizers~\eqref{eq:stabilizers},  providing  an error syndrome that is then fed into a decoder, such as  a look-up-table for small-distance codes, in order to identify the most-likely low-weight error that may have occurred and the operation or Pauli frame update that would revert it. 
Several quantum operations must be performed to obtain this error syndrome without affecting the logical information stored in the code subspace, which leads to what is called a syndrome extraction circuit.  
Minimal syndrome extraction circuits are optimal to minimize additional noise introduced either through gates, idle periods, or circuit operations, and thereby maximize logical fidelities. In this section, we compare the performance of different syndrome extraction circuits designed to  preserve the FT level of the QEC code.

The standard syndrome extraction gadget for a weight-4 stabilizer, depicted in Fig.~\ref{fig:xxxx}, measures a   plaquette operator $S^X_{p}$ by entangling the  corresponding data qubits   with a single ancillary syndrome qubit, which is subsequently measured projectively~\cite{doi:10.1098/rspa.1996.0136}. For $S^Z_{p}$ stabilizers, one needs to exchange the control-target roles  of the CNOTs, and conjugate them by a pair of Hadamard gates acting on the ancillary syndrome qubit, which effectively changes the initial state and the measurement basis. However, both of these circuits face an important limitation: certain single-qubit errors occurring  in the ancilla qubit between the first and last pair of CNOTs  propagate to half of the data qubits in the stabilizer, leading to the so-called hook errors that potentially reduce the code distance.
For the $d=3$ color code, all weight-2 error are homologically nonequivalent to weight-1 errors, so weight-2 errors are uncorrectable. Hence, using bare-ancilla circuits invalidates the error correction capabilities altogether for the $d=3$ color code.
This trend also generalizes to larger distances, where higher-weight stabilizers need to be measured,  and thus require the use of more complex  syndrome extraction circuits. In contrast, in the case of the surface code, a specific CNOT schedule can be chosen to force the error propagation to go in a different direction with respect to the logical operators, which we recall correspond to  Pauli strings connecting opposite boundaries of the direct and dual  lattices. This prescription allows to use  unverified bare ancilla qubits,  while simultaneously preserving the code distance during the syndrome extraction~\cite{PhysRevA.90.062320}, making the overhead   minimal.

There are several alternatives  to preserve fault tolerance in color-code syndrome extraction, such as using flag qubits~\cite{Chao_prl_2018,Chao2018,Reichardt_2021} as in Fig.~\ref{fig:flagged} or following a Steane-style QEC~\cite{PhysRevLett.78.2252}. For larger distances, although a  careful CNOT scheduling with bare ancillas  can  minimize the error propagation~\cite{landahl2014quantum},  this still  reduces the effective code distance as some correlated errors  can still propagate to the same logical block. Numerical studies have demonstrated  that a suppression of the error  with  code distance $d$ can still be  approached for intermediate code distances using these schedules. However, it is important to look for genuinely-FT approaches  that can exploit the full  potential of the color  code, particularly in the low-distance regime.

To preserve fault tolerance with the $d=3$ color code, any undetected fault by the QEC circuit should propagate to at most one data qubit. This is not the case for the traditional syndrome extraction circuit from Fig.~\ref{fig:xxxx}, where a single bit-flip error  propagates according to the red circuit trajectories. However, the dangerous weight-2 error that propagates into the logical block  becomes detectable if an additional ancilla qubit is used as a flag, as indicated in Fig.~\ref{fig:flagged}. When the dangerous error happens, the extra CNOTs that surround the intermediate ones and connect the syndrome qubit to the flag qubit lead to additional error propagation path,  such that a projective measurement on the flag detects the possible occurrence of this error when the  output is $1$, i.e. the flag is raised.   Instead of discarding the run, the QEC procedure can recover from such correlated errors by performing further rounds of QEC and using the combined flag and syndrome qubit information to perform the decoding. Conversely, if no dangerous error occurs, the combination of the two surrounding flag-syndrome CNOTs ensures a correct mapping of the stabilizer information into the syndrome qubit. For the small $d=3$ code, the flag-qubit decoding can be performed with a look-up table that depends on the combined syndrome and flag qubit measurements, and the specific QEC round in which the flag is raised~\cite{PhysRevA.99.022330}, which is discussed in detail in Sec.~\ref{sec:decoders}.

We can discuss alternatives towards the full FT syndrome extraction, which depend on how we schedule the stabilizer readout for the full code and how many ancillary qubits are at our disposal. We arrange them in terms of the required qubit count as summarized in Table~\ref{tab:stab_d_3_overheads}.

\begin{table}
    \centering
    \begin{tabular}{c|c|c|c}
      \hline
        \hline
        Overhead & Sequential  & Simultaneous  & Superdense  \\
        \hline
        \hline
        $\#$qubits & 20 & 28  & 28 \\
        (data) & 7+7 & 7+7  & 7+7 \\
        (syndrome) & 1+1 & 3+3  & 6+6 \\
          (flag) & 1+1 & 3+3  & - \\
          (surgery) & 1 & 2  & 2 \\
          \hline
        $\#$ CNOTs flagged & 36+36  &36+36 & 30+30 \\
        (un-flagged) & 24+24  & 24+24 & -\\
           \hline
        Depth flagged& 24/48 & 8/16 & 7/10 \\
           (un-flagged) & 18/36  & 6/12 & -\\
        \hline
    \end{tabular}
        \caption{{\bf QEC resources for $d=3$ color codes}:  total qubit number in logical teleportation, and number of CNOT gates and circuit depth for the different  syndrome extraction strategies. The circuit depth is indicated both for a half-round (measurement of either X or Z stabilizers) and a full round.}\label{tab:stab_d_3_overheads}
\end{table}

\begin{table}
    \centering
    \begin{tabular}{c|c|c|c}
      \hline
        \hline
        Overhead & Sequential & Simultaneous  & Superdense  \\
        \hline
        \hline
        $\#$qubits & $\sim\frac{3}{2}d^2$ & $\sim3d^2$ & $\sim3d^2$ \\
        (data) & $\sim\frac{3}{2}d^2$ & $\sim\frac{3}{2}d^2$ & $\sim\frac{3}{2}d^2$ \\
        (syndrome) & 1+1 & $\sim\frac{3}{4}d^2$ & $\sim\frac{3}{2}d^2$ \\
          (flag) & 1+1 & $\sim\frac{3}{4}d^2$ & - \\
          (surgery) & $1$ & $d-1$ & $d-1$ \\
          \hline
        $\#$ CNOTs flagged & $\sim12d^2$  & $\sim12d^2$ & $\sim\frac{21}{2}d^2$ \\
        (un-flagged) & $\sim9d^2$  & $\sim9d^2$ & -\\
           \hline
        Depth flagged& $\sim15d^2$ & $20$ & $10$ \\
        (un-flagged) & $\sim12d^2$  & $16$ & -\\
        \hline
    \end{tabular}
        \caption{{\bf QEC resources scalings with distance  $d$ for  color codes}: total qubit number in logical teleportation, and number of CNOT gates and circuit depth for the different  syndrome extractions.}\label{tab:stab_d_overheads}
\end{table}

\subsubsection{ Sequential stabilizer extraction}

The sequential stabilizer readout is the lowest-overhead strategy, requiring the use of a single syndrome qubit and a single flag qubit that can be both reset. This assumes that there is a programmable connectivity of CNOT gates such that the data qubits in the various plaquettes can be coupled to the syndrome qubit when needed. Otherwise, for QPUs with a fixed local connectivity, a syndrome and flag qubit per plaquette would be required. As discussed in App.~\ref{app:stab_readout_sequential}, one proceeds by mapping the stabilizers sequentially $\{S_1^X\to S_2^X\to S_3^X\to S_1^Z\to S_2^Z\to S_3^Z\}$, and adapting the circuits according to the mid-circuit flag-qubit outcomes. Flagged (Fig.~\ref{fig:flagged}) or un-flagged (Fig.~\ref{fig:xxxx}) circuits are chosen as the QEC proceeds, leading to dynamic circuits.  

The price one pays for reducing the overhead in qubit count (see Table~\ref{tab:stab_d_3_overheads}) is that the whole syndrome extraction circuit is longer, as we need more qubit measurements and resets, such that the amount of idling errors     can  increase substantially. In Table~\ref{tab:stab_d_3_overheads}, we also provide the distribution of qubits between data, syndrome, flag and surgery qubits, and the overhead in terms of the number of CNOT gates for the full $X$- and $Z$-type stabilizer readout, as well as the associated circuit depths.

\subsubsection{Simultaneous stabilizer extraction}

A more compact strategy  maps the information of all three stabilizers of the same type $\{S_1^X, S_2^X, S_3^X\to S_1^Z, S_2^Z, S_3^Z\}$, using a FT circuit with 3 pairs of syndrome and flag qubits, and the simultaneous measurement of those followed by reset. As discussed in App.~\ref{app:stab_readout_simultaneous}, we also adapt the
circuits according to the mid-circuit flag-qubit outcomes  to minimize the depth.
The corresponding circuits are constructed by a sequence of  either un-flagged (Fig.~\ref{fig:xxxx}) or flagged (Fig.~\ref{fig:flagged}) circuits, deferring the measurements of all six syndrome and flag qubits to a final simultaneous step. The qubit distribution and the overhead in CNOTs and circuit depth is also given in Table~\ref{tab:stab_d_3_overheads}, showing that the latter is  reduced with respect to the  sequential scheme.

Both the sequential and simultaneous protocols inherently rely on {dynamic circuits}, where mid-circuit measurement outcomes dictate the subsequent steps. The detection of a flag or syndrome flip triggers an adaptive branching in the measurement sequence, switching from flagged to un-flagged stabilizer extractions to efficiently identify and correct errors. This in-sequence logic allows the protocol to minimize resource usage by avoiding unnecessary repetitions of flagged circuits when no errors are detected, while ensuring robustness against potentially dangerous correlated errors when faults occur. Such dynamic adaptability is crucial for practical FT quantum computing, enabling real-time decision making based on measurement results within a single QEC cycle. These FT measurement protocols with dynamic control form a key component of our modular teleportation architecture, providing both scalability and reliability in the presence of realistic noise.

\subsubsection{Superdense stabilizer extraction}

Further parallelization can be achieved using super-dense stabilizer readout, an optimized circuity for the color code~\cite{gidney2023newcircuitsopensource,Lacroix2025}. That implementation achieves an enhanced resource efficiency by  arranging qubits on a hexagonal lattice embedded within a square grid, and also by employing entangled ancillary qubits to simultaneously measure both 
$X$- and $Z$-type stabilizers, thereby reducing measurement rounds and ancillary qubit overhead (see Table~\ref{tab:stab_d_3_overheads}). For every plaquette, two ancilla qubits are entangled to form a Bell state $\ket{\Phi^+}=(\ket{00}+\ket{11})/\sqrt{2}$. In a Bell state, the first ancilla qubit is prepared and measured in the $X$ basis, and the second one uses the $Z$ basis. This is leveraged in the superdense scheme to map the $X$ and $Z$ syndrome respectively to each ancilla qubit. The superdense circuit, displayed in Fig.~\ref{fig:superdensecircuit}, achieves fault tolerance since the propagation of a dangerous two-qubit error during the mapping of an $X$-type stabilizer to one of the ancilla qubits triggers the remaining conjugate ancilla qubit, which thus acts as a flag, and vice versa for $Z$-type stabilizers. A subsequent syndrome extraction round is sufficient to distinguish between single and two-qubit errors on data qubits.

\begin{figure}
\includegraphics[width=\linewidth]{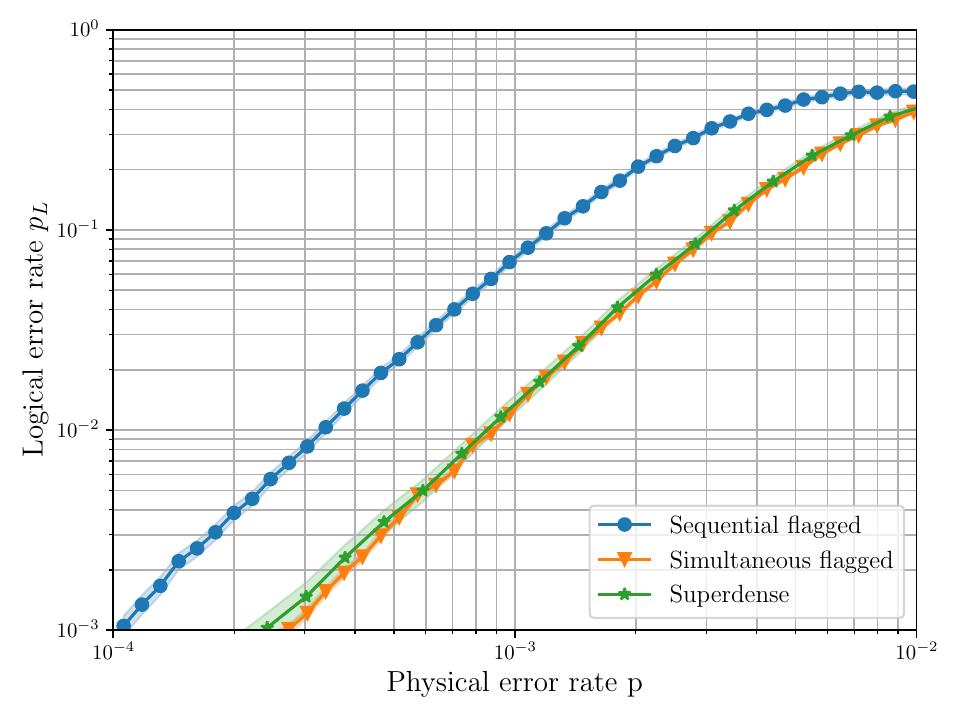}
\caption{{\bf Small color code syndrome extraction performance}: performance of different syndrome extraction circuits for the $d=3$ color code. Logical error rates are computed under standard circuit level noise, using a lookup table decoder. Sequential stabilizer readout has a huge performance impact due to idle errors, and it is advisable to measure all stabilizers in parallel if enough ancilla qubits are available. For small $d=3$ color codes, the superdense syndrome extraction scheme provides no advantage to the simultaneous flagged readout, due to the latter protocol switching to an un-flagged syndrome extraction as soon as an error is detected.}\label{fig:memd3}
\end{figure}

To provide a validation of FT  for these syndrome extraction approaches, and also compare their performance, we perform numerical simulations under the SCEM for depolarizing circuit-level noise with a single error rate $p$. We perform a numerical simulation of a color-code memory experiment, initialized in either $\ket{0}_L$ or $\ket{+}_L$, using the  flag-type verification  considered previously in Fig.~\ref{fig:steaneprep}, which gets degraded due to the depolarizing noise. We  numerically simulate  the resulting logical error rate after after a subsequent round of  flag-based QEC using our sparse Pauli-frame simulator, which samples over the circuit-level noise during the syndrome extraction, and counts the logical error events in which the quantum memory ends in either $\ket{1}_L$ or $\ket{-}_L$, respectively. In Fig.~\ref{fig:memd3}, the blue circles, orange triangles and green stars represent the performance of a $d=3$ color-code quantum memory using the sequential, simultaneous and superdense syndrome extraction, respectively, together with  the subsequent Pauli-frame update based on a FT lookup table decoder discussed in more detail below. By plotting the corresponding logical error rate $p_L$ as a function of the depolarizing error rate $p$ in log-log scale, we see that all schemes display similar slopes for sufficiently low errors, which indicates their FT nature. One also sees that the performance of the sequential extraction is much worse than that of the simultaneous or the superdense extractions, again  by roughly one order of magnitude at $p=0.1\%$. This indicates that  it is preferable to  switch to the simultaneous syndrome extraction scheme if the additional ancillary resources are available.

Let us close this section by also noting that, as can be seen in figure \ref{fig:memd3}, for  physical error rates on the order of $p_{\rm th}\sim0.1\%$, these small-distance codes  still perform worse than the bare physical bare qubits. Even if the error threshold of the color code lies around this range of physical error, as will be discussed below in more detail, the benefits of the QEC encoding would only become apparent for larger code distances. In this case, the pseudo-threshold at which $p^{d=3}_L/p=1$ lies well below $p\sim 0.01\%$, but will tend towards $p\to p_{\rm th}$ as $d$ increases. By steadily increasing the code distance provided that $p<p_{\rm th}$, one can reduce the logical error rate exponentially when working below threshold.

\subsection{Caveats with fault tolerance in flag-based QEC }\label{sec:ftcaveats}

In this section, we discuss some caveats with the fault tolerance of certain flag-qubit based approaches to QEC which, to the best of our knowledge,  have passed unnoticed in earlier QEC studies.
Several of those target quantum memory benchmarks, where the information encoded in the logical qubit(s) remains idle, and the physical qubits are only acted upon to extract the syndrome information and perform the pertinent corrections. In most cases,  one starts by just focusing on the performance of the ${Z}_L$ eigenstates $\ket{0}_L$ and $\ket{1}_L$ or,  in some other cases, also considers the ${X}_L$ basis $\ket{+}_L$ and $\ket{-}_L$. However, we have found that some of the flag-based protocols  used in the literature~\cite{Reichardt_2021}, which preserve fault tolerance in these memory benchmarks, fail to correct against certain type errors if used as  building blocks or QEC gadgets of a larger computation involving further operations on the logical qubit(s).

To protect against arbitrary qubit errors, a QEC code must be able to simultaneously correct bit and phase flip errors on every qubit. CSS codes satisfy this requirement, as bit flips and phase flips are decoded independently from each other. Moreover, this means that distance $d$ CSS codes can also simultaneously correct $(d-1)/2$ bit flips and $(d-1)/2$  phase flips for $d$ odd, even if these  act on different physical qubits. From this perspective, CSS codes such as the $d=3$ color code  can also correct certain  weight-2 errors in the form $X_aZ_b$. Thus, some  FT  protocols designed for CSS codes may allow for the propagation of such correctable weight-2 errors, and this is perfectly valid for a quantum memory benchmark.

\begin{figure*}
\subfloat[\label{subfig:nftprepare}]{\includegraphics[width=.42\linewidth]{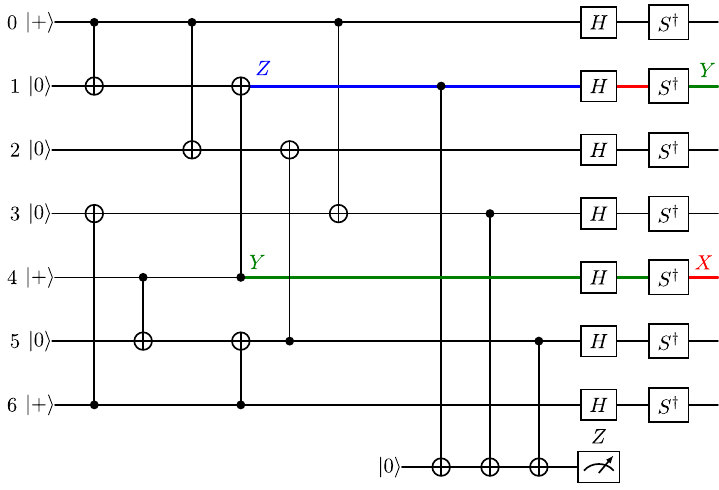}}\qquad
\subfloat[\label{subfig:nftsynd}]{\includegraphics[width=.48\linewidth]{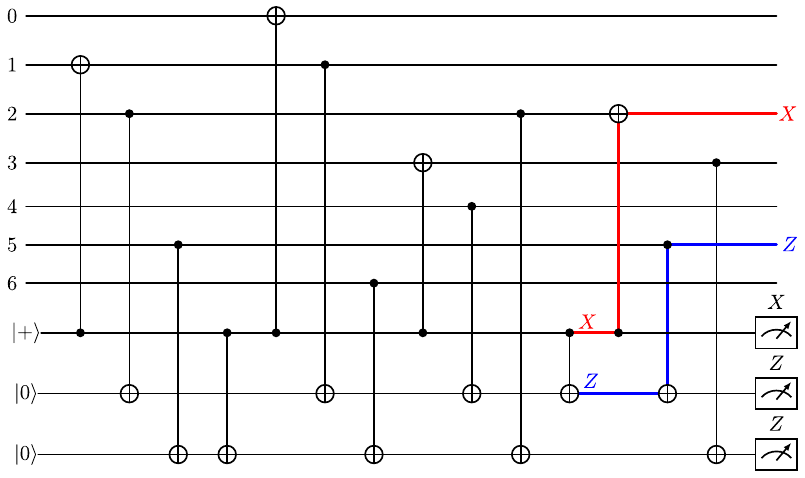}}
\caption{{\bf Uncorrectable errors in QEC gadgets}:
a) State preparation from figure~\ref{fig:steaneprep} is not FT when preparing the $\ket{\rm +i}_L$ state. The depicted error $Z_1\otimes Y_4=\ii X_4(Z_1\otimes Z_4)$ flips the red $X$ plaquette and the green $Z$ one in the left panel of Fig.~\ref{fig:colorcodes}. This error  would be correctable  when preparing a $\ket{0}_L$: the prescribed correction is $Z_0\otimes X_4$ which, taking into account that ${Z}_L=Z_0\otimes Z_1\otimes Z_4$ acts trivially on $\ket{0}_L$, is equivalent to the original $Z_1Y_4$ and thus reverts the error. On the other hand, applying transversal gates to obtain $\ket{\rm +i}_L=SH\ket{0}$, will convert such an error to $Y_1\otimes X_4$. The correction for that is $Z_1\otimes X_0$, which differs from the original error by a logical operator $X_L=X_0\otimes X_1\otimes X_4$, thus causing a logical error $\ket{\rm +i}_L\to\ket{\rm -i}_L$ and compromising the FT nature of the QEC protocol. b) The compressed syndrome extraction circuit , which removes the need for using flag qubits by using $Z$ syndrome qubits to flag $X$ errors and vice-versa, ensures that at most one $X$ and one $Z$ errors remain undetected. This is usually enough to preserve fault-tolerance, while reducing the qubit overhead. However, if a transversal $S_L$ gate is applied afterwards, it would convert the depicted $X_2\otimes Z_5$ error into an uncorrectable $Y_2\otimes Z_5$ logical error.}\label{fig:nftqec}
\end{figure*}

The problem arises when one aims at processing the logical information, as a subsequent  encoded gate may require applying a  gate to one of the faulted physical qubits. In fact, this physical gate can transform the innocent $X_aZ_b$ error to a different type of error, compromising in this way the FT of the whole approach. As an example, for the 2D color code, the logical $S$ gate is transversal: ${S}_L=\bigotimes_qS_q^\dagger$. Being transversal, one naively assumes that it will not change the FT nature of the whole QEC protocol. Even if  errors do not spread under this operation, one finds  upon closer inspection that an apparently mild weight-2 error  $X_a\otimes Z_b$ that could be corrected through the independent $Z$ and $X$ syndromes of a CSS code,  becomes instead $Y_a\otimes Z_b=\ii X_a(Z_a\otimes Z_b)$ after the application of this transversal gate, leading in this way to  a weight-2 phase-flip error that can no longer be corrected.
This contrasts with the use of a transversal Hadamard ${H}_L$, as it would simply exchange the role of this weight-2 error $Z_aX_b$. Thus, the use of QEC gadgets that allow some weight-2 errors to propagate should be analyzed with caution if followed by  transversal gates.  

Fig.~\ref{fig:nftqec} shows examples of QEC  protocols that preserve fault tolerance for a quantum-memory benchmark, but fail when a transversal $S$ gate is applied afterwards, e.g. when preparing a $\ket{\rm +i}_L$ state following~\cite{Goto2016}, or when compressing the circuits for syndrome extraction following~\cite{Reichardt_2021}. Let us focus on the encoding of a $\ket{\rm +i}_L$ state, simulating the effects of noise in the circuit of Fig.~\ref{subfig:nftprepare} modeled by the SCEM with a varying depolarizing error rate $p$. The orange triangles  in Fig.~\ref{fig:preparationcomparison} display the resulting logical error for such state preparation, which displays a different slope in comparison to the two FT strategies to prepare logical eigenstates in the $X$ and $Z$ basis. These simulations clearly show the aforementioned caveat with FT, and the special care that must be taken when applying additional operations on FT gadgets that allow for certain cascading of errors within the same logical block. If those protocols are used, a fully FT QEC step must always be inserted before any transversal $S$ gate is applied. This ensures that all single faults affect at most one qubit, correcting $X_aZ_b$ errors before the transversal $S$ gate is applied.

As a result, special care must be taken when performing QEC experiments beyond a quantum memory. When qubits are idling, these gadgets can be used while preserving fault tolerance, as they provide better performance. However, when performing some logical computations, they must be replaced by strictly FT protocols, which do not allow the propagation of a single fault to more than one data qubit.

\subsection{Color-code decoding strategies}
\label{sec:decoders}
Another crucial step in QEC is decoding, which converts the classical bits storing the error syndrome into a list of operations that have to be applied to the physical qubits. In this work, we will assess the performance of different software-based classical decoders, weighing their advantages and limitations for the 2D color code. We also note recent approaches, which have shown that it also possible to implement the decoder entirely as a quantum circuit, eliminating the need for repetitive syndrome measurements~\cite{PhysRevA.108.062426,PRXQuantum.5.010333,PhysRevResearch.6.043253,doi:10.1126/sciadv.adv2590,brechtelsbauer2025measurementfreequantumerrorcorrection,butt2025demonstrationmeasurementfreeuniversalfaulttolerant}.

\subsubsection{Flag-qubit-based decoding lookup table}
A lookup table is the simplest decoding protocol, which exhaustively maps every syndrome outcome to the most-likely correction consistent with the syndrome. After each QEC round, a search is performed and the matching correction is applied. When considering circuit-level noise, also the syndrome from previous rounds has to be taken into account to protect against measurement and hook errors~\cite{10.1063/1.1499754} (errors that appear in the middle of the syndrome extraction circuit), increasing the size of the lookup table. The drawback of lookup tables is that, for higher-distance stabilizer codes, the number of syndrome bits increases, and the size of the lookup table becomes exponentially large. Nevertheless, the procedure is very appealing for small $d=3$ codes, which only aim to correct single-qubit errors, as the decoding with a lookup table is  quick, easily allowing real-time decoding.

Lookup tables are particularly interesting when dynamic circuits can be executed in the physical device, that is, the gates to apply can be determined in real time depending on the measurement outcomes. This enables the conditional execution of syndrome extraction circuits only when they are required. For example, if a syndrome extraction round provides a trivial measurement outcome, no further round is required, but otherwise a new round is required to distinguish between data qubit errors, measurement errors and mid-circuit hook errors, and a correction circuit is executed afterwards. An extra advantage of this approach is that it makes QEC rounds self-contained, as the correction for a QEC round is performed exclusively using information from that round.

In Table~\ref{tab:lookup}, we show the lookup table for the flagged syndrome extraction of the 7-qubit color code~\cite{PhysRevA.99.022330}. This table has already been used in our quantum-memory simulations of Fig.~\ref{fig:memd3}. Additionally, we have also built a tool that finds the lookup table for a given syndrome extraction. It works by placing errors in all possible locations in the circuit, and propagating them to the end, checking the affected data qubits and  the flipped stabilizers. This automates the process of using  syndrome-extraction variants with different lookup tables.

\begin{table}
\begin{tabular}{|c|cccc|}

\hline
\backslashbox{Syndrome}{Flag} & \multicolumn{1}{c|}{None} & \multicolumn{1}{c|}{1st flag} & \multicolumn{1}{c|}{2nd flag} & 3rd flag \\ \hline \hline
000 & \multicolumn{4}{c|}{None} \\ \hline
001 & \multicolumn{1}{c|}{6} & \multicolumn{1}{c|}{6} & \multicolumn{1}{c|}{4, 5} & 6 \\ \hline
010 & \multicolumn{1}{c|}{4} & \multicolumn{1}{c|}{2, 3} & \multicolumn{1}{c|}{4} & 5, 6 \\ \hline
011 & \multicolumn{4}{c|}{5} \\ \hline
100 & \multicolumn{4}{c|}{0} \\ \hline
101 & \multicolumn{4}{c|}{3} \\ \hline
110 & \multicolumn{4}{c|}{1} \\ \hline
111 & \multicolumn{4}{c|}{2} \\ \hline \hline
\end{tabular}
\caption{{\bf Lookup table for the  flagged syndrome extraction.} In the $d=3$ color code, if one flag is triggered, all stabilizers are measured again, using them to obtain the final syndrome. The correction to apply, using the labeling of Fig~\ref{fig:colorcodes}, depends on whether a flag was triggered for the stabilizer measurement of the conjugate basis in the first round. The  table is valid both for bit- and phase-flip errors.}\label{tab:lookup}
\end{table}

Scalability is a key ingredient towards increased protection and FT algorithms, as it provides the tools to achieve arbitrarily low error rates by using larger codes. Even though lookup-table decoding can be scaled using code concatenation, in practice this scaling would require a high amount of data qubits, and typically compromises the code locality. Therefore, it is advisable to use alternative decoders, which allow a resource-efficient scaling of the codes.

\subsubsection{Restricted minimum-weight perfect matching}

In QEC, the error  syndrome can be understood as a pattern of detection events encoding information on  where errors may have occurred. For topological codes, such as the surface code, such errors typically create pairs of defects (or detection events) in the syndrome, which can be naturally represented as nodes in a graph. The minimum-weight perfect matching (MWPM) algorithm~\cite{edmonds1965maximum} constructs a matching graph where the nodes correspond to these detection events, which appear in pairs. Edges connect pairs of nodes, and represent potential error chains of physical phase- or bit-flip errors that could explain the observed  defect distribution; each edge is assigned a weight based on the likelihood or cost of the corresponding error. MWPM finds a set of edges forming a perfect matching that pairs all defects, such that the sum of the edge weights is minimized. This matching corresponds to the most-likely error configuration consistent with the observed syndrome, and  considers a simplified noise model in which the errors are assumed to be independent.

In realistic circuit-level noise models, errors can occur not only on data qubits, but also during measurements and other operations. This causes the observed syndrome to be unreliable, which requires upgrading to a space-time decoding graph where detection events correspond to changes in syndrome over time, i.e., spacetime defects. Edges in this extended graph represent not only data errors but also measurement errors, connecting defects both spatially and temporally. Applying MWPM to this space-time graph enables an efficient decoding strategy that can be applied to stabilizer codes such as the surface code~\cite{10.1063/1.1499754},  increasing the code distances in search for the specific FT error threshold~\cite{PhysRevA.89.022321,PhysRevA.80.052312,10.5555/2011362.2011368}.

\begin{figure*}
\centering
\subfloat[\label{subfig:decoders}]{\includegraphics[width=.45\linewidth]{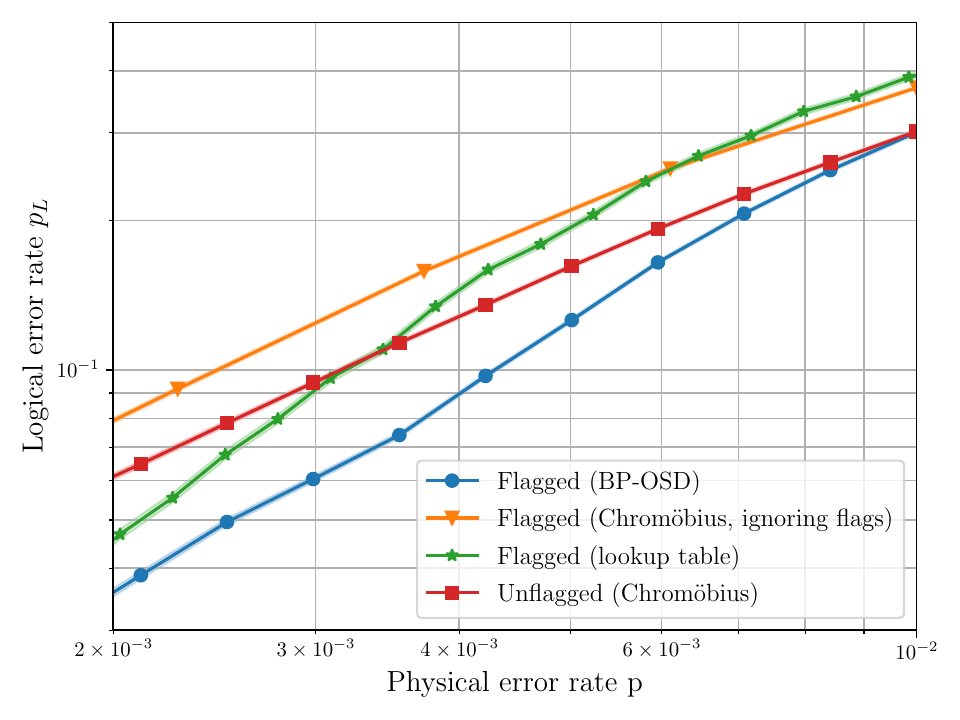}}
\subfloat[\label{subfig:effective_d3}]{\includegraphics[width=.45\linewidth]{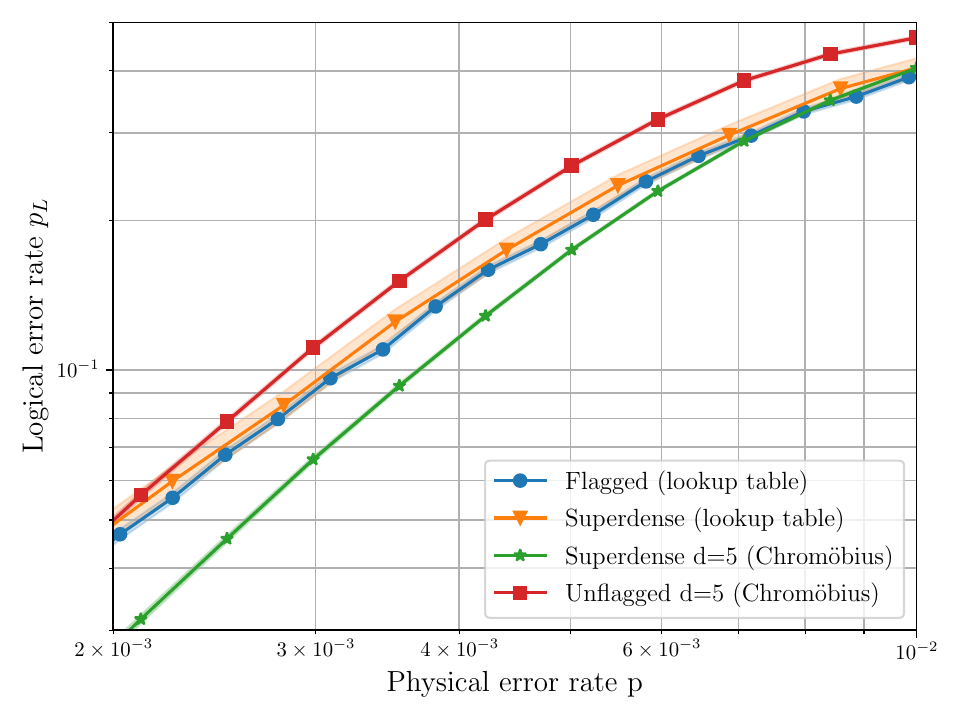}}
\caption{{\bf Decoding strategies for the Steane code}: logical error rates achieved by multiple Steane code decoders under standard circuit-level noise. The Chrom\"{o}bius decoder does not take into account information from flag qubits, and thus its performance can be improved by removing all flags from the circuit, showing a reduced code distance in both situations. By contrast, the BP-OSD decoder can use information from flag qubits to improve decoding, but the information is used in a suboptimal way, not being able to reach the full code distance. Nevertheless, with error rates above $p>10^{-2}$, the performance of those decoders is still superior to the lookup table decoding, even though the later method ensures the preservation of the code distance, as a more complex decoder may be able to correct some higher-weight errors.
}
\label{fig:decoders}
\end{figure*}

In order to describe more sophisticated decoders applicable to the color code, we use the terminology from stim~\cite{Gidney2021stimfaststabilizer, McEwen2023relaxinghardware}: nodes in the matching graph correspond to {detectors}, a combination of qubit measurements which is deterministic in absence of noise. Additionally,  edges correspond to individual errors that join the specific detectors they trigger, and they are assigned a weight depending on its error rate. This generalized description naturally captures the different error rates present under circuit-level noise, potentially giving the decoder additional information to reduce logical error rates.

As noted in the surface code, if $X$ and $Z$ errors are decoded independently, a single error flips at most two detectors, ensuring that the matching graph is actually graph-like. That is, the errors are represented by edges connecting either two nodes, or one node to a boundary. 
In that situation, the syndrome can be decoded using MWPM decoders such as PyMatching~\cite{higgott2023sparseblossomcorrectingmillion}. However, in the trivalent color code, each qubit is shared by three neighboring plaquettes (one from each color) in the bulk, meaning that a single error can trigger three detectors simultaneously. Therefore,  color-code decoding becomes a more complicated hypergraph that is not matchable, as it violates the pairing assumptions required for perfect matching decoders. One possible solution for this problem is to restrict the matching graph to only include detectors for  two of the three colors. Multiple restricted matching graphs can then be  built and decoded using a projected MWPM algorithm, and then merged back to estimate the most-likely error~\cite{Kubica2023efficientcolorcode}. Instead, a more sophisticated  lifting procedure can be  performed, which introduces auxiliary nodes and edges to encode hidden syndrome correlations and resolve possible degeneracies, such that  MWPM can accurately identify the  errors inherent to color codes. Although this lifting procedure is more demanding than the MWPM, the whole decoding strategy is still efficient, yielding  the so-called Chrom\"{o}bius decoder~\cite{gidney2023newcircuitsopensource}. 

Another complication for color-code decoding  is that, under circuit-level noise, the syndrome-extraction circuits discussed so far are capable of  exploiting the full QEC potential only if the flag qubits are correctly used to preserve fault tolerance, as the look-up table decoders achieve for the $d=3$ code. Depending on their location, qubit errors can trigger a flag in addition to a syndrome flip, making the code again unmatchable. Potential solutions around this problem  involve the modification of the un-flagged matching graph, adding and removing nodes and edges, as well as changing edge weights depending on which flags have been triggered, thus creating a matching graph which is specific to only one particular shot~\cite{Chamberland_2020, PRXQuantum.5.030352}.  Losing the ability of performing a batch decoding, that is, generating the matching graph once and reusing it for multiple shots, incurs in a significant performance impact when running numerical simulations, but it should not be a limiting factor for real-time decoding. Building a flag-correctable decoder capable of systematically decoding any color code protocol, and determining the performance of such decoder is left for future research. We will focus instead on the restricted MWPM  decoders, where benchmarks have shown that  color-code  decoding is actually not capable of reaching the full distance~\cite{Chamberland_2020, Lee2025colorcodedecoder} due to specific  errors that  cannot be corrected. 
This effect is only relevant under circuit-level noise if flag information is taken into account, otherwise the distance reduction is dominated by the uncorrectable hook errors. This is consistent with the results of our Pauli-frame simulations displayed in Fig.~\ref{subfig:decoders}. The green stars represent the fully-FT flag-based approach that uses a lookup-table decoder for the $d=3$ color code acting as a memory, and thus account for the logical effect of repetitive rounds of QEC. When using the Chrom\"{o}bius decoder,  we see that the red squares (un-flagged circuits) and orange triangles (flagged circuits) show the same slope, differing from the FT one of the previous lookup-table decoder. This clearly shows that the Chrom\"{o}bius decoder is not reaching the full QEC potential of the code as it disregards the flag information. In fact, it is preferable at this stage to use un-flagged circuits reducing the whole complexity  and depth of the QEC circuits, as the red curve lies below the orange one for the depolarizing rates studied.

In Fig.~\ref{subfig:effective_d3}, we show in green stars and red squares how the results of simultaneous syndrome extraction of the $d=3$ color code with a lookup table decoder compare to those of Chrom\"{o}bius for a larger distance $d=5$ code. As evidenced by the figure, the effective code distance is consistent $d_{\rm eff}=3$ for these larger codes, which reflects the fact that the decoder is not coping correctly with hook errors. Even if superdense circuits preserve fault tolerance, the restricted Chrom\"{o}bius decoder acts on $X$ and $Z$ error separately, and thus does not deal with the flag information arising in an opposite basis. However, it is interesting to remark that, in spite of not being optimal, increasing the distance and redundancy can indeed  reduce the logical error rates, as both curves get below the orange one in a certain range of depolarizing error rates.

\subsubsection{Belief propagation with ordered statistics}

Belief propagation (BP) is a classical decoder that is based on a message-passing algorithm~\cite{10.5555/534975}, in which messages are our current belief about the probability distributions of specific Pauli errors affecting each qubit. When combined with ordered statistics decoding (OSD), the algorithm can be used to decode quantum low-density parity check (qLDPC) codes~\cite{PhysRevResearch.2.043423}. This  provides a versatile   approach based on iterative message passing over the Tanner graph, which reflects the code parity check matrix and the physical qubits that participate on a given check. BP-OSD excels in handling codes with complex syndrome structures where a single physical error may trigger multiple syndrome bits, a scenario common in high-distance and high-weight stabilizer codes such as color codes. Unlike MWPM, which inherently assumes a graph-like  structure where errors flip pairs of detectors, BP-OSD  can incorporate diverse error correlations that do not comply with this constraint, including those arising from flag qubits. This flexibility makes BP-OSD particularly suitable for other qLDPC codes~\cite{PRXQuantum.4.020332,Bravyi2024,Old_2024} beyond the  surface-code family.

While competitive,  BP-OSD decoders typically fall short in comparison to the highly-optimized MWPM-based ones, in particular for  codes where the MWPM assumptions hold approximately, such as the low-distance color codes. The main limitation arises from the heuristic nature of BP’s iterative updates.
While BP-OSD has worse performance than MWPM, the interest lies in the fact that it can be used for general qLDPC codes, even if a single error triggers more than 2 detectors. In our particular context,  this decoder is also suitable for $d>3$ color codes, and can also incorporate information from flag qubits. As  shown by the blue circles of Fig.~\ref{fig:decoders}, the  BP-OSD decoder based on flagged syndrome-extraction circuits shows a similar slope as the fully-FT lookup-table decoder for a certain depolarizing error range.  In fact, it provides a reduced logical error rate showing that we are correcting the errors more accurately.  However, for sufficiently-low error rates, this trend   changes, and the slope becomes more similar to that of the Chrom\"{o}bius decoders.  In this regime, we only find  a small improvement over Chrom\"{o}bius, as BP-OSD is still not capable of reaching the full QEC distance of the code.

\subsection{QEC footprint for a scaled quantum memory }

We have  discussed  various aspects of color-code QEC, focusing on FT encoding, syndrome readout and decoding for the lowest-distance color codes. An essential metric for evaluating the practicality of QEC,  and comparing the effectiveness of various codes and  decoders is the QEC footprint: the number of physical qubits \(N(p, p_L)\) required to achieve a target logical error rate \(p_L\) for a specific QEC protocol, given a physical error rate \(p < p_{\text{th}}\), after $d$ rounds of error correction. Here, \(p_{\text{th}}\) denotes the code’s error threshold, and the metric requires that one increases the code distance $d$, and thus the number of physical data qubits, until the resulting logical error rate is reduced to the target value, capturing the resource overhead necessary to suppress logical errors below a desired level.

Using the simulation and decoding toolkit already described, we can efficiently estimate the error threshold and  QEC footprint.
In this section, we analyze the scaling behavior of these qubit footprints needed to enter a regime of  QEC advantage, defined here as achieving a logical error rate at least ten times lower than the  physical error rate in current high-fidelity  regimes \(p = 0.1\%\). We thus aim for $p_L=0.01\%$,  focusing on a representative  QEC protocol: a color-code quantum memory. We compare the resource requirements to those of the surface code,  restricting to the SCEM of depolarizing  noise.

\begin{figure}
\includegraphics[width=\linewidth]{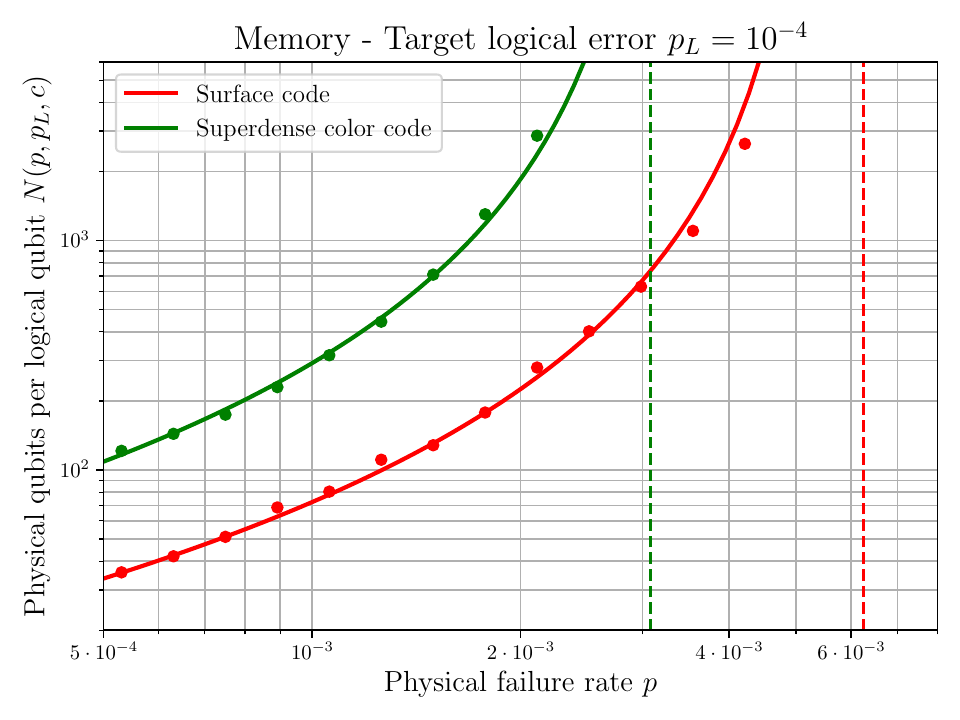}
\caption{{\bf  QEC-advantage footprint for a logical memory}: 
number of physical qubits $N(p,p_L)$ required to achieve a logical error rate of $p_L=10^{-4}$ after $d$ QEC rounds by scaling surface and color codes under standard circuit level noise with varying error rate $p$. Solid lines indicate the number of physical qubits required for different physical error rates, obtained from a power-law fit~\cite{benito2024comparative}. Vertical dashed lines represent the error threshold, which equals $0.3\%$ for the color code, and is lower than that of the surface code ($\sim0.7\%$).}
\label{fig:QEC_footprint}
\end{figure}

In Fig.~\ref{fig:QEC_footprint}, we present in red the results for the QEC-advantage footprint $N(p,10^{-4})$ as a function of the depolarizing error rate $p$. This footprint clearly asymptotes at the $p=p_{\rm th}$, which we find to be $p^{\rm sc}_{\rm th}=0.7\%$ consistently with previous studies of circuit-level noise under the SCEM. It should be noted that we are using the MWPM PyMatching decoder~\cite{higgott2023sparseblossomcorrectingmillion}. The resource overhead for a ten-fold QEC-advantage at $p=0.1\%$ is thus $N(0.1\%,0.01\%)\approx 60$ physical qubits per logical qubit encoded in the surface code. Turning to the color-code memory, we use the superdense syndrome extraction protocol discussed above, and decode the projected MWPM problems using the lifting procedure of the Chrom\"{o}bius decoder~\cite{gidney2023newcircuitsopensource}.   The results for the QEC-advantage footprint  are represented in green in this figure, showing an asymptote that is consistent with the color-code error threshold under depolarizing noise $p^{\rm cc}_{\rm th}\approx0.3\%$. This results in a $5\times$ increase in the QEC-overhead  $N(0.1\%,0.01\%)\approx 300$ at current physical error rates of $p\approx0.1\%$. We believe that the superiority of the surface code at this task is due to the more-involved decoding of the color code which does not fully exploit the flag-qubit information leading to an effective smaller distance. Until better color-code decoders become available, these results quantify  the resource tradeoff that one must assume in order to have a smaller number of physical qubits for a target code distance and, additionally, a fully transversal implementation of the  Clifford group. For  benchmarking of  quantum memories, the ability to prepare any of the six color-code cardinal states on the same footing, as opposed to the surface code in which $Y$-basis states are much more demanding~\cite{Gidney2024inplaceaccessto}, may allow for the first rigorous demonstration of sub-threshold scaling~\cite{Lacroix2025} based on the entanglement fidelity~\cite{vezvaee2025surfacecodescalingheavyhex}. In addition to the fact that Clifford subroutines in the color code can be executed within a single QEC cycle, color-code injection protocols can also exploit the more symmetric stabilizer structure to realize high fidelity state injection protocols.

Let us note that, even if the quantum memory experiment is the simplest QEC protocol, thanks to this color-code transversality, and the negligibly-small errors of  single-qubit Clifford operations on physical qubits, these QEC footprints give indicative estimates of the resource requirements for other  information processing tasks. In next section, we  show that  the QEC footprints for other processing goals, such as logical  teleportation or logical entangling gates  require similar footprints.

\section{\bf Color-code  lattice-surgery teleportation }
\label{sec:teleportation}
In order to complete the set of logical Clifford operations beyond single-qubit gates, an entangling two-qubit gate has to be defined. Being a CSS code, the color code permits a transversal logical CNOT, providing a qubit-wise protocol that  requires acting on pairs in parallel  by applying the corresponding  physical operation on each physical qubit pair. However, applying such a  qubit-wise CNOT between physical qubits in both logical blocks actually imposes strong connectivity requirements that may not available in  the specific  QPU, or may  become increasingly complicated when scaling up the  codes. For example, a transversal CNOT has been realized in trapped ion QPUs for small-distance color codes~\cite{Postler2022}, and used  to perform Steane-type QEC \cite{PRXQuantum.5.030326}. These results have exploited the programmable connectivity in  ion crystals, using addressed laser beams to perform the desired sequence of qubit-wise CNOTs. However, it becomes impractical for larger codes as the individual crystals cannot host arbitrarily-large number of ions. Following a shuttling-based QCCD approach~\cite{ryananderson2022implementingfaulttolerantentanglinggates}, one   only works with smaller ion crystals that are held in different regions of a segmented surface trap, and one eventually  splits, shuttles and merges them in  the pairs that must be entangled by the action of the corresponding CNOTs. To perform the fully-transversal CNOT, we note that the number of such crystal-reconfiguration operations would scale  with the code distance at least as $d^2$,  yielding an  overhead in terms of idling periods and  clock time due to the shuttling and re-cooling operations.

Instead of aiming at a transversal CNOT gate, it is also possible to generate entanglement between two logical qubits by performing a joint logical measurement such as $X_{L,1}X_{L,2}$, and also use this as a QEC gadget that allows for the teleportation of logical information~\cite{Nielsen_Chuang_2010}. To perform the logical measurement, only $d$ qubits from each logical block need to be accessed, reducing  both the connectivity constraints and the shuttling requirements. We note that this counting argument is  not entirely precise since, in order to preserve fault tolerance,  the joint measurement  must  be repeated  $d$ times to account for possible readout errors. In practice, the total gate count for both protocols (transversal CNOT and joint logical measurement) scales similarly, and the final choice of a preferred strategy will depend on the device connectivity, as well as several other details of the transpilation of the corresponding circuits into the native trapped-ion gates, which should also include the costs and errors due to the additional crystal reconfigurations~\cite{PhysRevA.99.022330}. 

In this work, we focus on the shuttling-based QCCD method, as it is more suited for implementation in a modular device. Modularity allows acting individually or in  parallel on the logical qubits, and then proceed to entangle them using  LS,  also allowing for logical-qubit teleportation. In comparison to previous studies on lattice-surgery CNOTs~\cite{PhysRevA.99.022330}, we will focus here on  the logical teleportation, providing a more realistic account of two possible modular trapped-ion architectures, and including a more detailed transpilation of the circuits and the associated error modeling. 
We leverage scalable decoding protocols going beyond the lookup table  used in~\cite{PhysRevA.99.022330},  carrying on larger-scale  simulations to estimate the QEC footprint for the lattice-surgery teleportation. Before getting there and focusing on   trapped-ion details, we discuss some platform-independent features of the lattice-surgery method.

\begin{figure}
\includegraphics[width=\linewidth]{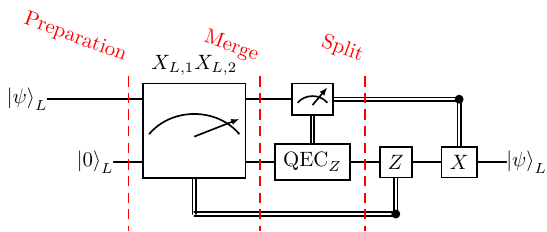}
\caption{{\bf Logical teleportation circuit using LS.} State teleportation moves any input logical state $\ket{\psi}_L$ to another logical qubit, without a direct interaction between qubits. The first step involves preparing a logical qubit in the $\ket{0}_L$ state, to which the initial state will be teleported. Afterwards, both logical qubits are merged into a single code block to measure the joint logical operator $X_{L,1}X_{L,2}$. Then, the code block has to be split back into two, and the first qubit must be  measured projectively. To complete the teleportation, the Pauli frame of the second qubit is updated taking into account the two types of measurement outcomes. }
\label{fig:teleportcircuit}
\end{figure}

\subsection{ Lattice-surgery  teleportation scheme }\label{sec:ls-teleportation-scheme}

We consider that a generic modular QEC architecture shall consist of modular units, each hosting the required data, syndrome, and flag qubits to perform the various  operations of FT QEC. These modules must be designed to be independently controllable and, ideally,  leverage a certain amount of programmable connectivity within their qubit registers to interconnect them, executing for instance the entangling operations required for LS. Multiple such units could then be arranged and replicated at scale, supporting a variety of concatenated distance-3 codes, or splitting larger topological codes in modular sub-registers  facilitating robust logical teleportation schemes. We envision a central processing region acting as a controllable  interface hosting additional ancillary qubits, often called surgery qubits, used to implement quantum teleportation at the logical level. These ancillary qubits are sequentially coupled to the modules to realize the necessary measurements of the joint  logical operators required for teleportation. We thus assume that they can be measured and reset within the interface, ensuring operational flexibility. This design fulfills   a divisibility requirement, avoiding  joint  operations between physical qubits of separate logical blocks,   thus minimizing  crosstalk errors  from imperfect addressing.

The modular logical teleportation depicted in Fig.~\ref{fig:teleportcircuit} then consists of three main steps.\textit{ (i) Preparation:} The source logical qubit \(\ket{\psi}_{L,1} = \alpha \ket{0}_{L,1} + \beta \ket{1}_{L,1}\) is initialized using a FT scheme in one of the six cardinal states \(\{\ket{0}, \ket{1}, \ket{\pm}, \ket{\pm \ii}\}\), while the destination logical qubit is prepared in \(\ket{0}_{L,2}\).
\textit{ (ii) Merge:} The two logical qubits are merged via LS by measuring the joint operator \(X_{L,1} X_{L,2}\). This step combines the two codes into a single larger code hosting a shared logical qubit.
\textit {(iii) Split:} The codes are split by measuring appropriate \(Z\)-type stabilizers, followed by a projective measurement of the source logical qubit in the physical \(Z\)-basis. Based on the measurement outcomes, conditional logical Pauli corrections are applied to the destination qubit through a Pauli frame update, completing the teleportation.

The joint measurement of $X_{L,1}X_{L,2}$ must be FT, ensuring that  single faults that would propagate to  correlated  errors affecting pairs of data qubits from the same logical block are forbidden, and only allowed 
if the two qubits actually belong to different logical blocks. A first possibility to perform such a joint measurement would be  to use a standard many-body measurement circuit with (physical) ancilla and flag qubits~\cite{ryananderson2024highfidelityfaulttolerantteleportationlogical}, similar to the one used for flagged syndrome extraction. However, such a circuit would have at least depth-$2d$, with a negative impact on performance. In fact, as already advanced in the introduction, special care must be taken in order to achieve FT, requiring the repetition of  the weight-2$d$ joint measurements, losing any practical advantage already for small distances. Instead, joint logical measurements can be performed using LS~\cite{Landahl2014,PhysRevA.99.022330}, where two logical qubits are merged into a single larger logical qubit by creating new  stabilizers at the shared boundaries. During the merge, the joint logical operator becomes a stabilizer of the larger code, whose value  can be obtained via syndrome extraction. The advantage of this method is that this joint operator can in fact be expressed as the product of multiple lower-weight stabilizers such that, instead of measuring a single weight-$2d$ operator, simpler stabilizer measurements can be combined to obtain the desired logical  outcome. Due to the joint stabilizer being broken in parts, some of the original stabilizers that would anti-commute with the new lower-weight stabilizers, must be combined to ensure that all stabilizers commute. After the measurement, those anti-commuting stabilizers are measured to split the code back to the original two data qubits.

\begin{figure}
\includegraphics[width=0.8\linewidth]{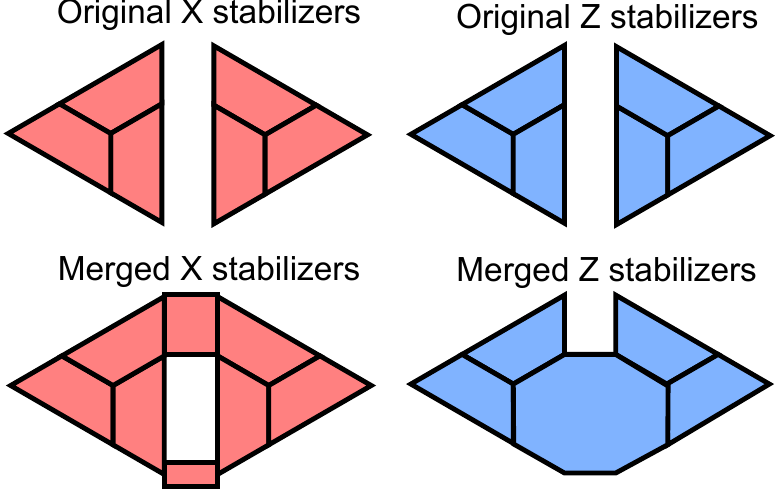}
\caption{{\bf Merged color-code stabilizers:} The original two $d=3$ color codes contain 14 data qubits and 12 $X$ and $Z$-type stabilizers, depicted in the top row as red and blue plaquettes respectively. The counting $k=n-G$ leads to 2 independent logical qubits. After the LS merging step, we get 2 more $X$-type stabilizers and  a less $Z$-type stabilizer, as depicted in the bottom row, so 13 in total . The counting $k=n-G$ now yields a single logical qubit as a result of the merging.   }
\label{fig:splitmerge}
\end{figure}

For the $d=3$ color code, the merging requires decomposing the measurement of a weight-6 Pauli operator supported on the boundaries into a pair of lower-weight operators acting on 2 and 4 qubits, respectively. A FT scheme to achieve this  uses bare ancilla qubits, and is  enabled by a careful scheduling of the entangling gates to prevent dangerous  error propagation. As a result of this joint measurement, the two  $d=3$ color codes merge into a combined \([[14,1,3]]\) code that temporarily hosts a single logical qubit. The set of \(X\)-type stabilizers gets enlarged by adding two plaquettes that span the boundary, while the individual \(Z\)-type stabilizers merge into a single weight-8 stabilizer across the boundary. Fig.~\ref{fig:splitmerge} illustrates the change in stabilizers from the original to the merged code, both for X and Z stabilizers.

Subsequently, the combined logical qubit is split back into the original two logical qubits by measuring the weight-4 \(Z\)-type stabilizers of each code block. This splitting operation requires the use of flag qubits, and repeated quantum error correction (QEC$_Z$) cycles (see Fig.~\ref{fig:teleportcircuit}) to maintain fault tolerance. Depending on the  outcome \(a \in\{ 0,1\}\) of the joint  measurement, corresponding to eigenvalues \(m_{X_{L,1} X_{L,2}} \in\{+1,-1\}\), the two logical qubits end up in the conditional entangled state
\begin{equation}
\label{eq:state_after_merge}
\begin{split}
\ket{\Psi_a} =
\frac{1}{\sqrt{2}}\ket{0}_{L,1} \otimes \left(\alpha \ket{0}_{L,2} + (-1)^a \beta \ket{1}_{L,2} \right) \\+ \frac{1}{\sqrt{2}} \ket{1}_{L,1} \otimes \left(\beta \ket{0}_{L,2} + (-1)^a \alpha \ket{1}_{L,2} \right).
\end{split}
\end{equation}
During the split, the source logical qubit is projectively measured in the physical \(Z\)-basis, yielding an outcome \(b \in \{0,1\}\)  corresponding to  \(m_{Z_{L,1}} \in\{ +1,-1\}\).  A conditional Pauli correction  completes the teleportation deterministically
\begin{equation}
\ket{\Psi_{\rm f}} = X_{L,2}^b Z_{L,2}^a  \frac{1}{2} \left(\mathbb{I} + (-1)^b Z_{L,1} \right) \ket{\Psi_a}  = \ket{b}_{L,1} \otimes \ket{\psi}_{L,2}.
\end{equation}
Importantly, these conditional logical operations can be implemented efficiently via Pauli frame updates, avoiding the need for explicit gate operations. Let us also note that when the input is e.g.\(\ket{\psi}_{L,1} = \ket{0}_{L,1}\), the  state prior to the split corresponds to a maximally-entangled logical Bell pair, enabling the use of lattice-surgery gadgets to distribute entanglement between logical qubits without any direct interaction.

\subsubsection{Teleportation error vs decoding strategy}

\begin{figure}
\centering
\includegraphics[width=\linewidth]{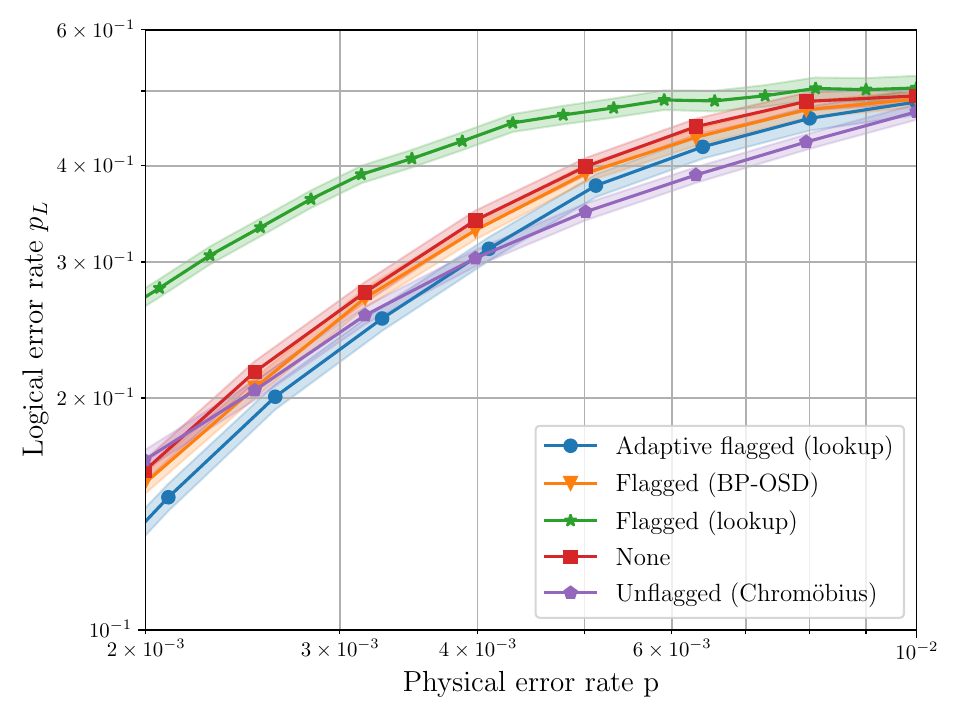}
\caption{{\bf Steane logical qubit teleportation}: performance for the teleportation of a Steane-encoded qubit, following different strategies, under standard circuit level noise. The adaptive protocol uses flagged syndrome extraction, but it switches to un-flagged syndrome extraction as soon as an error is detected, decoding in real time using a lookup table. The rest of methods in the plot use fixed (deterministic) circuits that do not depend on previous measurement outcomes, and different decoders are used to determine the corrections, which are tracked as a change to the final Pauli frame.}
\label{fig:steaneteleportation}
\end{figure}

\begin{figure*}

\subfloat[\label{fig:direct_mxx_comp}]{\includegraphics[scale=0.54]{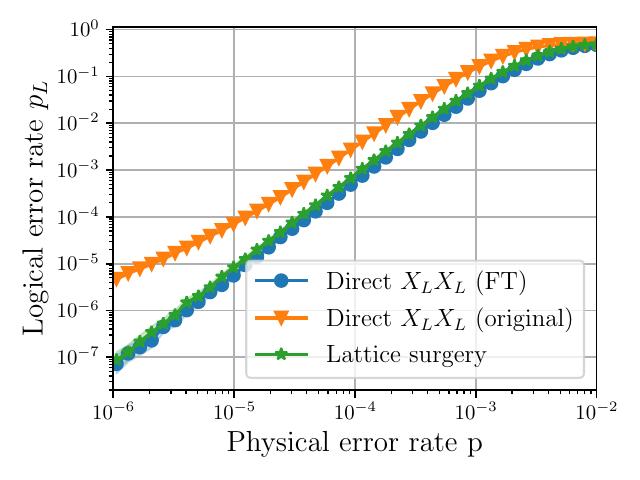}}
\subfloat[\label{fig:direct_mxx_scaling}]{\includegraphics[scale=0.54]{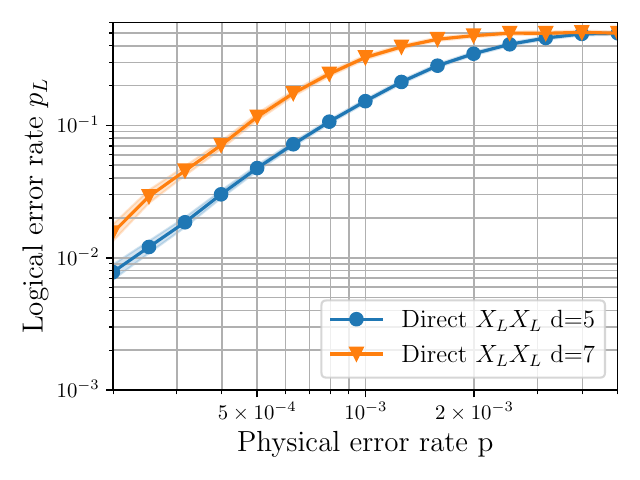}}
\subfloat[\label{fig:direct_mxx_ls}]{\includegraphics[scale=0.54]{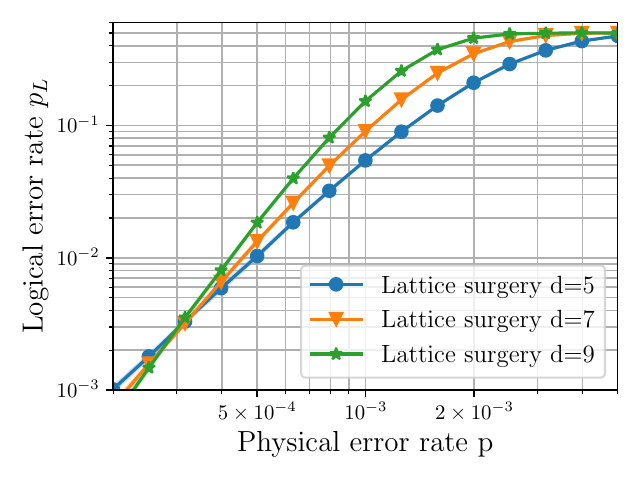}}
\caption{{\bf Lattice surgery or direct joint measurement}: Comparison of a logical qubit teleportation using two approaches to perform the joint measurement, lattice surgery and direct measurement using an ancilla and flag qubits. a) Sequential QEC syndrome extraction protocol for a $d=3$ code, under standard circuit level noise and decoded with a lookup table. For the direct measurement of the joint logical operator, we plot the original protocol described in~\cite{ryananderson2024highfidelityfaulttolerantteleportationlogical}, as well as a modified version that we designed to ensure fault-tolerance. Both the lattice surgery and the FT version of the direct measurement provide similar performance for small $d=3$ codes. b) Teleportation using a direct measurement of the joint logical operator for larger distance codes, using $\frac{d-1}{2}$ flag qubits. The circuit depth of the logical measurement increases with the code distance, worsening the performance for the explored error rates. c) Teleportation using lattice surgery. As the new surgery stabilizers can be measured in parallel, the circuit depth of a single logical joint operator measurement can be made distance-independent, and the protocol exhibits a threshold.
} 
\label{fig:directmxx}
\end{figure*}

The crucial steps for the LS teleportation are split and merge operations, since the code stabilizers change during the process, requiring special handling to preserve fault tolerance. If using dynamic circuits to minimize resources while maintaining fault tolerance, they are also the steps that involve the more complex circuits and in-sequence logic. In App.~\ref{app:lattice_surgery} we discuss the tree-like structure of the LS circuits, which require a specific set of dynamic circuits and in-sequence logic (see Fig.~\ref{fig:circbranch}). Leveraging the branching feature in our Pauli frame simulator, we  assess the performance of various logical teleportation schemes in  Fig.~\ref{fig:steaneteleportation}, also considering multiple decoders and different syndrome extraction techniques. 

Interestingly, the best performance is achieved by taking advantage of dynamic circuits, executing QEC rounds conditionally on previous syndrome measurements, in combination with a lookup table decoder, as represented by the blue dots. If we do not leverage the use of in-sequence logic and dynamic circuits, the code performance with the lookup table (green stars) is significantly degraded. Still, performance can be improved while sticking to a deterministic protocol by using more advanced decoders such as Chrom\"{o}bius (red squares) and BP-OSD (orange triangles). For the physical error rates simulated in the figure, the BP-OSD decoder is able to use information from flag qubits to get logical error rates which almost match the performance of dynamic circuits. Thus, there is a trade-off between requiring device support for dynamic circuits and having to use a more complex decoding algorithm.

\subsubsection{Caveats with fault tolerance in teleportation by a direct measurement  of joint logical operators}\label{sec:directmxx}

We have seen in the previous section that the right choice of a decoder can be important in order to  optimally benefit from the FT nature of a logical teleportation scheme. In this section, paralleling our discussion of Sec.~\ref{sec:ftcaveats}, we want to discuss some caveats in the fault-tolerance of the teleportation protocols that, to the best of our knowledge, have not been discussed carefully  in previous literature. In this case, instead of being related to subtleties of a flag-based approach, the caveats on FT focus on recent implementations of color-code teleportation~\cite{ryananderson2024highfidelityfaulttolerantteleportationlogical} based on a scheme that directly measures the joint logical operator instead of breaking it in smaller weight stabilizers, which deforms the code while preserving fault-tolerance if one adheres to the protocol discussed above.

In the standard LS approach, the  weight-6  joint operator for the teleportation on two $d=3$ color-code qubits is typically envisioned as a manipulation of the neighboring color-code boundaries using  LS merge and split operations. In the original formulation of color-code LS~\cite{Landahl2014},  the joint logical operator is split into smaller patches, leading to new higher-weight stabilizers along the boundaries that effectively merge the code into a single asymmetric logical block (see Fig.~\ref{fig:splitmerge}). This allows for constant-weight logical operations and bare-ancilla LS measurements even when increasing the code distance beyond $d=3$. In contrast, the direct measurement of the weight-6  Pauli operator~\cite{ryananderson2024highfidelityfaulttolerantteleportationlogical} operates within the logical-subspace level without 
manipulating the code's  lattice structure. Although this  simplifies the circuit complexity for the small $d=3$ codes, it  would require more ancilla-qubit resources when the distance is increased to attain fault-tolerance. Moreover, as we now show in detail, a numerical simulation of the original teleportation protocol reported in~\cite{ryananderson2024highfidelityfaulttolerantteleportationlogical} shows that it is not consistent with fault tolerance, even when restricting to $d=3$ color codes.  
Since the direct measurement of the joint logical $X_{L,1}X_{L,2}$ is not repeated, fault tolerance cannot be enforced, which would become apparent in the experiments~\cite{ryananderson2024highfidelityfaulttolerantteleportationlogical} if gate errors could be lowered to show the scaling with error rate. 

This can be easily in simulations by sweeping the rate of the depolarizing channel in the SCEM. In Fig.~\ref{fig:direct_mxx_comp}, the orange triangles show how scaling of the logical teleportation error  does not yield a single straight line on a log-log scale, but instead interpolates between two different slopes. For $d=3$ color codes, this amounts to the crossover form an $\mathcal{O}(p^2)$ scaling of the logical error to an $\mathcal{O}(p)$ one, revealing that weight-1 errors are not being completely  suppressed by the QEC rounds, losing fault tolerance. On the other hand, when the direct measurement is repeated at least twice (a third and definitive measurement may be required if both measurements disagree, or if an $X$ error is detected during syndrome extraction), 
the results in blue are consistent with  the FT scaling. We find that the teleportation error of this modified scheme  is very similar to our LS approach based on merging and splitting the color codes (green stars).

Let us finally comment that the small advantage of the direct measurement  FT scheme is  actually lost when moving to  larger distance codes. As shown in Fig.~\ref{fig:direct_mxx_scaling}, numerical simulations for $d=5,7$ suggest that there is no clear threshold in the teleportation error when the logical operator is measured directly. This outcome is, to some extent, expected: the FT measurement-circuit depth must grow with code distance $d$, and overhead of using flag qubits will also scale unfavorably with the weight of the joint operator. 
By comparison, our scaled LS scheme in Fig.~\ref{fig:direct_mxx_ls} shows a clear crossing and threshold behaviour, we conclude that the teleportation error is always smaller than for the direct measurement when $d>3$.

\subsubsection{ QEC footprint for scaled logical teleportation}

\begin{figure*}
\subfloat[\label{fig:superdense_teleport_threshold}]{\includegraphics[scale=0.54]{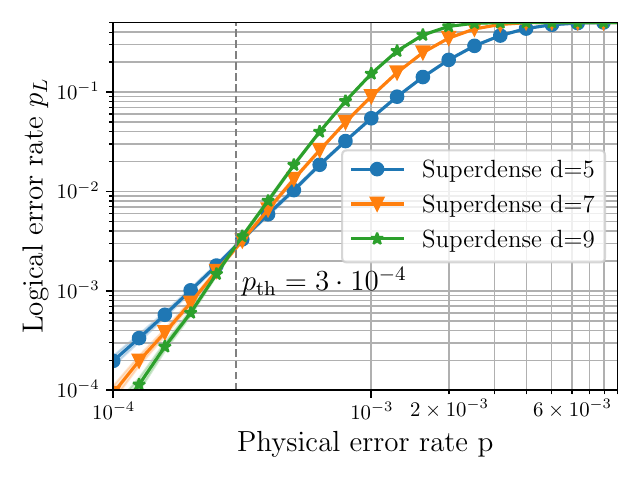}}
\subfloat[\label{fig:surface_teleport_threshold}]{\includegraphics[scale=0.54]{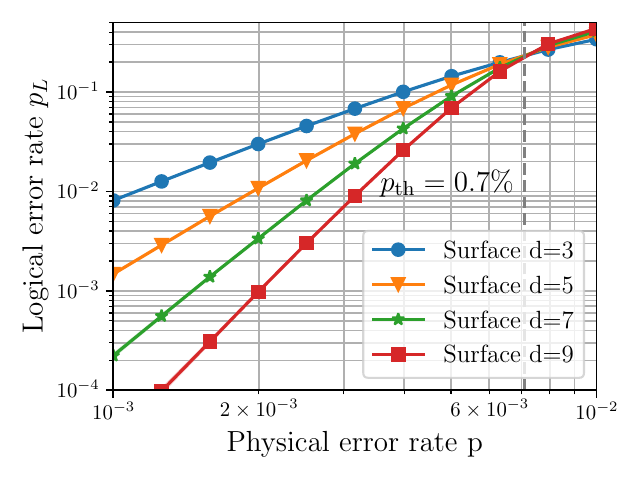}}
\subfloat[\label{fig:teleportation_footprint}]{\includegraphics[scale=0.54]{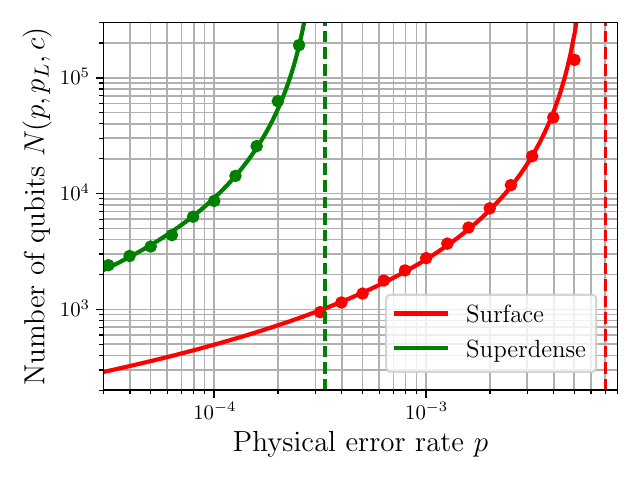}}
\caption{{\bf Logical qubit teleportation with higher code distances }: a) Threshold and error rates under standard circuit-level noise for the teleportation of a logical qubit using different color code lattice sizes, using the BP-OSD decoder. b) Same teleportation experiment using surface code logical qubits. The teleportation performance of the surface code stays close to that achieved as a quantum memory, with a threshold of $\sim0.7\%$, but for the color code the threshold is reduced to $\sim0.03\%$, an order of magnitude lower  than for the quantum memory. c) Comparison on the teraquop footprint, using the Chrom\"{o}bius decoder for the color code.}
\label{fig:scaledteleportation}
\end{figure*}

Let us close this subsection by showing the performance of the logical qubit teleportation under larger-scale simulations for higher-distance codes. In Fig.~\ref{fig:scaledteleportation}, we show the performance of different lattice sizes of the color code, and we compare them with the surface code. First of all, we observe the same effect as in Fig.~\ref{fig:decoders}: for sufficiently-small error rates, the BP-OSD decoder fails to achieve the full code distances, as can be seen for the $d=3$ color code (blue dots) at $p\approx0.05\%$. We did not explore physical error rates below $10^{-4}$, but we expect the same effect to appear for $d>3$ color codes when using BP-OSD. {Nevertheless, the loss of distance only happens for error rates that are well below those achieved by current devices.} We also observe that the threshold of the surface code $p^{\rm sc}_{\rm th}=0.7\%$ (crossing point for solid lines) remains the same as for the quantum memory, while it is significantly reduced from $p^{\rm cc}_{\rm th}=0.3\%$ to $p^{\rm cc}_{\rm th}=0.03\%$ (crossing point of dashed lines) for the color code. The reason is that, for the surface code, the merged code block preserves the surface code structure, and all code stabilizers remain of weight-4. However, for the color code, two weight-4 $Z$ plaquettes are merged to form a single weight-8 plaquette, which is not measured during the merge. As in the quantum memory case, this is a resource trade-off to transversally implement the full Clifford group.

In Fig.~\ref{fig:teleportation_footprint}, we show the \emph{teraquop} footprint~\cite{Gidney2021faulttolerant} for the logical qubit teleportation, that is, the number of physical qubits required to implement the teleportation protocol with a logical error rate below $p_L=10^{-12}$. For a physical error rate $p=10^{-4}$, the color code requires $N(p=10^{-4},p_L=10^{-12})\approx9000$, almost $\times20$ physical qubits more than the surface code ($N\approx500$ qubits) to achieve the same performance.
Due to the low threshold of the color code, physical error rates below $10^{-4}$ have to be explored, requiring a higher number of Monte-Carlo shots. This makes BP-OSD decoding impractical, as the decoding time per shot is significantly higher than when using Chrom\"{o}bius, limiting the accuracy of the logical error rates. Thus, we instead report the teraquop footprint for the Chrom\"{o}bius decoder.

\subsection{ Modular  trapped-ion   architectures }
\label{sec:architectures}

\begin{figure*}
\includegraphics[width=\linewidth]{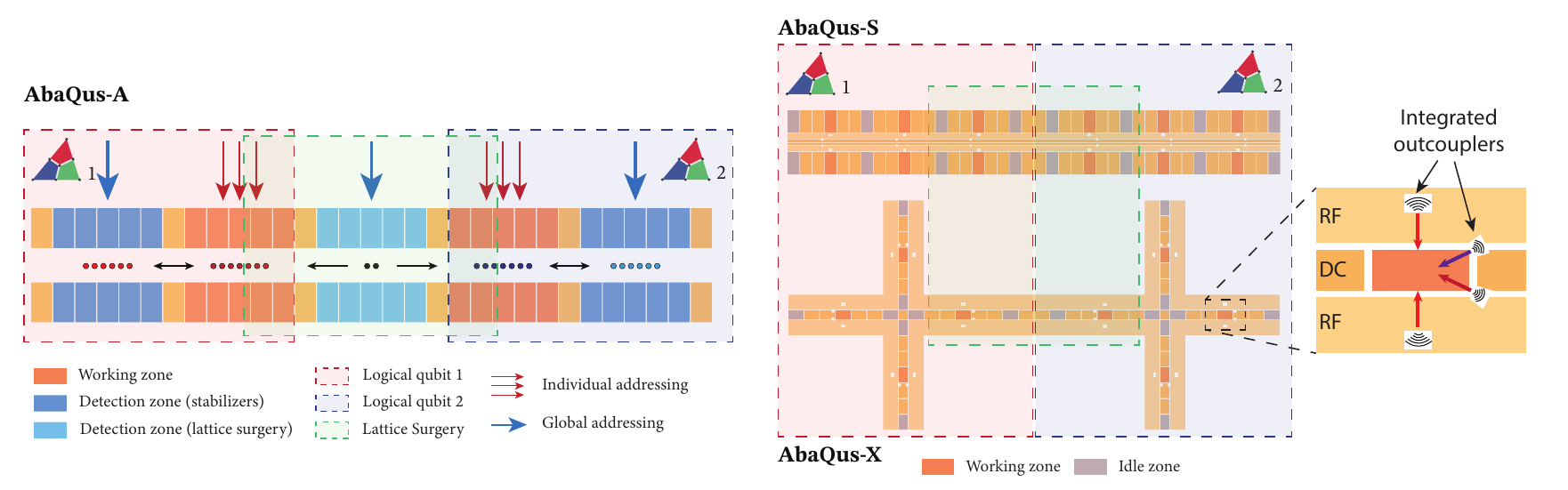}
\caption{{\bf Trapped-ion architectures for modular logical teleportation}: (Left) AbaQus-A linear segmented trap with free-space optics addressing, reduced shuttling, and dedicated regions for logical blocks, detection, and surgery ions. (Right) Integrated-photonics architectures: AbaQus-S (top) uses ion rotations to perform swaps in a linear trap, while AbaQus-X (bottom) employs a double-junction layout to allow for a more flexible ion braiding. Together with ion shuttling, split and merge, these allow for the required qubit connectivities to perform QEC and logical teleportation.
All architectures confine logical qubits to separate regions of the traps.}
\label{fig:all-architectures}
\end{figure*}

Demonstrating  FT logical teleportation towards the milestone of QEC advantage will require  architectures that integrate high-fidelity operations, modular scalability, and the ability to distribute and manipulate logical qubits  independently.
Such architectures would ideally allow to increase the code distance to decrease the logical errors exponentially when working below threshold. A key consideration for modularity is that operations between logical blocks should ideally be mediated through an interface, rather than direct interactions between their constituent physical qubits. This minimizes logical cross-talk and preserves the independent error protection of each block.
This constraint, motivated by experimental limitations on crosstalk errors, excludes the use of transversal multi-qubit gates such as the transversal CNOT gate, to advocate instead for approaches like the previous LS teleportation. In this section, we focus on trapped-ion QPUs~\cite{HAFFNER2008155,10.1063/1.5088164,10.1116/1.5126186}, and introduce specific trap architectures for  logical teleportation protocols that can meet with the above requirements.
We target teleportation fidelities above $95\%$~\footnote{ELQ program metrics: https://www.iarpa.gov/research-programs/elq} that capture the current experimental status for high-fidelity logical teleportation~\cite{Lacroix2025,ryananderson2024highfidelityfaulttolerantteleportationlogical}.

In the following sections, we present three different trap architectures that can be used to implement the teleportation protocol in a modular manner.
The first architecture, which we refer to as AbaQus-A, consists of a segmented trap hosting multiple medium-sized ion crystals.
In this setup, the ions can be individually addressed using free-space optics, allowing for all-to-all connectivities that reduce ion crystal reconfigurations.
The second and third architectures, which we refer to as AbaQus-S and AbaQus-X, feature integrated optical waveguides and gratings to direct light to specific regions hosting smaller ion crystals. We assume the two latter architectures to be operated in a QCCD fashion, with ions held in single or two-ion crystals and interconnected through shuttling. AbaQus-S employs a fully linear architecture, combining shuttling and crystal rotations for the crystal reconfigurations, while AbaQus-X makes use of cross junctions.

\subsubsection{AbaQus-A: Linear trap  with   beam-deflector all-to-all addressing}

For the first architecture we consider a linear segmented-electrode ion trap with multiple zones.
Compared to other linear traps used for previous demonstrations of QEC \cite{Nigg2014, Stricker2020, Erhard2021, egan2021faulttolerantoperationquantumerrorcorrection,egan2021faulttolerantoperationquantumerrorcorrection, Postler2022,PRXQuantum.5.030326,Pogorelov2025,butt2025demonstrationmeasurementfreeuniversalfaulttolerant}, in our proposed architecture segmented electrodes are used to create multiple potential wells, each containing a handful of ions.
The same electrodes can be used to split and merge neighboring crystals or transport crystals from one zone to another. 
  In this architecture we consider that the desired connectivity of CNOT gates within a logical block can be achieved using free-space optics \cite{Debnath2016,Figgatt2019, PRXQuantum.2.020343}.
  Since the two logical blocks are held in distant working zones, we can assume that crosstalk doesn't compromise the strict modularity and independent control of different logical qubits.
    We also assume that within a logical segment gate operations are applied sequentially to the ions.
    Thus this architecture balances the advantages of an all-to-all connectivity within each logical block with a reduced shuttling to achieve modularity and independent control.

  We consider dividing the trap layout into 5 zones: two manipulation or working zones and three detection zones as shown in the left panel of Fig.~\ref{fig:all-architectures}.
  The two sections of the trap in orange can each host the physical data ions that encode one logical qubit of the $d=3$ color code, as well as the corresponding set of ancillary syndrome and flag qubits required for syndrome extraction (see Table~\ref{tab:stab_d_3_overheads}).
  The two outer sections in blue correspond to detection zones, where the syndrome and flag qubits are shuttled after the stabilizer mappings to minimize residual photon scattering onto the data qubits during state-dependent fluorescence cycles for the syndrome readout ~\cite{Schindler_2013}. Note that the ion transport requires linear shuttling of selected ions~\cite{10.5555/2011477.2011478,Barrett2004,PhysRevLett.109.080502,PhysRevLett.116.080502,PhysRevLett.120.010501,PhysRevLett.128.050502,Fallek_2016}, together with crystal split and merge operations~\cite{Barrett2004,PhysRevLett.109.080502,PhysRevA.90.033410,doi:10.1126/science.aaw9415}.

   The central section represents an interface zone dedicated to the storage and readout of the surgery ions used for the merge operations. By shuttling ions between the central region and the logical qubit segments, the required entangling gates for LS are always performed on isolated physical qubits belonging to different logical blocks, thereby enforcing the divisibility constraint required for  modular QEC.
    Once these entangling gates have mapped the information of the merged $X$ stabilizers that conform the joint logical operator $X_{L,1}X_{L,2}$ to the surgery qubits, these can be shuttled back to the central detection zone, where they can be measured  with minimal crosstalk to the data qubits.
    
\subsubsection{AbaQus-S and AbaQus-X: Surface traps with integrated photonics}

We also consider two other architectures both incorporating integrated photonics~\cite{Mehta2016,Mehta2020,Niffenegger2020,PhysRevX.11.041033,Kwon2024,PhysRevLett.130.133201, PhysRevX.15.011040, xing_rapid_2025} into modular surface-electrode trap designs  \cite{10.5555/2011670.2011671, PhysRevLett.96.253003}.
A key distinguishing feature of these architectures is the use of integrated photonics to deliver light to the ions, replacing traditional free-space optics.
Each trap is divided into multiple individually addressable zones, with each zone hosting ion crystals composed of pairs such as data/syndrome, syndrome/flag, or data/surgery ions, along with sympathetic coolant ions.
In this approach, laser light is delivered to the zones using waveguides integrated into the trap structure and is focused onto the target ions via diffractive-grating outcouplers (see the rightmost zoomed region of Fig.~\ref{fig:all-architectures}) \cite{Mehta2016, Mehta2020, beck_grating_2024}.
This approach enables parallel quantum operations across multiple sites in a scalable manner while minimizing crosstalk between ions \cite{PhysRevX.15.011040, Kwon2024}.

Quantum operations on physical qubits require transporting them to designated working zones, while the remaining ions are held in idling zones.
These operations include single- and two-qubit gates, syndrome readout, and re-cooling.
We further assume that each operation zone can contain only one or two data qubits at a time.
Achieving all-to-all connectivity in these architectures requires reconfiguring ions into different working zones, which relies on extensive ion transport, including linear shuttling, merging, splitting, and physical swapping~\cite{Splatt_2009,PhysRevA.95.052319,van_mourik_coherent_2020, Hilder2022}.
Each information ion is accompanied by a re-cooling ion, which is used to sympathetically cool the ion crystals before any quantum operation is applied~\cite{PhysRevA.68.042302,doi:10.1126/science.1177077,Negnevitsky2018}.
This strategy upgrades the native nearest-neighbor connectivity of the integrated-photonics solution to the flexible swap-based connectivity required for QEC.

Integrated outcouplers are envisioned to drive the entangling gates required for QEC operations within individual logical-qubit zones, as well as for LS-based teleportation performed in intermediate interface zones.
Such modules can employ either a linear segmented or a double-junction architecture as shown in the right panel of Fig.~\ref{fig:all-architectures}.

The first integrated-photonics architecture, which we refer to as AbaQus-S, is based on a linear segmented trap with 14 distinct zones as shown in Fig.~\ref{fig:all-architectures}.
Of the 14 zones, 6 (shown in red) are designated for qubit operations and readout, while the remaining 8 (shown in blue) serve as idling zones.
The trap is divided into two large regions, each containing one logical qubit.
Each region contains three working zones and four idling zones.
The four central regions of the trap, comprising two working zones and two idling zones, can be repurposed to perform the LS procedure's required operations.

The second integrated-photonics architecture, termed AbaQus-X, features a double-junction electrode layout ~\cite{PhysRevLett.102.153002,PhysRevA.84.032314,Wright_2013, decaroli_design_2021,https://doi.org/10.1002/qute.202000028,Zhang_2022,PhysRevLett.130.173202,PhysRevX.14.041028, 10.1063/1.2164910}.
This layout features 24 zones, divided into 8 working and 16 idle zones. The zones are arranged into two X-shaped junctions, each containing 12 zones: 4 working and 8 idling.
We envision this architecture to be utilized such that each of the ions forming each logical qubit remains contained in one of the junctions,
As with the AbaQus-S architecture, LS can be implemented by re-purposing the two working zones located between the junctions.

This architecture offers greater transport flexibility compared to AbaQus-S, as ions can be shuttled along all directions of the X-junctions, enabling reconfiguration with fewer transport operations.
For example, transport can be designed to enable effective braiding of ion trajectories, bringing any desired pair of ions into proximity without requiring physical rotations ~\cite{PhysRevX.14.041028}.

Compared with the segmented AbaQus-A architecture, integrated-photonics solutions require more frequent shuttling: data, syndrome, flag, and sympathetic coolant ions must be transported or swapped repeatedly across segments or through junctions. However, operating with smaller ion crystals in each zone is essential to minimize laser crosstalk, motional heating, and other sources of technical noise. This approach enables laser-based entangling gates with world-leading fidelities~\cite{PhysRevLett.117.060504,PhysRevLett.127.130505}. The benefits extend to other operations, including microwave~\cite{PhysRevLett.113.220501,42w2-6ccy} and laser-driven~\cite{sotirova2024highfidelityheraldedquantumstate} single-qubit gates, as well as state preparation and measurement~\cite{PhysRevA.104.L060402,PhysRevLett.129.130501}. Furthermore, a design featuring independently controlled logical qubit zones interconnected by dynamically routed ions offers a clear path toward scalable FT logical teleportation. These architectures could demonstrate a QEC advantage in the near future, as it is not constrained by scaling limitations; larger QEC codes can be distributed across additional surface-electrode regions.
 
Collectively, the three architectures encompass the design space for  trapped-ion QPUs for implementing  logical qubit teleportation with small codes considered in this work. Identifying the most promising approach requires a detailed transpilation of the teleportation circuits, microscopic modeling of such transpilation  including a realistic description of noise processes, and advanced numerical simulations of teleportation performance under circuit-level noise. Such a comprehensive comparative assessment can guide the optimization of transport protocols, gate scheduling, and error correction strategies tailored to each hardware platform. In the following subsections, we present the results of performing Monte Carlo simulations on circuits transpiled to these architectures, utilizing an adapted noise model that incorporates different gate mechanisms, noise sources, durations of shuttling reconfigurations, and re-cooling dynamics. These simulations enable direct comparison of logical error rates. For further details on the simulator, see App.~\ref{sec:sparsesim}.

\subsection{Assessment of trapped-ion modular teleportation}

To compare these three architectures, we start by transpiling the LS teleportation protocol into a set of specific primitive operations that form the building blocks of the higher-level description presented in the previous section. The analysis considers not only the translation of CNOTs and Hadamards into the native trapped-ion gate set, but also the specific ion transport mechanisms, including linear shuttling, ion swaps, junction crossing, and split/merge operations, as well as sympathetic re-cooling and measurement/reset procedures. This transpilation  allows us to estimate the total duration of these operations, which is then used to account for errors of idling qubits.
With this method, we can realistically assess the performance of the trapped-ion color-code teleportation.

\subsubsection{Primitive trapped-ion operations }
\label{sec:primitives}

We define primitive operations as indivisible actions forming the building blocks of the teleportation protocol. These include:
\begin{itemize}
   \item {\it Gate operations:} Two-qubit entangling gates and single-qubit rotations, both of which are driven by laser light \cite{HAFFNER2008155, Bruzewicz2019}.
   
    \item {\it Transport operations:} Linear ion shuttling \cite{PhysRevLett.109.080502, walther_controlling_2012, PhysRevX.15.011040},  splitting/merging  of two different ion crystals \cite{10.5555/2012086.2012087, lancellotti_low-excitation_2024}, and junction crossings (restricted to one ion or ion pair at a time) \cite{moehring_design_2011, Zhang_2022, PhysRevLett.130.173202}.
    \item {\it Physical Swaps:} Implemented by merging two ion crystals, each containing a single information ion, and performing a physical exchange of the ions \cite{van_mourik_coherent_2020, PhysRevA.95.052319, Splatt_2009}.
    \item {\it Sympathetic re-cooling:} Prior to  entangling gates in which previous transport operations have incurred  heating above a desired target, sympathetic cooling is applied to cool the ions near the ground-state \cite{doi:10.1126/science.1177077, Negnevitsky2018, moses_race-track_2023}.
    \item {\it Measurement and reset:} State detection via state-dependent fluorescence, and re-initialization via optical pumping \cite{HAFFNER2008155, Bruzewicz2019}.
\end{itemize}

In this work, we focus on single-qubit gates implemented through focused laser-beam addressing, which drive rotations in the qubit Bloch' sphere of the form  
\begin{align}
R_{P_i}(\theta) = \exp\!\left(-\ii \tfrac{\theta}{2} P_i\right), \qquad P_i \in \{X_i,Y_i,Z_i\},
\label{eq:single_qubit}
\end{align}  
where   $P_i$ is any of the possible Pauli operators acting on the $i$-th qubit. While rotations around an axis in the equatorial plane $X(\theta)=R_X(\theta),Y(\theta)=R_Y(\theta)$ require laser radiation that couples resonantly to the qubit carrier transition~\cite{RevModPhys.75.281}, local ac-Stark shifts enable phase rotations around the $Z$ axis $Z(\theta)=R_Z(\theta)$~\cite{Schindler_2013}. Alternatively, the latter ones can be implemented via in-software phase advances.
In all cases, $\theta$ is controlled through the radiation intensity and the pulse timing.

\begin{figure}
\centering
\includegraphics[width=\linewidth]{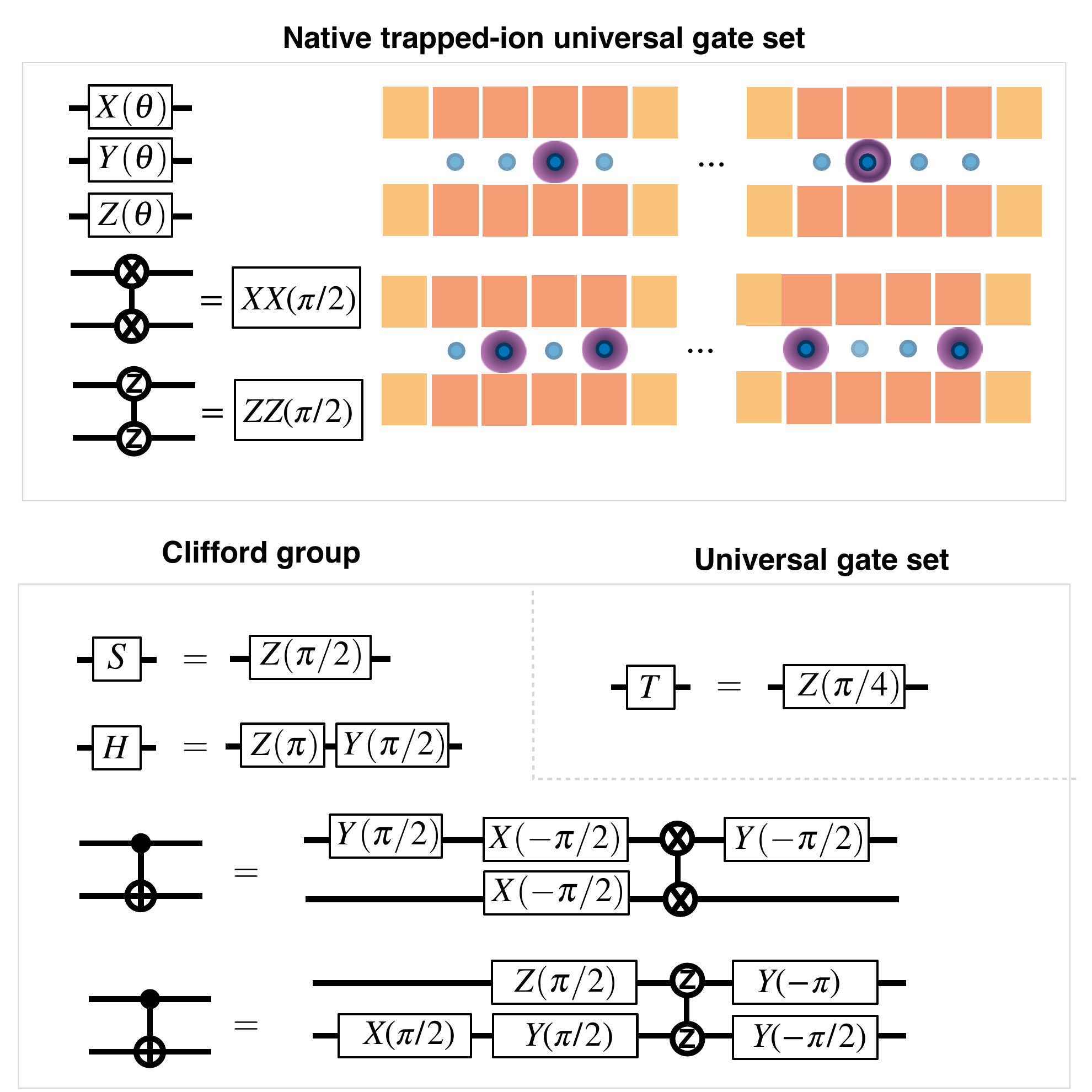}
\caption{{\bf Native trapped-ion gates}: (Top panel) Elementary single- and two-qubit operations available in a segmented trap, including local rotations $X(\theta)$, $Y(\theta)$, $Z(\theta)$ and maximally entangling gates $XX(\pi/2)$ and $ZZ(\pi/2)$. (Lower panel) Decomposition of single-qubit Clifford and $T$ gates into native rotations, where the phase gate $S$ corresponds to $Z(\pi/2)$, the $T$ gate to $Z(\pi/4)$, and the Hadamard $H$ to a sequence of $Y(\pi/2)$ and $Z(\pi)$ rotations. The construction of the CNOT gate from either an $XX(\pi/2)$ or a $ZZ(\pi/2)$ entangling operation combined with single-qubit rotations.}
\label{fig:native_gates}
\end{figure}

Two-qubit entanglement can be generated through two distinct methods, both of which involve applying spin-dependent forces to a selected pair of ions.
The Mølmer–Sørensen interaction~\cite{PhysRevLett.82.1835, Sorensen:1999,Sackett2000,Roos_2008,Benhelm:2008} produces $XX$-type qubit-qubit couplings by addressing these ions with a bichromatic beam that has frequencies close to the red and blue motional sidebands. Under optimal conditions, the resulting entangling gate reads
\begin{align}
XX_{ij}(\theta) = \exp\!\left(-\ii \tfrac{\theta}{2} X_i X_j \right).
\end{align}  
Alternatively, two-qubit operations can be achieved using light that off-resonantly couples to the qubit transition causing ac-Stark shifts. Modulating the gradient of these Stark shifts near the trap frequency can result in spin-dependent motional trajectories in phase space. With appropriate calibration, these trajectories result in an entangling operation ~\cite{Leibfried2003,PhysRevA.76.040303,PhysRevA.103.012603, clark_high-fidelity_2021, sawyer_wavelength-insensitive_2021}.
This light-shift entangling gate can be expressed as
\begin{align}
ZZ_{ij}(\theta) = \exp\!\left(-\ii \tfrac{\theta}{2} Z_i Z_j \right).
\end{align} 
Both gates are maximally entangling when $\theta = \pi/2$.
Together with the single-qubit rotations~\eqref{eq:single_qubit}, either family of entangling gates suffices to compile standard universal gate sets at the data-qubit level, such as $\{H,S,T,\mathrm{CNOT}\}$ (see Fig.~\ref{fig:native_gates}). This provides flexibility in compiling QEC circuits depending on whether $XX$- or $ZZ$-type primitives are employed. 
The exact choice of gate is motivated by a particular qubit encoding, e.g. optical qubits versus hyperfine or Zeeman qubits, the available narrow-linewidth lasers and the chosen ion species  \cite{10.1063/1.5088164}. 

Ion transport in segmented ion traps is controlled by applying time-dependent dc voltages to the segmented electrodes. 
The collection of such voltages is commonly referred to as {waveforms}.
Careful design and operation of such waveforms allow one to reshape the confining potentials, enabling linear shuttling, merge/split of crystals and  rotations of two-ion crystals~\cite{PhysRevLett.109.080502, walther_controlling_2012, PhysRevX.15.011040, 10.5555/2012086.2012087, lancellotti_low-excitation_2024, moehring_design_2011, Zhang_2022, PhysRevLett.130.173202,van_mourik_coherent_2020, PhysRevA.95.052319, Splatt_2009, kaushal2020shuttling}

In this manuscript we consider four transport primitive operations.
Linear shuttling is used to transport an ion crystal from one trapping zone to another.
These zones can be used as operation zones or idling zones.
Merge and split operations require reshaping the trapping potential from two axial wells into one, or split one well into two.
These operations open the possibility of reconfiguring data and ancilla qubits such that ions undergoing an entangling operation sit in the same potential well.
Since this operation results in the instantaneous configuration of vanishing  quadratic axial trapping, e.g. single to double-well quartic potential, careful control is needed to minimize the resulting motional excitation ~\cite{10.5555/2012086.2012087, PhysRevA.90.033410}. 
Crystal rotations which allow for a physical swap between qubits are realized by gradually tilting and reshaping the confinement potential so its principal axes rotate in a controlled fashion.
\cite{van_mourik_coherent_2020, PhysRevA.95.052319, Splatt_2009}.
Finally, dc waveform  control can also be used to transport ions through junctions which allow to increase the dimensionality of the qubit register from one to two dimensions and can be used to lower the time required for ion crystal reconfigurations.

After a set of transport operations, the transported ions will likely result in some undesired motional excitation.
Different physical mechanisms can induce such excitations.
These can include coherent excitations caused by imperfect timing of the waveform \cite{PhysRevLett.109.080502, walther_controlling_2012}, thermal population exchange between different motional modes caused by mode crossings \cite{lancellotti_low-excitation_2024}, increased heating rates caused by lowering of trapping frequencies \cite{bruzewicz_measurement_2015, brownnutt_ion-trap_2015}, as well as uncontrolled potentials caused by the presence of dielectric materials in the traps with integrated photonics \cite{PhysRevX.15.011040}.

To achieve the high gate fidelities required for the logical teleportation, ion crystals need to be re-cooled before driving any  gate after the associated crystal reconfigurations.
To protect the information stored in data/syndrome/flag/surgery qubits during cooling, the ion crystals can be sympathetically cooled down near the ground state.
One option to perform the sympathetic cooling is to use mixed-species ancilla ions where one ion species is used to store information while the other one is used for re cooling \cite{Negnevitsky2018, doi:10.1126/science.1177077, moses_race-track_2023}.
This approach leverages large spectral isolation between the coolant and information ions thus reducing errors induced by off-resonant scattering during re-cooling.
Nevertheless, mass mismatch between the two species causes the participation ratio of the two ions in the different modes to be largely uneven, thus slowing down cooling \cite{home_quantum_2013}.

Alternative approaches use ions of the same species both as coolant and information ions.
One possibility is to use two different isotopes from the same species, thus achieving a better mass relation between coolant and information ions, at the expense of reduced spectral isolation. 
Another possibility is to use different encodings within the same species \cite{shi_long-lived_2025, allcock_omg_2021}.
This approach, however, is limited by the lifetime of the excited states used to shelve the information during re-cooling.

Ion crystals can be sympathetically re-coooled in a number of ways depending on the amount of vibrational excitation. Doppler cooling is fast but leaves ions at motional excitations at temperatures which compromise gate fidelities.
Methods such as  Sisyphus or electromagnetic induced transparency (EIT) cooling can bring ions to excitations below 1 quanta in a broadband manner, allowing also for the cooling of many modes. While current demonstrations of Sisyphus cooling reach temperatures of $\bar{n}\sim$1 phonons \cite{clements_sub-doppler_2024, joshi_polarization-gradient_2020, ejtemaee_3d_2017}, EIT cooling can reach the percent level~\cite{PhysRevA.93.053401, xing_rapid_2025}.
Sideband cooling \cite{reed_comparison_2024} is commonly used for achieving $\overline{n} < 0.01$ for high fidelity gates \cite{ballance_high-fidelity_2016, Sorensen:1999, kirchhoff_correction_2025}.

\subsubsection{Operational assumptions and timings}
\label{sec:operational-assumptions}
We apply the assumptions and constraints detailed in Sec.~\ref{sec:primitives} to transpile the QEC circuits into the specific transport steps required for each architecture described in Sec.~\ref{sec:architectures}.
To enable comparison of QEC performance across architectures, we assign a specific timing to each basic operation, facilitating translation into a quantitative error model.
These timings allow us to map transport operations onto idle periods, during which qubits are susceptible to environmental errors modeled as pure dephasing.
Evaluating these timings during each circuit transpilation is necessary to determine the overall error rates within the microscopic noise model.

For AbaQus-A, we assume all-to-all connectivity for two-qubit gates among pairs of ions within a chain of up to 13 ions.
Each chain consists of 7 data qubits and 6 syndrome and flag qubits, enabling the simultaneous syndrome extraction scheme described in Sec.~\ref{sec:syndrome_extraction}. To achieve modularity and operational independence, two such chains are maintained and operated separately in distinct zones of AbaQus-A.
Since each chain shares common vibrational modes, entangling gates for each stabilizer must be applied sequentially. This constraint is addressed by introducing idling gaps for qubits not involved in the CNOT gates.
In contrast, parallel operations during readout enable simultaneous measurement of all stabilizers of a given type. This approach reduces idling errors compared to a fully sequential measurement strategy.
In the AbaQus-A architecture, required transport operations are limited to splitting the 13-ion chain into two separate chains, one containing 7 data qubits and the other 6 ancilla qubits, or merging these chains back into a single chain.
We also consider linear shuttling of the partial chains between gate zones and readout zones.

For the integrated photonics architectures, i.e. AbaQus-S and AbaQus-X, operations are subject to the following constraints and assumptions:
\begin{enumerate} 
\item Each working zone can host at most two qubit ions at a time.
\item All working zones execute the same operation simultaneously. Qubits not involved in the gates are transported and stored in dedicated idling zones until required at a later stage of the protocol.
\item We assume there is no optical cross-talk between operations on ions located in different zones.
    \item Two ions can only be swapped by first merging them into a single crystal, performing the physical rotation, and then splitting them again.
    \item Only two information ions can be swapped at a time.
    \item Merges are restricted to combining exactly two wells into one, while splitting splits one well into two.
\end{enumerate}

These constraints, based on realistic assumptions about trapped ion operation, may initially limit protocol flexibility. However, because transport and gate operations can be executed simultaneously with minimal interference, these limitations are offset by the time advantages and scalability provided by parallelism. When multiple primitive operations are performed in parallel across different zones, the duration of the most time-consuming operation, including re-cooling, determines the overall time required for that step.

\begin{figure*}
\includegraphics[width=\linewidth]{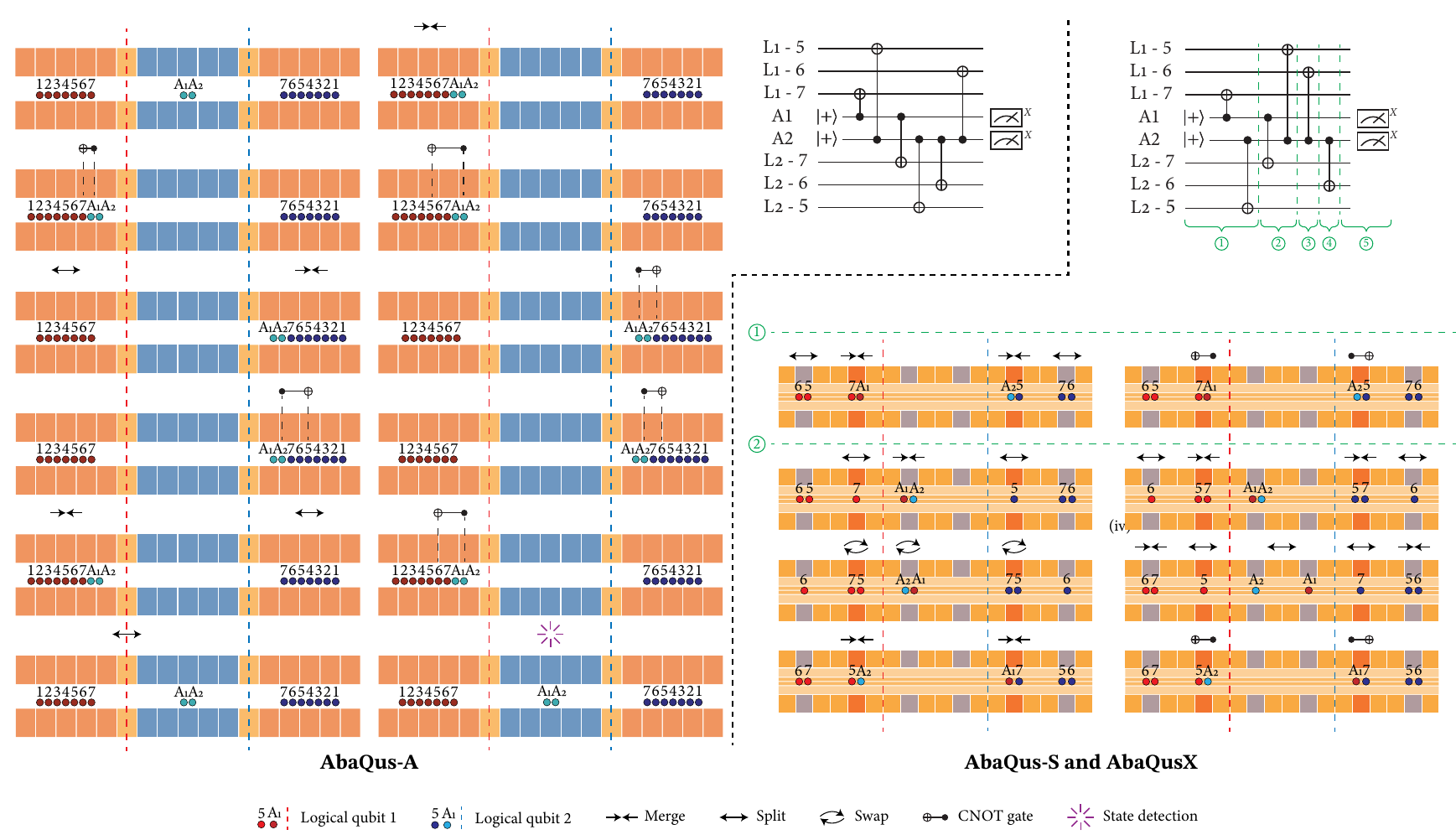}
\caption{{\bf Trapped-ion transpilation for lattice surgery} (Left) Circuit and transport sequence required to measure the joint logical X operator in the AbaQus-A architecture. (Right) Circuit and section of the transport sequence required for the joint measurement in the AbaQus-S and AbaQus-X architectures. In this case, only the primitive steps of sections 1 and 2 of the circuit (marked with green dashed lines) are shown.}
\label{fig:surgery_transpilation}
\end{figure*}

For each architecture, we estimate 3 different sets of durations for the primitive operations.
One of these sets aligns with the current state of the art, while the other two are based on plausible intermediate and optimistic improvement projections.
In addition, we estimate the re-cooling times needed to achieve high-fidelity gates.
For the AbaQus-A architecture, we consider a simple model in which cooling is performed only after measurements, thereby increasing the total readout time.
This approach is justified because transport operations are expected to have minimal impact on the excitation of the radial motional modes used for entangling operations. 
The operational timings for this architecture, including gate durations and cooling times, are presented in Table~\ref{tab:times-for-primitives}.

\begin{table}[ht]
  \centering
    \begin{tabular}{lccccc}
    \hline
    \hline
    \rowcolor[rgb]{ .91,  .91,  .91}       & \textbf{Split/merge} & \textbf{Linear} & \textbf{Gate} & \textbf{Readout} & \textbf{Re-cooling} \\
    \hline
    \rowcolor[rgb]{ .937,  .714,  .714} 
    \hline
    \textbf{Duration }($\mu$s) & 300  & 300   & 300   & 400 & 2000 \\
    \rowcolor[rgb]{ .984,  .886,  .835} 
    \hline
    \textbf{Duration }($\mu$s) & 200   & 200   & 250   & 300 & 1000 \\
    \rowcolor[rgb]{ .855,  .949,  .816} 
    \hline\textbf{Duration} ($\mu$s)& 100    & 100   & 150   & 200 & 500 \\
    \hline
    \hline
    \end{tabular}
      \caption{{\bf Estimated timings for AbaQus-A:} Current operations (red)  and projected intermediate (orange) and optimistic (green) improvements.}
  \label{tab:times-for-primitives}
\end{table}

For the integrated-photonics architectures, we instead account for the total accumulated excitation induced by transport and measurements, deferring cooling until a single- or two-qubit gate is applied \cite{Pino2021, moses_race-track_2023}.
We assume the excitation to have a coherent component, independent of the operation duration, and a thermal component, calculated as the heating rate multiplied by the operation duration.
We further assume the heating rates to be [$10^3$, $10^2$ and 1] quanta/s for the three timing scenarios. 
Table~\ref{tab:times-for-primitives_integrated} presents the assumed durations for the various primitive operations and the resulting excitation of the crystal following each operation.
The values in the table are informed by recently published results \cite{Pino2021,van_mourik_coherent_2020, PhysRevLett.130.173202,PhysRevX.14.041028}.

We use these values to estimate re-cooling times after reconfiguration of the qubit registers.
Here, we assume a cooling model in which the instantaneous cooling rate is proportional to the mean phonon occupation of the crystal.
Therefore, the total recooling time is
\begin{equation}
t_\text{cool}=\frac{1}{W_c}\log{\frac{\bar{n}}{\bar{n}_0}},
\label{eq:cooling-time}
\end{equation}
where $\bar n$ denotes the total excitation following a given set of transport operations, $\bar n_0$ is the target excitation after cooling, and $W_c$ represents the cooling rate.
For the calculations, we assume $\bar n_0 = 0.01$ quanta.
We estimate $\bar{n}$ as the sum of the total thermal and coherent components during the transport sequence.
We assume $W_c$ = [$1$, $3$,  $5$] $\times10^{4}$ quanta/s for the three timing scenarios.

\subsubsection{Circuit transpilation into primitive operations}

We now  illustrate the architecture-dependent trapped-ion transpilations of the  QEC gadgets required for logical teleportation with $d=3$ color codes.
The presented transpilations supersede previous efforts~\cite{PhysRevX.7.041061,PhysRevA.100.062307,PhysRevA.99.022330}, where  limiting constraints on the transport primitives were not considered in detail. For example, prior work considered crystal rotations with more than two ions and did not address limitations on parallelizing operations. Additionally, previous analyses did not account for increased phonon numbers per operation or incorporate cooling and heating rates to estimate the time required to reach a target mean phonon number for subsequent entangling gates.
We generate transport sequences for the complete protocol, divided into three main components: logical state preparation, stabilizer readout, and lattice surgery. 
For brevity, we exemplify the transpilation procedure only for the LS section of the protocol in this section.

\begin{table*}[htbp]
  \centering
    \begin{tabular}{lccccccc}
    \hline
    \hline
    \rowcolor[rgb]{ .851,  .851,  .851}       & \textbf{Split} & \textbf{Merge} & \textbf{Junction transport} & \textbf{Swap} & \textbf{Linear transport} & \textbf{Gate} & \textbf{Readout} \\
    \rowcolor[rgb]{ .937,  .714,  .714} 
    \hline
    \textbf{Duration} ($\mu$s)& 300   & 300   & 300   & 300   & 300   & 300   & 250 \\
    \rowcolor[rgb]{ .937,  .714,  .714} \textbf{Coherent excitation after operation} & 3 & 3 & 3 & 4 & 2 & 0 & 0\\
    \rowcolor[rgb]{ .937,  .714,  .714} \textbf{Thermal excitation after operation} & 0.3 & 0.3 & 0.3 & 0.3 & 0.3 & 0 & 10\\
    \rowcolor[rgb]{ .937,  .714,  .714} \textbf{Total excitation after operation} & 3.3 & 3.3 & 3.3 & 4.3 & 2.3 & 0 & 10\\
    \rowcolor[rgb]{ .984,  .886,  .835} 
    \hline
    \textbf{Duration} ($\mu$s)& 200   & 200   & 200   & 200   & 200   & 200   & 200 \\
    \rowcolor[rgb]{ .984,  .886,  .835} \textbf{Coherent excitation after operation} & 1 & 1 & 1 & 2 & 0.5 & 0 & 0\\
    \rowcolor[rgb]{ .984,  .886,  .835} \textbf{Thermal excitation after operation} & 0.02 & 0.02 & 0.02 & 0.02 & 0.02 & 0 & 5\\
    \rowcolor[rgb]{ .984,  .886,  .835} \textbf{Total excitation after operation} & 1.02 & 1.02 & 1.02 & 2.02 & 0.52 & 0 & 5\\
    \rowcolor[rgb]{ .855,  .949,  .816} 
    \hline
    \textbf{Duration} ($\mu$s)& 100   & 100   & 100   & 100   & 100   & 50    & 150 \\
    \rowcolor[rgb]{ .855,  .949,  .816} \textbf{Coherent excitation after operation} & 0.2 & 0.2 & 0.2 & 1 & 0.1 & 0 & 0\\
    \rowcolor[rgb]{ .855,  .949,  .816} \textbf{Thermal excitation after operation} & 0 & 0 & 0 & 0 & 0 & 0 & 1\\
    \rowcolor[rgb]{ .855,  .949,  .816} \textbf{Total excitation after operation} & 0.2 & 0.2 & 0.2 & 1 & 0.1 & 0 & 1\\
    \hline
    \hline
    \end{tabular}
  \caption{{\bf Estimated times, motional excitation, and cooling overheads for the AbaQus-S and AbaQus-X architectures.} 
  Duration, coherent excitation, thermal excitation and total excitation for all the primitive operations.
  All values are shown under current estimates (red) and intermediate (orange) and optimistic (green) improvements.}
   \label{tab:times-for-primitives_integrated}
\end{table*}

Lattice surgery, central to the teleportation protocol, involves a joint measurement of two logical operators, such as the weight-6 $X_{L,1} X_{L,2}$ operator. This measurement can be decomposed into two measurements: a weight-4 operator and a weight-2 operator acting on boundary qubits of the logical blocks as discussed in~\ref{sec:ls-teleportation-scheme}. Since this measurement cannot be performed by direct operations on the two logical qubits together to comply with the modularity and independence conditions for logical teleportation, the corresponding measurement circuits involve controlled operations and detailed scheduling between the physical and ancillary ions  that are sequentially shuttled, merged and split, as shown in Fig.~\ref{fig:surgery_transpilation}.

The left panel of the figure shows the steps needed to measure the logical $X_{L,1} X_{L,2}$ operators using the AbaQus-A architecture.
The all-to-all connectivity present within the working zones of each logical qubit
can be leveraged to reduce the complexity of ion transport operations.
To perform the measurements, the ancilla is first merged into the first logical qubit, where two CNOT gates are applied. It is then separated from the first logical qubit and merged into the second logical qubit, where three additional CNOT gates are applied. Finally, the ancilla is separated from the second logical qubit and merged again into the first logical qubit, where the last CNOT gate is applied.
The ancilla ions are subsequently transported to the intermediate detection zones, where their states are measured and reset. Re-cooling of the ions can also be performed at this stage.

The right panel of the figure shows the circuit used for the joint measurement and the corresponding transport sequence required to implement sections 1 and 2 of the circuit for the integrated trap architectures, subject to the constraints outlined in~\ref{sec:operational-assumptions}.
We assume the LS sequence for both the AbaQus-S and AbaQus-X architectures to be identical since the intermediate region of both architectures features the same layout, consisting of two working zones and six idling zones.
However, the state-preparation and stabilizer readout sections of the protocol differ between the two architectures.
Step 1 of the circuit consists of two parallel CNOT operations, each between an ancilla and a data ion for each logical qubit. Step 2 involves swapping the ancillas and performing another set of two parallel CNOT operations between the ancillas and the data ions.
We transpile the remaining circuit and other sections of the protocol using a similar approach.
Based on the values from Table~\ref{tab:times-for-primitives_integrated}, we estimate the total motional excitation for step 2 to be  [27,9, 2.4] quanta, necessitating approximately [0.79, 0.23, 0.11] ms of re-cooling for the three respective timing scenarios.

\begin{figure*}
    \centering
    \includegraphics[width=\linewidth]{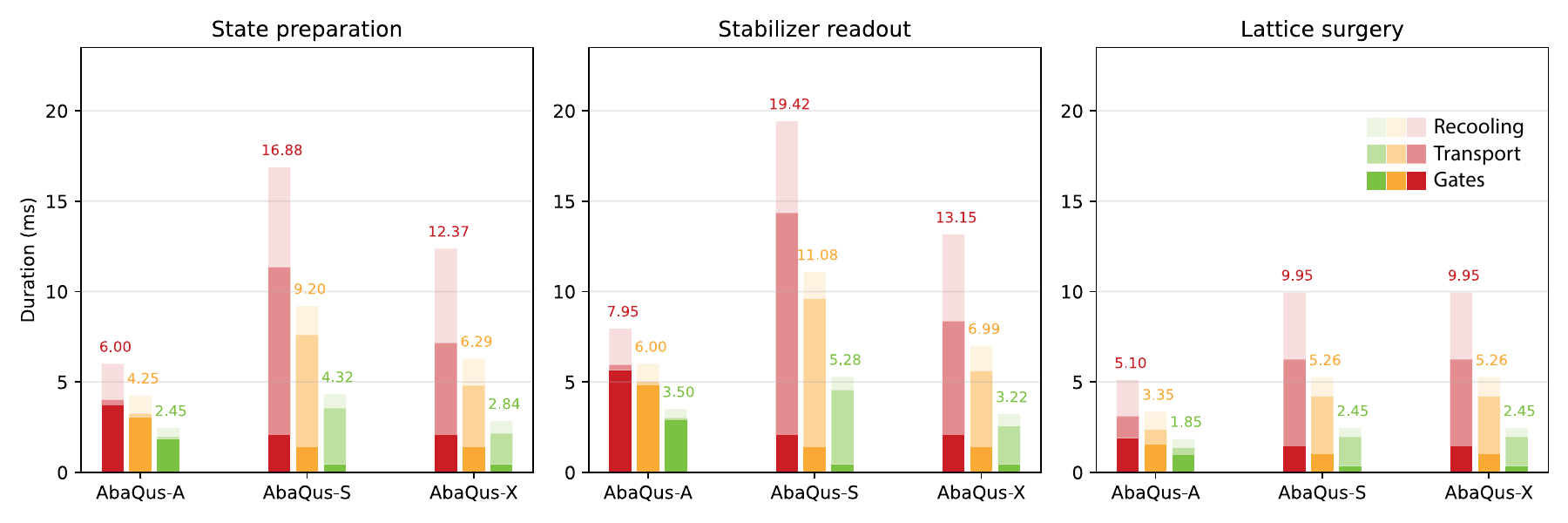}
    \caption{{\bf Duration for the transpiled QEC gadgets:} Duration of the gates, recooling and transport for the logical state preparation, stabilizer readout and lattice surgery of the three architectures under the three timing scenarios. The durations of the stabilizer readout only include one type of stabilizer sub-round (X or Z) and the durations of the lattice surgery only include one measurement of the joint logical X operator.}
    \label{fig:architecture-timings}
\end{figure*}

The next step is to work out the idle times associated to the transpiled circuits, including the sympathetic re-cooling, and translate those times to the microscopic error model. In 
Fig.~\ref{fig:architecture-timings}, we present  these timings, and how they distribute among actual quantum gates, ion transport, and re-cooling. We include the current (red), intermediate (orange) and optimistic (green) estimates based on the numbers discussed in Tables~\ref{tab:times-for-primitives} and~\ref{tab:times-for-primitives_integrated}, for all three architectures.

This comprehensive transpilation and timing analysis enables us to describe in the following subsection  a realistic multi-parameter noise model at the circuit level for the LS-based logical teleportation protocols. By incorporating detailed ion transport, gate durations, and re-cooling overheads across these different architectures, we can efficiently predict  the expected teleportation fidelities and directly compare the three approaches. 

\subsubsection{Multi-channel  noise model with microscopically-estimated error rates}

For the simulations presented in previous sections comparing  different QEC strategies and  decoders, 
we considered the SCEM with a single-parameter  depolarizing  rate $p$, accounting for a common noise that affects  the gates, measurements and idle errors equally.
In the following sections, we upgrade this model to simulate realistically the $d=3$ color-code teleportation protocol for the three  trapped-ion architectures.
For those simulations, we introduce a more-detailed multi-channel error model, where we include various quantum channels to discuss the deviations from the ideal  behavior of the transpiled circuits.
This model includes different error rates for the different quantum operations and incorporates other constraints such as the different crosstalk/transport overhead of the three architectural approaches.
We consider all channels to be Pauli type~\eqref{eq:pauli_channel}, and are built from:

\begin{itemize}
\item {\it Single-qubit gates}: depolarizing noise with rate $p_\text{1q}$.
\item {\it Two-qubit gates}: two-qubit depolarizing noise with rate $p_\text{2q}$. Since the native gate is not a CNOT (e.g. M\o lmer-S\o rensen or light-shift gates), the gate-error overhead $p_\text{1q}$ of the single-qubit rotations required to implement a CNOT is included in $p_{2q}$.
\item {\it Cross-talk error}: two-qubit depolarizing noise with rate $p_\text{ct}$ acting on all qubit pairs $(t,n)$, where $t$ is one of the two target qubits and $n$ is a nearest-neighbor qubit of either target. 
\item {\it Measurement error}: measurement outcome flipped with probability $p_\text{m}$.
\item {\it Reset error}: Bit-flip  $X$ error channel with probability $p_\text{r}$ after preparation.
\item {\it  Idle error}: Dephasing $Z$ error channel with probability $p_\text{id}(t)=\frac{1}{2}(1-\ee^{-t/T_2})$.
 These errors happen when a qubit is idling due to an ongoing operation on other qubits, during reconfiguration of the qubits using transport operations, or during re-cooling after the latter ones.
 The error rate depends on the duration of the operation during which the qubit remains idle, which we denote as $t$. We assume qubits to be stored in the ground state during idling and transport periods, switching to the optical encoding before applying a gate. Thus, we consider the relaxation time $T_1$ as infinite, while we set $T_2$ to the {effective} coherence time of the ground state qubit considering dynamical decoupling.
\end{itemize}

\begin{table}[htbp]
  \centering
    \begin{tabular}{ccccc}
    \hline
    \hline
    \rowcolor[rgb]{ .851,  .851,  .851} \textbf{1Q gate} & \textbf{2Q gate} & \textbf{Cross-talk} & \textbf{Readout} & \textbf{Init} \\
    \rowcolor[rgb]{ .851,  .851,  .851} $p_\text{1q}$ & $p_\text{2q}$ & $p_\text{ct}$ & $p_\text{m}$ & $p_\text{r}$ \\
        \hline
        \hline
        \multicolumn{5}{c}{AbaQus-A}\\
    \rowcolor[rgb]{ .937,  .714,  .714}  $3.6\cdot10^{-3}$  & $2\cdot10^{-2}$   & $2\cdot10^{-4}$   & $2\cdot10^{-3}$   & $7\cdot10^{-4}$  \\
    \rowcolor[rgb]{ .984,  .886,  .835}   $2\cdot10^{-3}$   & $1\cdot10^{-2}$   & $1\cdot10^{-4}$   & $1\cdot10^{-3}$   & $5\cdot10^{-4}$  \\
    \rowcolor[rgb]{ .855,  .949,  .816}   $1\cdot10^{-3}$   & $5\cdot10^{-3}$   & $5\cdot10^{-5}$   & $5\cdot10^{-4}$   & $2.5\cdot10^{-4}$  \\
    \hline \hline
    \multicolumn{5}{c}{AbaQus-S and AbaQus-X}\\
    \rowcolor[rgb]{ .937,  .714,  .714}  $4\cdot10^{-3}$   & $1.2\cdot10^{-2}$ & - & $2.4\cdot10^{-4}$   & $2.4\cdot10^{-4}$  \\
    \rowcolor[rgb]{ .984,  .886,  .835}   $2\cdot10^{-3}$   & $6\cdot10^{-3}$  & - & $1.2\cdot10^{-4}$   & $1.2\cdot10^{-4}$  \\
    \rowcolor[rgb]{ .855,  .949,  .816}   $5\cdot10^{-4}$   & $1\cdot10^{-3}$  & - & $5\cdot10^{-5}$  & $5\cdot10^{-5}$  \\
        \hline
        \hline
    \end{tabular}
      \caption{{\bf Gate, readout and reset performance.} Depolarizing probabilities for single- and two-qubit gates as well as readout and reset errors for all architectures. Values in red show the current values \cite{PRXQuantum.5.030326,PhysRevA.107.042422} while orange and green show intermediate and optimistic improvements. }
  \label{tab:system-performance}
\end{table}

\subsection{Comparative analysis of different architectures}
\begin{figure*}[t!]
\subfloat[\label{subfig:longchains_gs_vs_T2}]{\includegraphics[width=.49\linewidth]{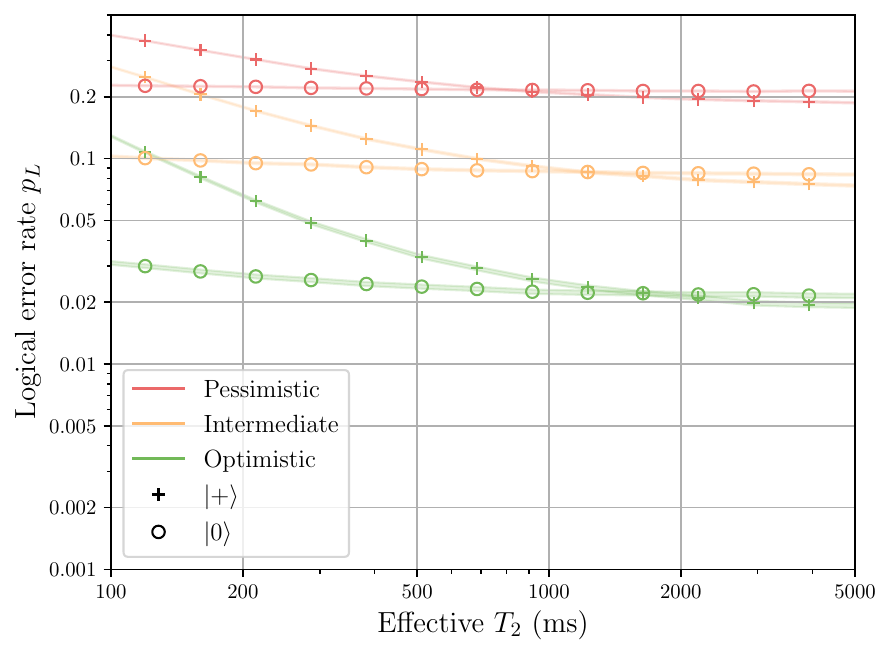}}
\subfloat{\includegraphics[width=.49\linewidth]{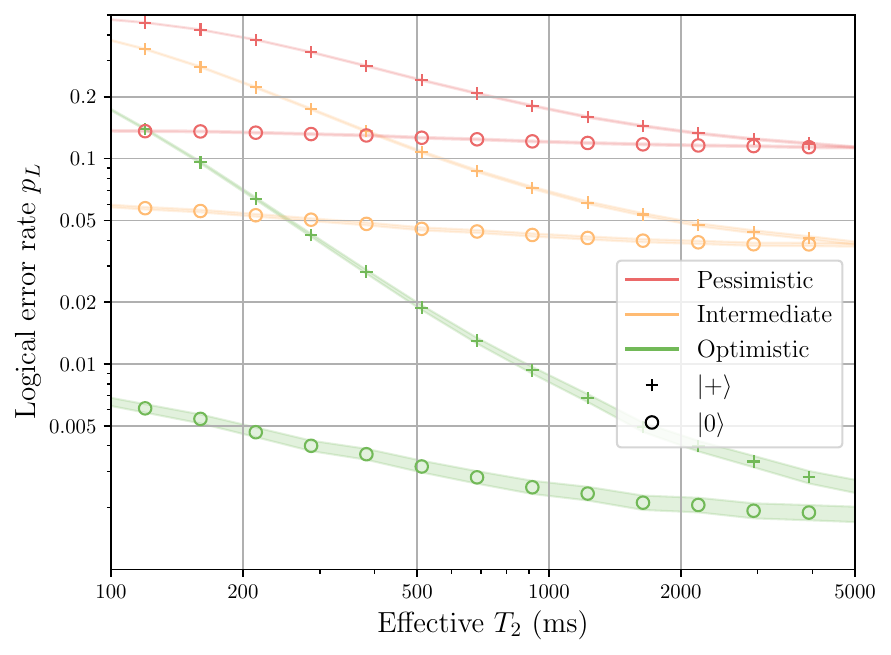}}\\
\subfloat{\includegraphics[width=.49\linewidth]{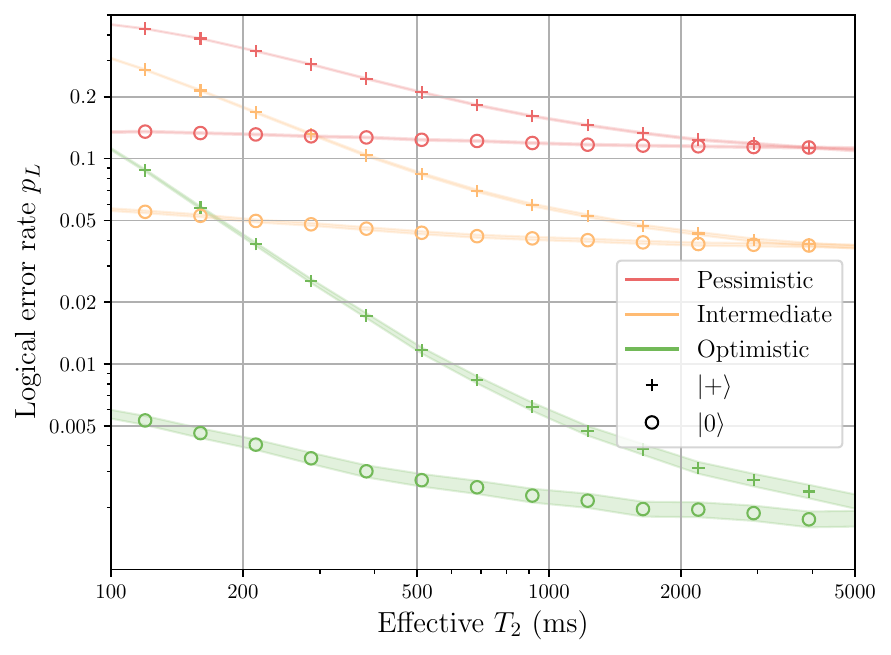}}
\subfloat[\label{subfig:trap_comparison}]{\includegraphics[width=.48\linewidth]{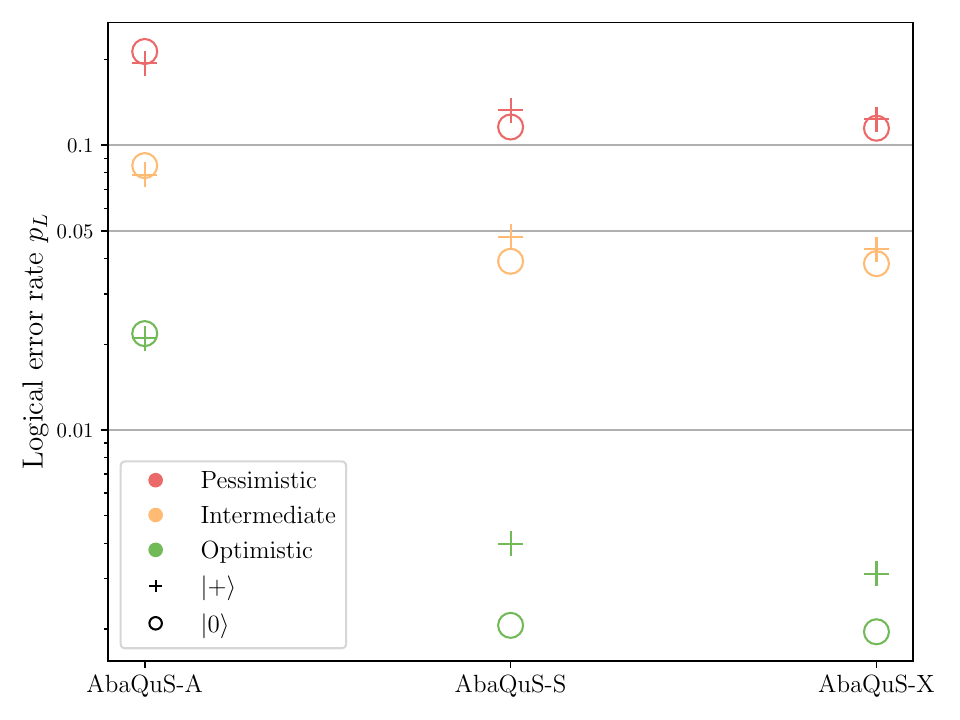}}
\caption{{\bf Logical teleportation in trap-ion devices}: simulated performance of the teleportation protocol for a logical qubit encoded in the Steane code for the AbaQus-A (a), AbaQus-S (b) and AbaQus-X (c) architectures as a function of the coherence time $T_2$. (d) Architecture comparison for $T_2=2\,\mathrm{s}$.
}
\label{fig:trap-performance-improvements}
\end{figure*}

Based on the architectural capabilities described in the previous sections, the transpilation and the multi-channel error model, and our sparse Pauli frame simulator with adaptive dynamic circuits, we can now perform numerical simulations of the logical teleportation protocol  adapted to the three trapped-ion AbaQus architectures. A simple comparison with an error threshold as in the the previous teleportation using the SCEM is no longer possible, as we have now  multiple noise parameters.
However,  representative plots can be achieved by changing one of those parameters.  Fig.~\ref{fig:trap-performance-improvements} shows the performance of the logical teleportation  as a function of the microscopic $T_2$ time, which essentially captures the sensitivity of the various architectures towards idling errors.

For each architecture, in addition to varying the effective $T_2$ time,  we consider the performance of individual operations in the three different regimes (current, intermediate, optimistic), using the corresponding error rates and timings outlined in Tables~\ref{tab:times-for-primitives}, ~\ref{tab:times-for-primitives_integrated} and ~\ref{tab:system-performance}.
We observe that the logical error rate for the AbaQus-A  architecture (Fig.~\ref{subfig:longchains_gs_vs_T2}) exhibits minimal dependence on the system's $T_2$  time.
This trend is especially evident in the current and intermediate error rate regimes, indicating that physical gate errors are the primary contributors to logical infidelity. In particular, the simplifications  offered by the all-to-all connectivity on intermediate-size crystal comes with limits in the gate fidelities, particularly  for two-out-of-many entangling   gates.
Achieving lower logical teleportation errors in this architecture will require substantial improvements in two-qubit gate operations. In addition,  measurement errors are also a limiting factor, so further improvements would be required to get well above the $95\%$ teleportation fidelity in the future, which cannot be achieved with the current performance levels even for long coherence times exceeding the second.

Panels (b) and (c) of Fig.~\ref{fig:trap-performance-improvements} show the performance of the integrated-photonics  architectures.
In contrast to the AbaQus-A architecture, the attainable logical error rate in these architectures can be much lower, and depends strongly on the system $T_2$ time.
The explanation for such behavior is two-fold.
First, the modularity provided by smaller ion crystals enables higher operational fidelities.
Second, the requirement for ion transport and re-cooling operations to achieve necessary connectivity leads to extended idling periods during which ions are susceptible to dephasing.
Under the optimistic timing scenario, these architectures can achieve teleportation fidelities for the logical $\ket{0}$ state exceeding 95\% target regardless of the specific value of the $T_2$ time. 
In the intermediate  scenario, the logical $\ket{0}$ state can be teleported with fidelity approaching the 95\% target, whereas in the pessimistic scenario, fidelity remains below 90\%. For the $T_2$ values under consideration, the fidelity of the teleported $\ket{+}$ state remains below the $95\%$ target considering   the current timing scenario.
The $\ket{+}$ state can be teleported with fidelities exceeding the 95$\%$ target when $T_2$ is approximately 200 ms in the optimistic scenario and 1000 ms in the intermediate scenario.
We note that such coherence times are achievable in state-of-the-art ion systems \cite{ruster2016long}.

Let us close this subsection by providing a more visual inter-architecture  comparison of the logical teleportation performance. In Fig.~\ref{subfig:trap_comparison}, we plot the logical teleportation error for all architectures using present-day experimental error rates, as well as intermediate and optimistic improvement estimates for the reconfiguration timings and cooling conditions.  This plot shows the clear advantage of the integrated-photonics architectures, particularly for the envisioned improvements on shuttling and cooling. Since current gate fidelities are worse for long ion chains,  the teleportation performance of Abaqus-A  cannot reach the 95$\%$ illustrative target. On the other hand,  the integrated-photonics architectures, which only consider gates in one- and two-ion crystals, can indeed reach and surpass this target for all four cardinal states already for intermediate improvements. It is also interesting to  observe that both integrated-photonics architectures have a similar performance, with the double junction slightly outperforming the linear design. The underlying reason is that the parallelization of transport operations  is more efficient in the junction approach (see Fig.~\ref{fig:architecture-timings}).
\section{\bf Conclusions and outlook}
\label{sec:conclusions}

In this work, we have developed and benchmarked a modular logical teleportation protocol based on lattice surgery of triangular color codes. We began by identifying and addressing subtle caveats in FT design, outlining some design principles to incorporate a flag-based gadget inside a larger QEC protocol without compromising fault tolerance. We have emphasized the importance of dynamic circuit adaptation and flag-qubit-based techniques to effectively minimize overheads in circuit depths, and  introduced a sparse Pauli-frame simulation framework to numerically evaluate its performance. This has allowed us to efficiently simulate   in-sequence logic, as required for adaptive measurements and classical feed-forward, enabling in this way efficient large-scale numerical simulations of the noisy QEC circuits for lattice-surgery teleportation. This has lead to   benchmarks of various  QEC  strategies and decoding algorithms, including the scalable restricted minimum-weight perfect matching and belief propagation with ordered statistics, highlighting their relative performance and limitations when compared to small-distance lookup-table decoders. We have quantified the QEC footprint for both a quantum memory and a modular logical  teleportation, both simulated  under  a standard circuit-level depolarizing error model,  providing concrete resource comparisons of the scalability tradeoffs.

We have then moved to a more realistic modeling by explicitly  considering  architectural constraints and native  primitives of trapped-ion  QPUs. By performing detailed transpilation of the lattice-surgery teleportation circuits considering ion transport, gate operations, and flag-qubit-based FT gadgets, all combined with with detailed circuit-level noise modeling and efficient numerical simulation, we have provided a comprehensive assessment of the performance of three different  architectures for the lattice-surgery-based teleportation with two $d=3$ color codes.
Our analysis reveals that lattice surgery  enables the realization of a modular logical teleportation that can reach the high-fidelity regime with mild improvements in shuttling and intermediate cooling. The modularity here refers to the fact that  these methods do not  require applying direct interactions between distinct logical blocks, thereby mitigating crosstalk and preserving fault tolerance. While fully connected ion chains allow for flexible intra-block connectivity that might simplify some processing tasks, segmented modular traps with integrated photonics provide enhanced isolation and independent control of logical qubits, at the cost of increased ion shuttling and re-cooling overheads. Our numerical simulations show that the integrated-photonics approach will play a key role in the demonstration and scaling  of high-fidelity logical teleportation.

Overall, our work provides a quantitative methodology to guide experimental efforts and architecture design choices for modular trapped-ion  QPUs,  which we have illustrated with a   lattice-surgery logical teleportation   primitive that can be used in more complex fault-tolerant operations. This framework allows for quantitative architecture comparisons for quantum computation in near- and mid-term  trapped-ion  QPUs.
Looking ahead, further refinement of the noise models capturing transport-induced errors and laser control imperfections will be necessary to optimize protocol fidelity and runtime. Scaling the previous QEC footprint studies to larger code distances, but upgrading the previous depolarizing error model with these more sophisticated error models. Importantly,  the physics-aware transpilations will be crucial to extend this assessment  beyond small-distance  codes, which will be the focus of future work.
Additionally,  integrating advanced decoding methods and adaptive recalibration, including those leveraging machine learning, will be useful in the near future, as these are  expected to improve logical error thresholds. 

Moreover, hybrid approaches promise a path toward even larger universal FT trapped-ion quantum computing.
In trapped-ion systems, one natural direction is to embed larger-distance color-code blocks within segmented multi-zone traps, where logical patches are shuttled and merged via shared ancilla regions; alternatively, networked architectures could interconnect smaller blocks through photonic entanglement links or teleported stabilizer measurements. Another promising path involves hybrid modular arrays, in which local zones perform high-fidelity QEC while sparse interconnects enable occasional lattice-surgery operations between distant modules. These scenarios could reveal how trade-offs among transport overhead, crosstalk, and parallelism determine whether larger-distance codes provide a net gain in logical fidelity or latency. In the longer term, this line of investigation defines a coherent architecture roadmap: using the present teleportation benchmarks as calibration points for progressively refined  QEC performance analysis,  ultimately guiding the design of modular fault-tolerant processors capable of scaling logical QEC  to the regime of practical quantum computing  advantage.

\acknowledgments 
We gratefully acknowledge support by the Office of the Director of
National Intelligence (ODNI), Intelligence Advanced Research Projects Activity (IARPA), under the Entangled Logical Qubits program through Cooperative Agreement Number W911NF-23-2-0216.  We  acknowledge support from the European Union’s Horizon Europe research and innovation
program under Grant Agreement Number 101114305
(“MILLENION-SGA1” EU Project). CB and AB  acknowledge support from
PID2024-161474NB-I00 (MCIU/AEI/FEDER,UE), and from
QUITEMAD-CM TEC-2024/COM-84, from the
Grant IFT Centro de Excelencia Severo Ochoa CEX2020-
001007-S, funded by MCIN/AEI/10.13039/501100011033, and
from the CSIC Research Platform on Quantum Technologies PTI-001.

\appendix

\section{\bf Sparse Pauli frame simulator}\label{sec:sparsesim}

If restricting to Pauli noise and Clifford circuits, noisy simulations can be seen as the propagation of a Pauli frame through the circuit \cite{Gidney2021stimfaststabilizer}. This has the downside that it cannot provide measurement results, but only whether they have been flipped due to errors. It is possible to obtain the actual measurement results by sampling an additional shot using a full stabilizer simulation. However, for the specific QEC circuits that we will consider, all measurement results will either be 0 or random, which means that we can skip the reference sample by assuming instead that all measurement outcomes are 0 in absence of noise, since it is always a valid outcome for the circuits. A Pauli frame is simply a Pauli string indicating for every qubit if it suffered a bit-flip ($X$) and/or a phase flip ($Z$) error.

Updating the Pauli frame after each operation has a constant $O(1)$ overhead per shot, which means that taking $N_S$ shots of a circuit with $L$ operations has total $O(N_SL)$ overhead. However, if the physical error rate is low, many shots will not suffer any error and their Pauli frame will be trivial. For those shots, the Pauli frame remains unchanged during the circuit until an error happens. Thus, we only store in memory the shots for which an error has already happened. At the beginning of the circuit, no shot is stored in memory since all of them are error-free. Then, we simultaneously update Pauli frames from every shot advancing through the circuit:
\begin{itemize}
\item After a Clifford gate $U$ is applied, we update the Pauli frame $f$ by conjugation $f \to C^\dagger f C$. If the frame is trivial, there is nothing to update, so error-free shots have no computational nor storage cost. Shots with errors are updated in $O(1)$ operations.
\item After a noise instruction, we sample from a geometric distribution with rate $p$ equal to the error rate (for multi-component noise, this is the total probability of getting an error). Updating all shots in parallel allows sampling from a geometric distribution instead of a Bernoulli distribution, reducing the computational overhead of noise sampling from $O(N_\text{shots})$ to $O(N_\text{shots}p)$:
\begin{algorithmic}[1]
\State faulty\_shots $\gets$ Dictionary where key $j$ contains the Pauli frame for shot $j$ 
\State $i \gets \mathsf{sample\_geometric\_distribution}(p)$ 
\While{$i < N_\text{shots}$}
\If{ $i$ in faulty\_shots}
\State $f\gets$ \text{faulty\_shots}[i]
\Else
\State $f\gets$ Trivial Pauli frame
\EndIf
\State Update $f$ according to the noise operation (e. g. for depolarizing noise, randomly choose an error from $\{X,Y,Z\}$ and apply it to $f$).
\State $\text{faulty\_shots}[i]\gets f$ 
\State $i \gets i + 1 +\mathsf{sample\_geometric\_distribution}(p)$ 
\EndWhile
\end{algorithmic}
\item After a reset gate in qubit $q$, remove the $X_q$ term from the Pauli frame if it exists.
\item After a measurement in qubit $q$, if the Pauli frame contains a bit flip $X_q$ we store a $1$, otherwise we store a $0$.
\end{itemize}

After the $k$-th noise instruction in the circuit, there will be $O(\min\{kp,1\}N_\text{shots})$, so the overall cost of the circuit simulation is $O(\min\{n^2p,n\}N_\text{shots})$ if the circuit has $n$ instructions. As a comparison, a non-sparse shot representation has a $O(nN_\text{shots})$ overhead. Thus, the advantage of the sparse representation appears when the physical error rate is small $np<1$, so it is useful when exploring the very low error rate regime (e.g. to check the fault tolerance of a circuit). On the other hand, for realistic error rates $p\approx 10^{-3}$, the representation is not sparse anymore so both approaches become equivalent.

Using a sparse representation has one caveat: it does not check the validity of the noiseless circuit, as it bypasses the uncertainty principle (it allows non-commuting measurements). Thus, the circuit has to be validated first by other means, e. g. by randomizing the Z errors of all the qubits after qubit initialization and measurement, as is done in stim, or running a full stabilizer simulation. However, this only needs to be done once for each protocol.

The simulation supports in-sequence logic and feed-forward operations, where the protocol to be simulated is divided into multiple circuits following a tree-like structure (see Fig.~\ref{fig:circbranch}). When a circuit simulation is completed, we determine the next circuit to run depending on the measurement results from each shot. When the end of a circuit is reached, shots are sorted depending on the next circuit to be executed for each of them, and their respective Pauli frame is used as the initial state for the new circuit. By grouping shots together, their Pauli frames are updated simultaneously improving the performance.

The source code for our Pauli frame simulator is available at~\cite{framesim}.

\section{\bf QEC gadgets with adaptive dynamic circuits}

In this appendix, we detail the dynamic circuits used for various flag-qubit QEC gadgets, and how they adapt to mid-circuit measurements to reach FT while minimizing  depth. 

\subsection{Dynamic circuits for stabilizer measurements}
\label{app:stab_readout}

 We start by describing two adaptive schemes for FT syndrome extraction based on the flag-based circuit of Fig.~\ref{fig:flagged}. As already noted in the main text, these protocols employ flag qubits to detect and revert the propagation of dangerous correlated errors during syndrome extraction. We organize these building blocks to extract the full error syndrome according to two possible adaptive schemes, detailing the  tree-like  procedures that must be incorporated  in our sparse Pauli-frame  for a sequential and a simultaneous approach.
 
\subsubsection{Sequential FT stabilizer readout}
\label{app:stab_readout_sequential}

\begin{figure}
\begin{forest}
[Sequential QEC
[Flagged $S^X_1$,
    for tree={draw,calign=first, l sep+=10pt, s sep+=30pt}
  [Flagged $S^X_2$
    [Flagged $S^X_3$
      [Flagged $S^Z_1$
        [Flagged $S^Z_2$
          [Flagged $S^Z_3$
            [END,tier=end,draw=none]
            [Unflagged QEC,name=unflagged, edge label={node[midway,fill=white,font=\scriptsize]{[1]}}
              [END,tier=end,draw=none, edge label={node[midway,fill=white,font=\scriptsize]{[2]}}]
            ]
          ]
        ]
        {\draw[-] [out=south east,in=north]() to node[pos=0.3, fill=white, font=\scriptsize] {[1]} (unflagged);}
      ]
      {\draw[-] [out=south east,in=north]() to node[pos=0.3, fill=white, font=\scriptsize] {[1]} (unflagged);}
    ]
    {\draw[-] [out=south east,in=north]() to node[pos=0.3, fill=white, font=\scriptsize] {[1]} (unflagged);}
  ]
  {\draw[-] [out=south east,in=north]() to node[pos=0.3, fill=white, font=\scriptsize] {[1]} (unflagged);}
]
{\draw[-] [out=south east,in=north]() to node[pos=0.3, fill=white, font=\scriptsize] {[1]} (unflagged);}
]
\end{forest}
\begin{flushleft}
[1] Flipped flag or syndrome qubit.\newline
[2] Correct using the standard lookup table, using syndrome bits from the un-flagged circuit, taking into account whether a flag has been triggered.
\end{flushleft}
\caption{{\bf Tree-like structure  for  sequential  syndrome extraction}. We measure all $X$- and $Z$-type stabilizers one by one using flag qubits. When any syndrome or flag qubit is triggered (its measurement outcome is flipped), we are sure that an error has happened. In that situation, we have to re-measure all stabilizers again. For this stabilizer re-measurement, as an error has already happened, we can use un-flagged circuits, as a triggered flag in this second round would mean that two errors happened (the first one that forced us to remeasured, and the flagged error), which is a situation that cannot be corrected by the code. 
}\label{fig:sequential_qec}
\end{figure}
Figure~\ref{fig:sequential_qec} illustrates the sequential measurement protocol, where the \(X\)- and \(Z\)-type stabilizers are measured in a predetermined order using flagged circuits. Each flagged stabilizer measurement involves an ancillary syndrome qubit coupled to a flag qubit via additional CNOT gates designed to detect dangerous correlated errors. The protocol proceeds as follows:
\begin{enumerate}
    \item Measure the  stabilizers \(S_p^X,S_p^Z\), in plaquette order  $p=1,2,3$,  using the flagged circuits of Fig.~\ref{fig:flagged}.
    \item If any flag or syndrome qubit measurement indicates a fault (flag raised or syndrome flipped), immediately interrupt the sequence, and perform un-flagged stabilizer measurements using the circuits of Fig.~\ref{fig:xxxx} for all of the stabilizers as a corrective step.
    \item If no flags are triggered throughout the measurement sequence, the protocol proceeds to the next round of QEC using flag circuits, thus going back to step 1.
\end{enumerate}

The branching structure shown in Figure~\ref{fig:sequential_qec} encodes this conditional dynamic circuits, starting and the top and proceeding downwards. Each box named `Flagged $S_p^{\alpha}$' uses the  circuit in Fig.~\ref{fig:flagged} for the corresponding qubits,  resetting the syndrome and flag qubits, and also inverting the direction of the CNOTS and changing the basis of the ancillary qubits depending on the $X$ or $Z$ type measurement. The notation \([1]\) in the figure denotes branches in the logic taken when a flag or syndrome qubit indicates an error, prompting a switch to un-flagged measurements and thus an adaptation of the corresponding circuits. In this case, one uses the corresponding circuit of Fig.~\ref{fig:xxxx} without compromising FT-level $t=1$, as one knows that some type of error has already occurred. The notation \([2]\) represents the final error correction step using the standard lookup table~\ref{tab:lookup}, which incorporates syndrome information and flag outcomes to determine the appropriate correction.

\subsubsection{Simultaneous FT stabilizer readout}
\label{app:stab_readout_simultaneous}
\begin{figure}
\begin{forest}
[Simultaneous QEC
[Flagged $S^X$,
    for tree={draw,calign=first,l sep+=10pt, s sep+=20pt}
  [Flagged $S^Z$
    [END,tier=end,draw=none]
    [Unflagged QEC,name=unflagged, edge label={node[midway,fill=white,font=\scriptsize]{[1]}}
      [END,tier=end,draw=none, edge label={node[midway,fill=white,font=\scriptsize]{[2]}}]
    ]
  ]
]
{\draw[-] [out=south east,in=north]() to node[midway, fill=white, font=\scriptsize] {[1]} (unflagged) ;}
]
\end{forest}
\begin{flushleft}
[1] Flipped flag or syndrome qubit.\newline
\hspace{1ex}[2] Correct using the standard lookup table, using syndrome bits from the un-flagged circuit, taking into account whether a flag has been triggered.
\end{flushleft}
\vspace{-1ex}
\caption{{\bf Tree-like structure for simultaneous  syndrome extraction}: We begin by measuring all 
$X$-type stabilizers, followed by all 
$Z$-type stabilizers, using fault-tolerant flagged circuits. If none of the flag or syndrome qubits is triggered, the round is completed and the syndrome is directly used for decoding. However, if any syndrome qubit or flag qubit is triggered (i.e., its measurement outcome is non-trivial), this signals the possible occurrence of a fault that may have propagated to more than one data qubit. In this case, to preserve fault tolerance, we repeat the full stabilizer extraction, but this time using un-flagged circuits, as we are certain that at least one error has already occurred, even if we cannot yet unambiguously determine of which type. The combined information from the flagged and un-flagged rounds is then used to determine the appropriate correction.
}
\label{fig:parallel_qec}
\end{figure}

Figure~\ref{fig:parallel_qec} presents the simultaneous stabilizer measurement protocol. Here, flagged measurements of the \(X\)- or \(Z\)-type stabilizers are performed simultaneously in only two steps, reducing the overall QEC cycle time. The adaptive branching structure is simpler compared to the sequential case:
\begin{enumerate}
    \item Perform flagged measurements of all \(X\)-type or all \(Z\)-type stabilizer, using three copies of the circuit in Fig.~\ref{fig:flagged}, and deferring the projective measurements of  the 6 syndrome and flag qubits to a single step.
    \item If any flag or syndrome qubit measurement signals an error, immediately halt and perform un-flagged measurements of all code stabilizers using   the circuit in Fig.~\ref{fig:xxxx}.
    \item If no flags are triggered throughout the measurement
sequence, the protocol proceeds to the next round of
QEC using flag circuits, thus going back to step 1.
\end{enumerate}
Final error correction is performed as in the sequential protocol using the lookup table with flag information.
The parallel protocol offers reduced latency at the cost of potentially increased resources in qubit count,  but also maintains fault tolerance through the flag-based error detection and correction strategy.

\subsection{Dynamic circuits for state preparation}
\label{app:state_prep}
\begin{figure}
\begin{forest}
[State preparation
[Initialize $\ket{0}^{\otimes7}$,for tree={draw,calign=first,l sep+=10pt, s sep+=50pt}
[Flagged $S^X$
  [Flagged $S^X$
    [END,tier=end,draw=none, edge label={node[midway,fill=white,font=\scriptsize]{[3]}}]
    [Unflagged $S^X$ and $S^Z$,name=unflagged, edge label={node[midway,fill=white,font=\scriptsize]{[2]}}
      [END,tier=end,draw=none, edge label={node[midway,fill=white,font=\scriptsize]{[4]}}]
    ]
  ]
]
{\draw[-] []() to node[midway, fill=white, font=\scriptsize] {[1]} (unflagged);}
]
]
\end{forest}
\begin{flushleft}
[1] A flag has been triggered. We do not care about syndrome qubits as they will be random.\newline
[2] A flag has been triggered, or the syndrome at this round disagrees with the measured syndrome at the previous round.\newline
[3] The $X$ stabilizers are random when first measured. Thus, even if no error has been detected and both $S^X$ rounds agree, a correction has to be applied to some data qubits according to the lookup table, to force the state to the $+1$ eigenvalue of all $X$ stabilizers.\newline
[4] Correct using the standard lookup table, using syndrome bits from the un-flagged circuit, taking into account whether a flag has been triggered in either of the flagged rounds.
\end{flushleft}
\caption{Tree-like FT state preparation, valid for all cardinal states (including $\ket{\pm i}$) by appending a transversal gate.}\label{fig:stabilizer_prepare}
\end{figure}
Figure \ref{fig:stabilizer_prepare} presents a stabilizer measurement-based state preparation protocol. Here, data qubits are initialized in $\ket{0}$ and projected to a stabilizer eigenstate by performing flagged measurements of the $X$-type stabilizers. A second measurement of the $X$-type stabilizers is also required to correct hook errors. The adaptive branching structure is as follows:
\begin{enumerate}
    \item Perform flagged measurements of all $X$-type stabilizers, using three copies of the circuit in Fig.~\ref{fig:flagged}.
    \item If any flag qubit measurement signals an error, immediately halt and perform un-flagged measurements of all code stabilizers (including $Z$-type stabilizers) using the circuit in Fig.~\ref{fig:xxxx}.
    \item If no flag was triggered, perform flagged measurements of all $X$-type stabilizers again.
    \item If any flag qubit measurement signals an error, or syndrome differs from previous round, halt and perform un-flagged measurements, including both $X$- and $Z$-type stabilizers.
\end{enumerate}
The measured syndrome is random, but it should agree between rounds if no error happened. Thus, to project to a $+1$ eigenstate of all stabilizers, corrections have to be applied (even in absence of noise) according to the lookup table \ref{tab:lookup}, incorporating syndrome information and flag outcomes.
\subsection{Dynamic circuits for lattice-surgery merge-split}
\label{app:lattice_surgery}
As explained in the main text, merge and split steps are the core of the LS-based logical teleportation protocol, and they involve the more complex branching sequences of dynamic circuits to ensure fault tolerance. In Fig.~\ref{fig:circbranch}, we illustrate the protocol that we use for split and merge, including the full branching.

The joint measurement performed during the {\it merge step }  combines the two $d=3$ color codes by measuring weight-2 and weight-4 
$X$-type operators along the shared boundary. To ensure fault tolerance, both operators, which have random outcomes, must be measured twice to account for measurement errors, as depicted in the upper panel of Fig.~\ref{fig:circbranch}. 

We start measuring the weight-4 operator. If the first two measurements agree, no further repetition is necessary. On the other hand, if the outcomes  disagree, a third measurement is performed to confirm the syndrome.  Subsequent measurements after this confirmation, including those of the weight-4 operator, can be performed using un-flagged circuits. Importantly, the weight-4 operator FT  measurement is implemented with a single bare ancilla qubit, and  achieved by alternating CNOT gates between the two logical qubits to prevent error propagation within a single logical block, enabling the use of un-flagged measurement circuits. If both weight-4 measurements agree, the protocol proceeds to measure the weight-2 operators twice. Again, if the two measurements agree, no further measurement is needed. If the outcomes disagree, a third confirmation measurement is performed, and the protocol proceeds using un-flagged circuits.

Another key subtlety for maintaining fault tolerance is the need to measure the 
$X$-type stabilizers of both logical qubits after the joint measurements, denoted as 'flagged $S^X$' in the figure. This step identifies conjugate 
$Z$-type errors that may have occurred before or after the merge measurement. If such an error is detected before the joint measurement, the measurement outcome must be flipped during the decoding. This creates a complex in-sequence logic, requiring dynamic branching of the measurement protocol depending on intermediate flag and syndrome results.
The branching structure of the protocol, shown in the upper panel of the figure, describes the adaptive measurement sequence and corrective actions that minimize resource usage while fully preserving fault tolerance during the merging step.

\begin{figure*}
\begin{forest}
[MERGE
[$2\times$Measure $X_4^1X_5^1X_4^2X_5^2$,
  for tree={draw,calign=first,l sep+=10pt, s sep+=5pt}
  [$2\times$Measure $X_6^1X_6^2$
    [Flagged $S^X$
      [FT SPLIT,tier=split,draw=none,edge label={node[midway,fill=white,font=\scriptsize]{[2]}}]
      [Unflagged $S^X$,name=nftx, edge label={node[midway,fill=white,font=\scriptsize]{[3]}}
        [NFT SPLIT,tier=split,draw=none,edge label={node[midway,fill=white,font=\scriptsize]{[4]}}]
        [Measure $X_4^1X_5^1X_4^2X_5^2$, edge label={node[midway,fill=white,font=\scriptsize]{[5]}}
          [NFT SPLIT,tier=split,draw=none]
        ]
        [Measure $X_6^1X_6^2$, edge label={node[midway,fill=white,font=\scriptsize]{[6]}}
          [NFT SPLIT,tier=split,draw=none]
        ]
      ]
    ]
  ]
  {\draw[-] []() to node[midway,fill=white,font=\scriptsize]{[1]} (mxx3);}
  [Measure $X_4^1X_5^1X_4^2X_5^2$, edge label={node[midway,fill=white,font=\scriptsize]{[1]}}
    [Measure $X_6^1X_6^2$,name=mxx3
      [Unflagged $S^X$
        [NFT SPLIT,tier=split,draw=none, edge label={node[midway,fill=white,font=\scriptsize]{[7]}}]
      ]
    ]
  ]
]
]
\end{forest}
\begin{flushleft}
[1] Measurements disagree, measure a third time.\newline
[2] No error has been detected, the measure of $X_L^1X_L^2=(X_4^1X_5^1X_4^2X_5^2)(X_6^1X_6^2)$ is trustworthy.\newline
[3] Flipped flag or syndrome qubit, measure again without flags and trust this second syndrome measurement.\newline
[4] Error not in qubits 4, 5 or 6. Correct using the standard lookup table.\newline
[5] Measured syndrome compatible with error on qubit 4 or 5, determine if it happened before or after merging by measuring $X_4^1X_5^1X_4^2X_5^2$ a third time. If it agrees, the error happened before merging, and the stored value of $X_L^1X_L^2$ must be flipped with respect to the measured one. Correct the data qubits using the standard lookup table.\newline
[6] Syndrome compatible with error on qubit 6, proceed as in [5] but for $X_6^1X_6^2$.\newline
[7] Trust the last performed measurement of both $X_4^1X_5^1X_4^2X_5^2$ and $X_6^1X_6^2$. If the measured error syndrome gives an error in qubits 4, 5 or 6, flip the stored value of $X_L^1X_L^2$ with respect to the measured one. Correct the data qubits using the standard lookup table.
\end{flushleft}
\begin{forest}
[{},draw=none
[FT SPLIT,edge={draw=none}
  [Measure $Z^1_L$, for tree={draw,calign=first,l sep+=10pt, s sep+=29pt}
    [Flagged $S^Z$ in qubit 2
      [END,tier=end,draw=none, edge label={node[midway,fill=white,font=\scriptsize]{[9]}}]
      [Unflagged $S^Z$ in qubit 2,name=nftqecz, edge label={node[midway,fill=white,font=\scriptsize]{[8]}}
        [END,tier=end,draw=none, edge label={node[midway,fill=white,font=\scriptsize]{[9]}}]
      ]
    ]
  ]
]
[NFT SPLIT,edge={draw=none}
  [Measure $Z^1_L$, for tree={draw,calign=first,l sep+=10pt, s sep+=5pt}]
  {\draw[-] []() to (nftqecz);}
]
]
\end{forest}
\begin{flushleft}
[8] Flipped flag or syndrome qubit. During merge, $Z^2_2Z^2_3Z^2_5Z^2_6$ is not a stabilizer and will give a random outcome even in absence of noise. To determine if the related syndrome bit has been flipped, consider instead whether the product $(Z^1)_2(Z^1_3)(Z^1_5)(Z^1_6)(Z^2_2Z^2_3Z^2_5Z^2_6)\equiv Z_{W8}$ of measurement outcomes is flipped, as $Z_{W8}$ remains a stabilizer for the whole protocol. \newline
[9] After splitting, both logical qubits must be jointly decoded, due to the shared stabilizer $Z_{W8}$. If $Z^1_2Z^1_3Z^1_5Z^1_6=Z^2_2Z^2_3Z^2_5Z^2_6=-1$, an additional correction $X^1_6X^2_6$ has to be applied to flip both of them to $+1$ without affecting $Z_{W8}$. The same $X^1_6X^2_6$ correction has to be performed in the case where $Z_{W8}=-1$, the flipped plaquette is $Z^1_2Z^1_3Z^1_5Z^1_6$ and no more errors (i. e. triggered flags or other stabilizers) are flipped in qubit 1 (and likewise for qubit 2). Additionally, if coming from [3] or [8], a $X$ or $Z$ flag could have been triggered, respectively. Correct the measurement records of logical qubit 1 and data qubits of logical qubit 2 using the appropriate column of the lookup table. For the $Z_2Z_3Z_5Z_6$ syndrome bit, use the (random) measured ones, flipping the corresponding bit if the aforementioned $X^1_6X^2_6$ correction has been applied.
\end{flushleft}
\caption{Detailed branching structure of the split and merge steps of the logical teleportation protocol for the Steane code.}
\label{fig:circbranch}
\end{figure*}

Let us now move to the tree-like structure of the lattice-surgery {\it split step}. This is implemented as the projective measurement in the $Z$ basis of the source qubit, but must be accompanied by a \(Z\)-type QEC round on the remaining logical qubit that will encode the teleported state. This requires projective measurements of the two \(Z\)-type stabilizers that were combined during the merge. Although the individual measurement outcomes of these stabilizers appear random , their product corresponds to a higher-weight merged stabilizer that remains fixed in the absence of errors. Decoding must respect this constraint when interpreting syndrome data.

\begin{itemize}
    \item If an error was detected during the merge step, the split uses un-flagged \(Z\)-type stabilizer measurements. The decoding incorporates prior flag information to identify and correct errors.
    
    \item If no error was detected in the merge, two rounds of stabilizer measurements are performed during the split:
    \begin{enumerate}
        \item The first round employs FT flagged measurements.
        \item If errors are detected, a second round of un-flagged measurements resolves ambiguities.
    \end{enumerate}
\end{itemize}

In the lower panel of Fig.~\ref{fig:circbranch}, we detail the adaptive logic that must be applied to minimize resources while maintaining fault tolerance. Here, the rounds of QEC are labeled as `(un)flagged $S^Z$', discussing various fine details on how this adaptive branching protocol ensures fault tolerance during the split step while minimizing measurement overhead and resource consumption. 

\newpage
\newpage
\bibliography{biblio}
\end{document}